\documentclass[aps,reprint,longbibliography]{revtex4-2}
\usepackage[T1]{fontenc}
\usepackage{graphicx}
\usepackage{amssymb}
\usepackage{xcolor}
\usepackage[fleqn]{amsmath}
\usepackage[font=small,skip=3pt]{caption}
\usepackage{subcaption}
\usepackage{booktabs}
\PassOptionsToPackage{hyphens}{url}
\usepackage[breaklinks=true]{hyperref}
\usepackage{cleveref}
\usepackage{xurl}
\usepackage{bm}
\newif\ifdark
\ifdark
  \pagecolor{black}
  \makeatletter
  
  \AtBeginDocument{%
    \color{white}%
    \g@addto@macro\frontmatter@title@format{\color{white}}%
    \g@addto@macro\frontmatter@authorformat{\color{white}}%
    \g@addto@macro\frontmatter@affiliationfont{\color{white}}%
    \g@addto@macro\frontmatter@abstractfont{\color{white}}%
    \g@addto@macro\frontmatter@RRAP@format{\color{white}}%
    \let\rvtx@orig@maketitle\maketitle
    \def\maketitle{\rvtx@orig@maketitle\color{white}}%
  }
  \makeatother
\fi

\let\mrm\mathrm

\newcommand{\thalf}{\ensuremath{t_{1/2}}}
\newcommand{\ntn}{\ensuremath{(\mrm{n},2\mrm{n})}}
\newcommand{\nthn}{\ensuremath{(\mrm{n},3\mrm{n})}}

\newcommand{\ngamma}{\ensuremath{(\mrm{n},\gamma)}}

\newcommand{\betam}{\beta^{-}}


\begin{document}

\title{Scalable production of nuclear battery alpha emitters using fusion neutrons}

\author{J.~F.~Parisi}
\email{jason@marathonfusion.com}
\affiliation{Marathon Fusion, 150 Mississippi Street, San Francisco, CA 94107, USA}

\begin{abstract}
Nuclear batteries powered by alpha decay have been deployed successfully for over 60 years, on a worldwide ${}^{238}$Pu supply of kilograms per year. We show that the 14 MeV neutrons of a single deuterium-tritium fusion plant can produce alpha emitter battery fuels up to \textit{tons} per year, in three classes: fuels with completely new production pathways (${}^{236}$Pu, ${}^{227}$Ac, ${}^{210}$Pb), fuels proposed in the 1960s whose scarce feedstock the same pathways now breed at scale (${}^{232}$U, ${}^{228}$Th), and the established ${}^{238}$Pu. OpenMC simulations of actinide channels in a tokamak blanket give, per GW$_\mathrm{fus}$\,yr of fusion: 11 to 57 kg of ${}^{236}$Pu, whose chain releases 18 gigajoules per gram over a century, ending at stable ${}^{208}$Pb, plus up to 5.2 t of co-product ${}^{238}$Pu; up to 1.4 t of ${}^{231}$Pa from thorium, and, from channels loaded with ${}^{231}$Pa, up to $\sim$15 t of ${}^{232}$U or $\sim$122 kg of ${}^{210}$Pb, with ${}^{227}$Ac produced at 21 g yr${}^{-1}$ per tonne of ${}^{231}$Pa. Neutron capture also upgrades ${}^{241}$Am to a ${}^{242}$Cm/${}^{242\mathrm{m}}$Am/${}^{241}$Am/${}^{238}$Pu blend with up to 10 times higher power density. The same ${}^{236}$Pu and ${}^{232}$U used for batteries also function as proliferation safeguards: the ${}^{237}$Np, ${}^{232}$Th, and ${}^{231}$Pa channel products are self-protecting, the plutonium by ${}^{236}$Pu and ${}^{238}$Pu decay heat and the 2.6 MeV gammas from ${}^{208}$Tl content, and similarly the uranium from its ${}^{232}$U. Many of these fuels (${}^{236}$Pu, ${}^{232}$U, ${}^{228}$Th, ${}^{227}$Ac) have an order of magnitude higher power and energy density than current alpha emitters, and at human spaceflight-relevant doses the ${}^{227}$Ac and ${}^{210}$Pb chains need less shield mass than ${}^{238}$Pu or ${}^{241}$Am above a few hundred watts. Fusion neutrons could therefore enable nuclear batteries at the kilowatt to megawatt scale and unlock new possibilities for power sources requiring exceptionally high energy density.
\end{abstract}
\maketitle

\section{Introduction}

\begin{figure}[!tb]
\centering
\includegraphics[width=\columnwidth]{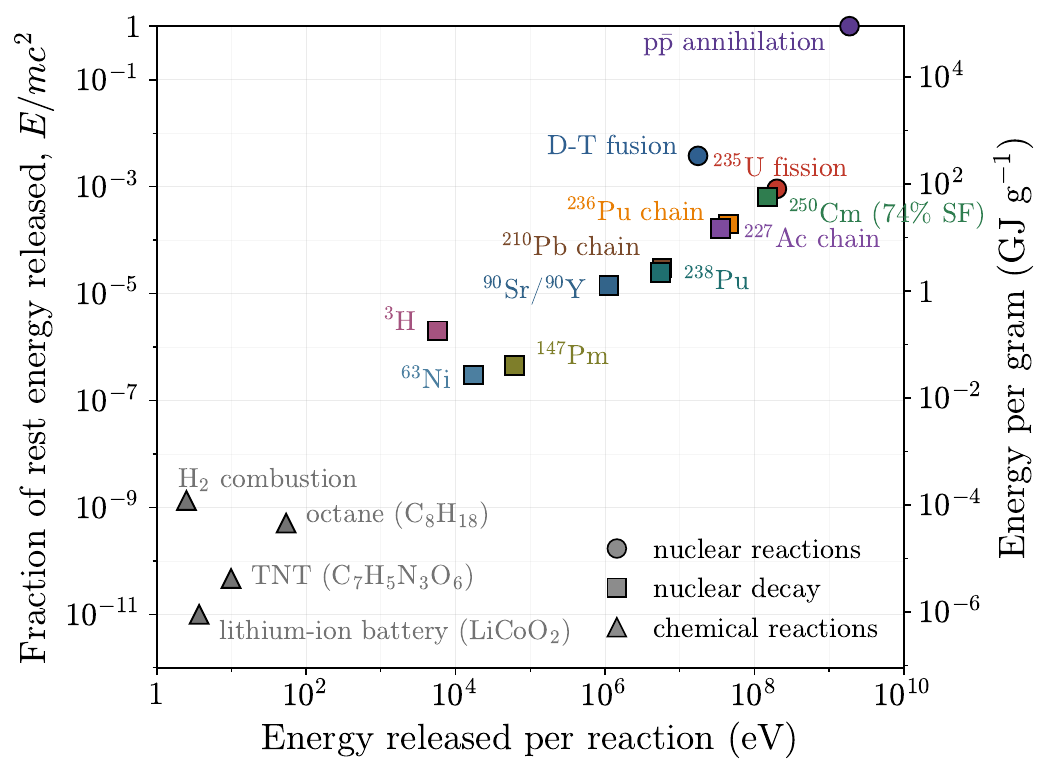}
\caption{Fraction of fuel rest energy released as usable heat versus energy released per reaction grouped as nuclear reactions, nuclear decay, and chemical reactions.}
\label{fig:mc2_ladder}
\end{figure}

The ability to store and harness energy at increasingly high energy and power density has unlocked new capabilities throughout human history~\cite{layton2008comparison,smil2015power}. At the upper end, the energy density of matter itself is $\sim$90 terajoules per gram, enough for a single gram to raise the temperature of 21.5 million tons of water by 1 degree Kelvin. While releasing such a high energy density is only known to be possible with matter-anti-matter annihilation, the next known highest tier of energy release is via nuclear reactions: ${}^{235}$U fission releases 82 gigajoules per gram of ${}^{235}$U, and deuterium-tritium (D-T) fusion releases 340 gigajoules per gram of DT. Chemical reactions, by comparison, release only $\sim$10 kilojoules per gram. Between these extremes lies radioactive decay, which releases $\sim$keV-MeV of nuclear energy per decay, but spontaneously, often over years to centuries. $\beta^-$ emitters sit at the lower end of energy, because the antineutrino released in each decay escapes with most of the energy: $\beta^-$ emitter ${}^{147}$Pm deposits 40 megajoules per gram as it decays to stable ${}^{147}$Sm over several years, while ${}^{90}$Sr, a legacy radioisotope thermoelectric generator (RTG) fuel, reaches 1.2 gigajoules per gram, already tens of thousands of times higher than any chemical fuel. Alpha decay emits no neutrino, so alpha emitters lose almost nothing, only the fraction lost at the occasional beta step of a chain, and can reach another order of magnitude higher energy density. \Cref{fig:mc2_ladder} compares these pathways by the fraction of fuel rest energy released and the energy per reaction.

\begin{figure*}[!tb]
\centering
\includegraphics[width=0.98\textwidth]{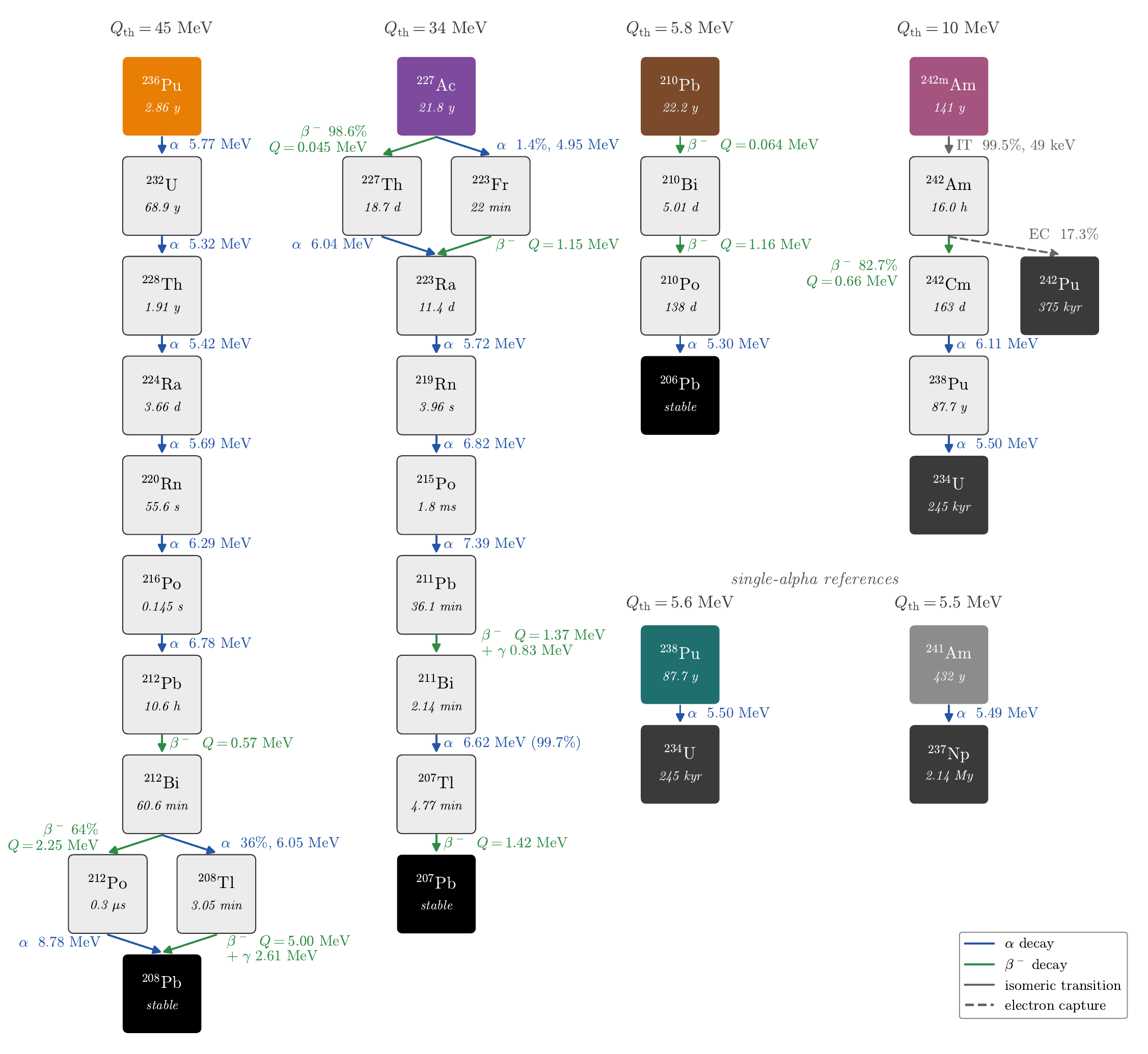}
\caption{The main decay chains considered in this paper, with average energy per chain $Q_\mathrm{th}$, excluding any radionuclide with a half-life greater than 500 years. Neutrino energy is excluded.}
\label{fig:decay}
\end{figure*}

The energy density of some alpha emitter chains is immense: $^{236}$Pu decays to stable $^{208}$Pb over $\sim$100 years, releasing 45 MeV of usable energy, corresponding to an energy density of 18 \emph{gigajoules} per gram of fuel. This is equivalent to releasing 0.020\% of the $^{236}$Pu rest mass into energy, roughly a fifth of the 0.091\% of rest mass released by ${}^{235}$U fission, so letting a gram of $^{236}$Pu decay over $\sim$100 years releases as much energy as fissioning a fifth of a gram of ${}^{235}$U. A $^{236}$Pu fuel can therefore access energy densities comparable to fission fuel by radioactive decay alone. Some alpha emitter chains therefore contain energy densities comparable to fissioning a significant fraction of fissile material, but without the nuclear waste and complexity of operating a fission reactor. The conventional alpha emitter for nuclear batteries $^{238}$Pu releases 5.6 MeV, converting 0.0025\% of its rest mass to energy, eight times lower than $^{236}$Pu, but still a million times higher than a lithium-ion battery, which stores $\sim$1 kilojoule per gram of battery material. 

In this work we describe scalable methods for producing battery fuels such as $^{236}$Pu, $^{210}$Pb, and ${}^{227}$Ac using fast neutrons from fusion reactions. To our knowledge, these production pathways have not been reported in prior literature. The same blankets also supply ${}^{232}$U and ${}^{228}$Th, whose production pathways were proposed at Hanford in 1960 but never supplied beyond gram scale for lack of neutron capacity~\cite{Rohrmann1960,Rohrmann1963}.

\begin{figure*}[!t]
\centering
\includegraphics[width=\textwidth]{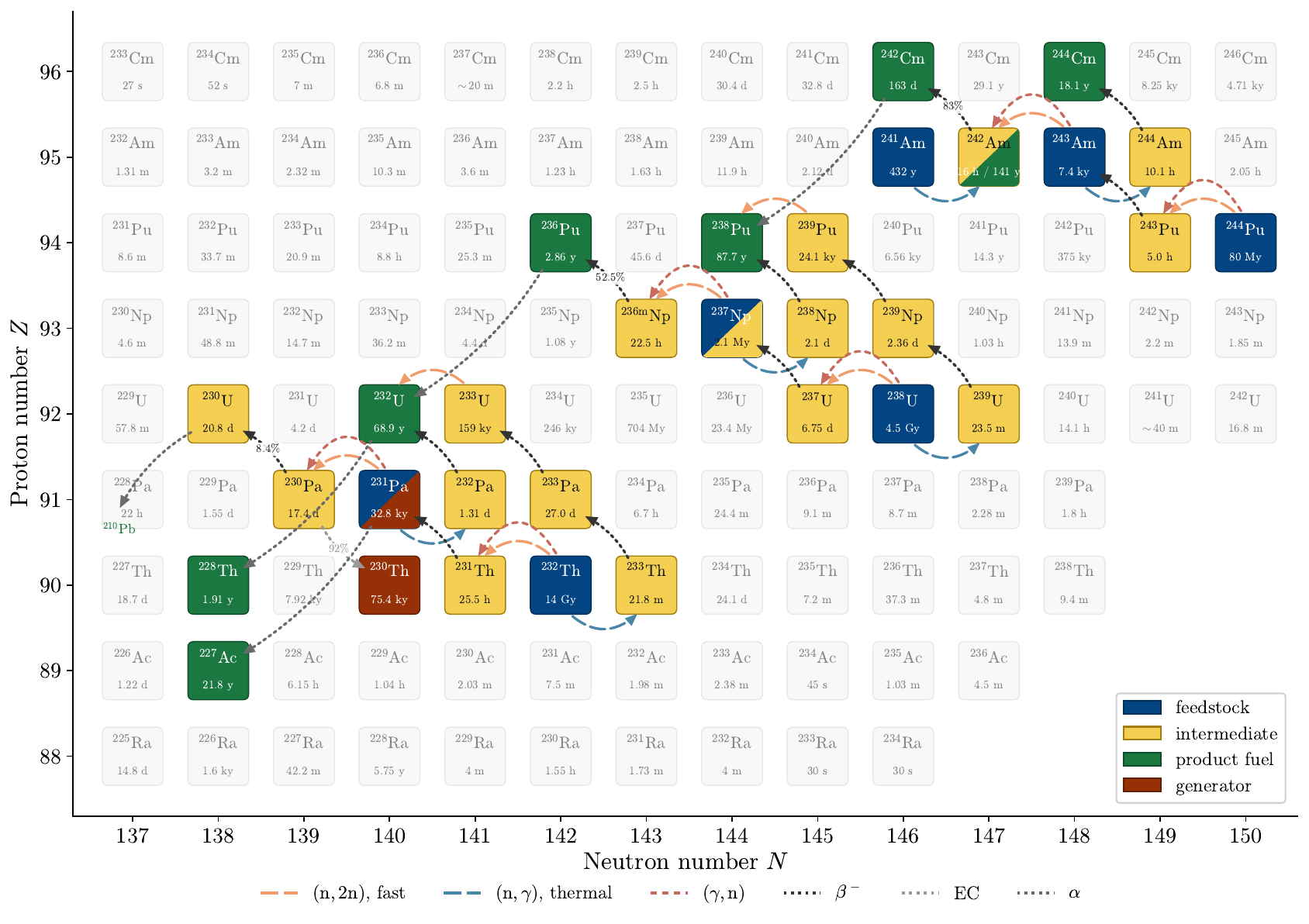}
\caption{Transmutation pathways of \Cref{tab:pathways} on the chart of nuclides. Boxes are colored by role (feedstock, intermediate, product fuel, generator); the outer arcs mark photonuclear ($\gamma$,n) alternatives to the \ntn\ steps.}
\label{fig:pathways_chart}
\end{figure*}

RTGs are the main nuclear battery qualified for multi-decade deep-space missions, long-lifetime maritime systems, and remote terrestrial sensors~\cite{Cataldo2011,Ambrosi2019,Prelas2016,Lange2008}, and commercial interest in nuclear batteries is accelerating~\cite{ChemWorld2024,IEEESpectrum2025,Arkenlight2024,CityLabs}. The standard RTG fuel, ${}^{238}$Pu, can maintain specific thermal power of $\sim$0.57 W/g for decades, and production is ramping toward the 1.5 kg per year requested by NASA~\cite{SpaceNewsPu238,ORNLPu238,INLPu238,DOEIsotope2025,Pan2025Pu238}. Both ${}^{238}$Pu and ESA's ${}^{241}$Am alternative \cite{Barco2019} emit one alpha per atom, with a daughter that produces essentially no further heat over the lifetime of a battery (\Cref{fig:decay}). ${}^{238}$Pu is obtained from neutron capture on ${}^{237}$Np, and ${}^{241}$Am is separated from spent nuclear fuel. For missions above roughly a kilowatt of electric power, studies have turned to compact fission reactors such as NASA's Kilopower~\cite{Gibson2017,Poston2020KRUSTY} and Fission Surface Power~\cite{oleson2022deployable,kaldon2023overview}, in large part because the ${}^{238}$Pu supply cannot sustain RTGs at that scale; the fuel supplies opened in this work remove that constraint and put kilowatt-and-above missions within reach of nuclear batteries.

One of our main focuses among alternative alpha emitters is the ${}^{236}$Pu decay chain, which provides an order of magnitude more energy and power density than existing RTG fuels.  ${}^{236}$Pu sits at the top of a nine-step decay chain to stable ${}^{208}$Pb (\Cref{fig:decay}) that emits seven alphas and two betas, depositing 45 MeV of thermal energy in the fuel, roughly eight times the 5.6 MeV per ${}^{238}$Pu atom. Fresh ${}^{236}$Pu has a specific power of $\sim$18 W/g; the output stays above 5 W/g for the first decade and then falls off slowly on the 68.9 yr ${}^{232}$U half-life. The main challenge with ${}^{236}$Pu is the ${}^{208}$Tl daughter at the bottom of the chain, whose 2.6 MeV gamma is hard to shield~\cite{Cataldo2011}. For this reason ${}^{236}$Pu is deliberately removed from ${}^{238}$Pu fuel during purification~\cite{Nelson2023}. While we believe it is challenging to produce isotopically pure ${}^{236}$Pu using fusion neutrons at scale due to ${}^{238}$Pu co-production, producing ${}^{236}$Pu and allowing it to decay to ${}^{232}$U provides much higher isotopic purity of ${}^{232}$U relative to other uranium isotopes. ${}^{232}$U's daughter, ${}^{228}$Th, can be produced with even higher isotopic purity, which is exceptionally useful for nuclear batteries and targeted alpha therapy~\cite{Yong2015,Kokov2022,Pretze2025}.

Despite significant shielding challenges for the ${}^{236}$Pu decay chain, two developments motivate a second look. First, in this paper we show that ${}^{236}$Pu can be produced at scale with fast D-T fusion neutrons, which drive ${}^{237}$Np\ntn${}^{236\mrm{m}}$Np in a blanket at up to $\sim$57 kg of ${}^{236}$Pu per GW$_\mrm{fus}$-yr (\Cref{sec:production}). This is a scalable alternative to fission-neutron irradiation of ${}^{237}$Np to make ${}^{238}$Pu \cite{urban2021initial}, and to charged-particle routes, which produce insufficient quantities for nuclear batteries~\cite{Aaltonen2003,Artun2020}. Second, the ${}^{236}$Pu decay chain includes the noble gas ${}^{220}$Rn, which is mobile and escapes from porous materials. The ${}^{236}$Pu decay chain could therefore be physically separated at ${}^{220}$Rn: the upstream four-alpha chain ${}^{236}$Pu $\to$ ${}^{232}$U $\to$ ${}^{228}$Th $\to$ ${}^{224}$Ra $\to$ ${}^{220}$Rn stays in the main fuel and releases 22 MeV per chain, while the daughters below ${}^{220}$Rn can be diverted to a shielded chamber, vented overboard in deep space where the exhaust can harm nothing, or piped to a chamber on a boom. This paper treats the $\alpha$ emitters; a recent companion paper~\cite{parisi2026betaemitters} surveys nuclear battery $\beta^-$ emitters producible in fusion blankets~\cite{Rutkowski2025}. Fusion neutron-driven medical radioisotope production has also been studied~\cite{engholm1986radioisotope,Bourque1988FAME,Ridikas2006_HybridWaste,Leung2018_CompactNG,pietropaolo2021sorgentina,li2023feasibility,Honney2023FusionNeutrons,pereslavtsev2024potential,Parisi2025,Evitts2025}.

Loading a fusion blanket with tonnes of actinides poses a proliferation risk. Work has shown that placing 5 to 50 t of ${}^{238}$U or ${}^{232}$Th in an ARC-class~\cite{Sorbom2015} FLiBe blanket breeds a significant quantity of ${}^{239}$Pu or ${}^{233}$U in weeks to months at better than 99\% isotopic purity~\cite{Ball2025,GlaserGoldston2012}. The blankets described here do not mix ${}^{238}$U or ${}^{232}$Th into the FLiBe but hold them in separate channels. Two features of these materials work against proliferation: the intense fission environment the material sits in and the 2.6 MeV ${}^{208}$Tl gamma that makes ${}^{236}$Pu difficult to handle as a battery fuel in the first place. However, neither removes the need to design with safeguards. We discuss this in Appendix \ref{app:u233}, showing that the proliferation concerns with a blanket containing ${}^{238}$U are significant, breeding several tons of high-purity ${}^{239}$Pu annually, but ${}^{232}$Th and ${}^{237}$Np blankets appear much more manageable due to self-protecting plutonium and uranium radioisotopes.

\begin{figure*}[!tb]
\centering
\includegraphics[width=\textwidth]{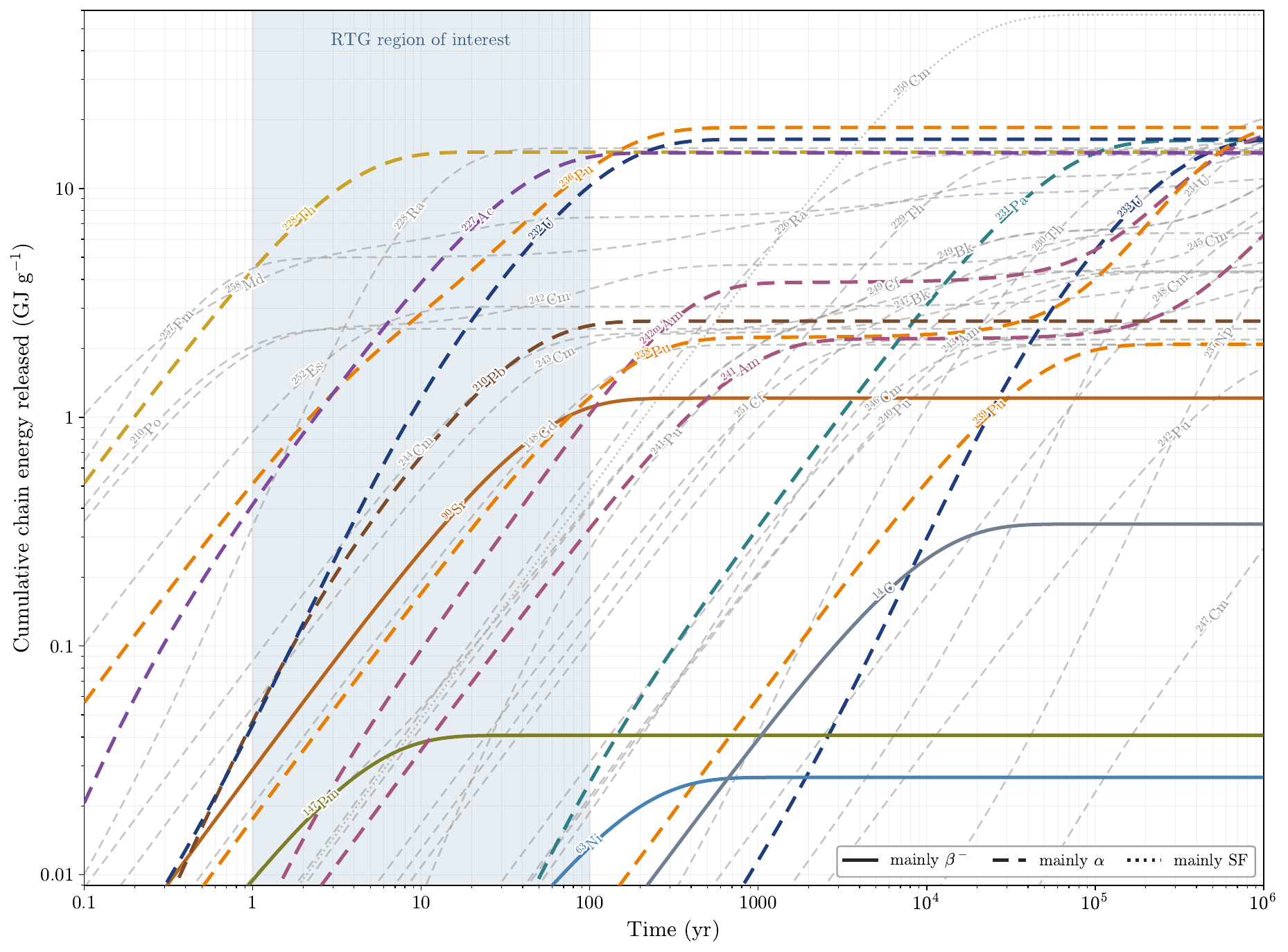}
\caption{Cumulative chain energy released per gram versus time for fuels starting from a given radionuclide and following the full downstream chain (radionuclides with chains releasing more than half their energy within 0.1 yr are omitted). Energies are locally deposited heat. Color marks fuels with scalable production pathways; gray curves are the remaining $\alpha$ emitters. Linestyle gives the dominant contribution: solid $\beta^-$, dashed $\alpha$, dotted spontaneous fission (${}^{250}$Cm).}
\label{fig:alpha_release}
\end{figure*}

We organize this work as follows. In \Cref{sec:chain} we compare several alpha emitter chains. In \Cref{sec:production} we present OpenMC blanket simulations for scalable alpha emitter production. In \Cref{sec:shieldmass} we compare the shield mass each fuel requires, and in \Cref{sec:radon} we describe radon separation to minimize the ${}^{208}$Tl dose from the ${}^{236}$Pu chain. We conclude in \Cref{sec:discussion}. 

Appendices contain additional information on the production pathways (\ref{app:pathways}), blended power profiles (\ref{app:blends}), curium from americium (\ref{sec:curium}), ${}^{210}$Pb (\ref{app:pb210}), millennial heat sources (\ref{app:pa231}), ${}^{236}$Pu isotopic purity (\ref{app:purity}), proliferation and safeguards (\ref{app:u233}), OpenMC depletion simulations (\ref{app:depletion}), a Mars logistics trip (\ref{app:mars}), and decay data (\ref{app:decaydata}).

As a big-picture summary, \Cref{fig:pathways_chart} shows every production pathway developed in this paper on the chart of nuclides. We will refer back to \Cref{fig:pathways_chart} often throughout this paper.

\section{Properties of Alpha Decay Chains} \label{sec:chain}

We first compare the decay properties of alpha emitter chains. We consider production of four main chains, headed by ${}^{236}$Pu, ${}^{210}$Pb, ${}^{227}$Ac, and ${}^{242 \mathrm{m} }$Am, each shown in \Cref{fig:decay}. For completeness, we also include chains headed by ${}^{238}$Pu and ${}^{241}$Am. ${}^{238}$Pu can be produced at high rate by thermalizing fast neutrons, and ${}^{241}$Am appears to be much better suited for production in fission reactors.

\subsection{Chain Overview}

We first summarize the properties of several decay chains. Detailed decay data for the six chains compared here are shown in Appendix \ref{app:decaydata}.

\emph{${}^{236}$Pu chain}: The ${}^{236}$Pu chain has the highest released energy with an average of 45 MeV, ending at stable ${}^{208}$Pb, and featuring the 2.6 MeV ${}^{208}$Tl gamma from the 36\% alpha decay branch of ${}^{212}$Bi. The 2.6 MeV ${}^{208}$Tl gamma requires significant shielding~\cite{Nelson2023}. However, the immense energy density of the ${}^{236}$Pu chain motivates searching for practical shielding solutions.

\emph{${}^{210}$Pb chain}: The ${}^{210}$Pb chain releases an average of 5.8 MeV, ending at stable ${}^{206}$Pb. Almost all of it is the 5.41 MeV ${}^{210}$Po alpha: the ${}^{210}$Pb and ${}^{210}$Bi betas average only 38 and 389 keV. It is nearly free of gamma lines: its only lines are the 46.5 keV photon of ${}^{210}$Pb (4\%) and the 803 keV ${}^{210}$Po line at $10^{-3}$\,\% intensity. Its shielding requirement is instead set by bremsstrahlung from the 1.16 MeV ${}^{210}$Bi beta. A spent source is stable ${}^{206}$Pb, reducing long-term radioactive waste issues compared with alpha chains such as ${}^{238}$Pu and ${}^{241}$Am that produce long-lived ${}^{234}$U and ${}^{237}$Np. The ${}^{210}$Pb can be thought of as a long-lived generator for ${}^{210}$Po, which has a much shorter half-life and has been considered as an RTG fuel~\cite{blanke1960nuclear}. This is somewhat analogous to ${}^{90}$Sr~\cite{kumar2015atomic} as a long-lived generator for ${}^{90}$Y. Milking the lead gives carrier-free ${}^{210}$Bi. While scalable ${}^{210}$Pb production has been previously declared inaccessible ~\cite{blanke1960nuclear}, in this work we will show a scalable production pathway.

\emph{${}^{227}$Ac chain}: The ${}^{227}$Ac chain releases an average of 35 MeV, ending at stable ${}^{207}$Pb through ${}^{207}$Tl. The hardest significant line is the 832 keV gamma accompanying the ${}^{211}$Pb decay, a factor of three below the ${}^{208}$Tl line from the ${}^{236}$Pu chain and much less penetrating. Unfortunately, the 4.0 s ${}^{219}$Rn half-life is likely too short for many of the radon separation techniques suggested in \Cref{sec:radon} for the ${}^{236}$Pu chain, meaning the chain cannot avoid gamma emitters. In \Cref{sec:u232} we will show that ${}^{227}$Ac is producible at scale by milking a ${}^{231}$Pa stockpile produced in large quantities from fast-neutron irradiation of ${}^{232}$Th.

\emph{${}^{242\mathrm{m}}$Am chain}: The ${}^{242\mrm{m}}$Am chain releases an average of 10 MeV, generating ${}^{242}$Cm and ${}^{238}$Pu over its 141 yr half-life before ending at long-lived ${}^{234}$U (245 kyr). The 17.3\% electron-capture branch produces long-lived ${}^{242}$Pu (375 kyr). Its photons are soft (the 49 keV isomeric transition and the 44 keV ${}^{242}$Cm line), so its shielding is set not by gammas but by the spontaneous-fission and ($\alpha$,n) neutrons of the ${}^{242}$Cm it breeds, which may require moderating layers rather than dense metal~\cite{Nelson2023}. Production routes for the americium and curium fuels are given in Appendix \ref{sec:curium}.

\emph{${}^{238}$Pu chain}: ${}^{238}$Pu releases an average of 5.6 MeV, a single alpha onto long-lived ${}^{234}$U. Its own photons are negligible (a 43 keV line at $4\times10^{-2}$\,\%). In flight-qualified fuel the dose is instead set by trace ${}^{236}$Pu feeding ${}^{208}$Tl and by ($\alpha$,n) neutrons in the oxide, so pure-${}^{238}$Pu shielding is neutron-limited rather than photon-limited~\cite{Nelson2023}. Given the very low dose rates, NASA's ${}^{238}$Pu RTGs are unshielded.

\emph{${}^{241}$Am chain}: ${}^{241}$Am releases an average of 5.6 MeV, a single alpha into ${}^{237}$Np (2.14 My). Its main photon is the 60 keV line (36\% per decay), soft enough that a fraction of a millimeter of lead attenuates it ten times. ${}^{241}$Am is the ESA's RTG fuel despite five times lower specific power than ${}^{238}$Pu~\cite{Ambrosi2019,AMPPEX}.

\subsection{Chain Power Profiles}

\begin{figure}[!tb]
\centering
\includegraphics[width=\columnwidth]{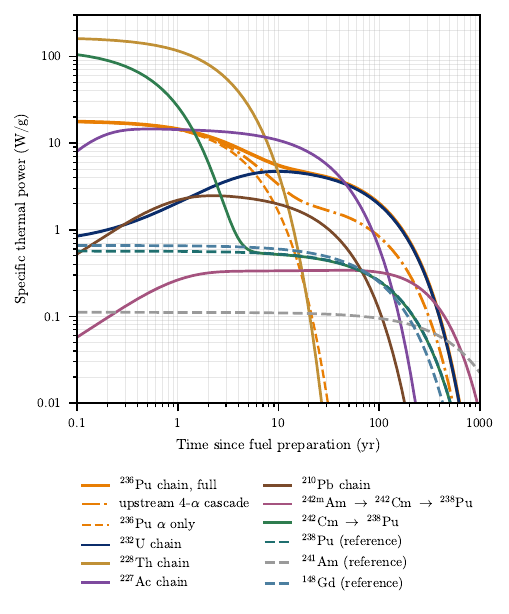}
\caption{Thermal power per gram of initial fuel versus time, for the ${}^{236}$Pu chain (full, its upstream four-alpha chain, and the ${}^{236}$Pu $\alpha$ alone), the ${}^{232}$U, ${}^{228}$Th, ${}^{227}$Ac, ${}^{210}$Pb, ${}^{242\mrm{m}}$Am, and fast-start ${}^{242}$Cm chains, and the ${}^{238}$Pu, ${}^{241}$Am, and ${}^{148}$Gd references (legend).}
\label{fig:specific_power}
\end{figure}

\begin{figure}[!tb]
\centering
\begin{subfigure}{\columnwidth}
\centering
\includegraphics[width=\columnwidth]{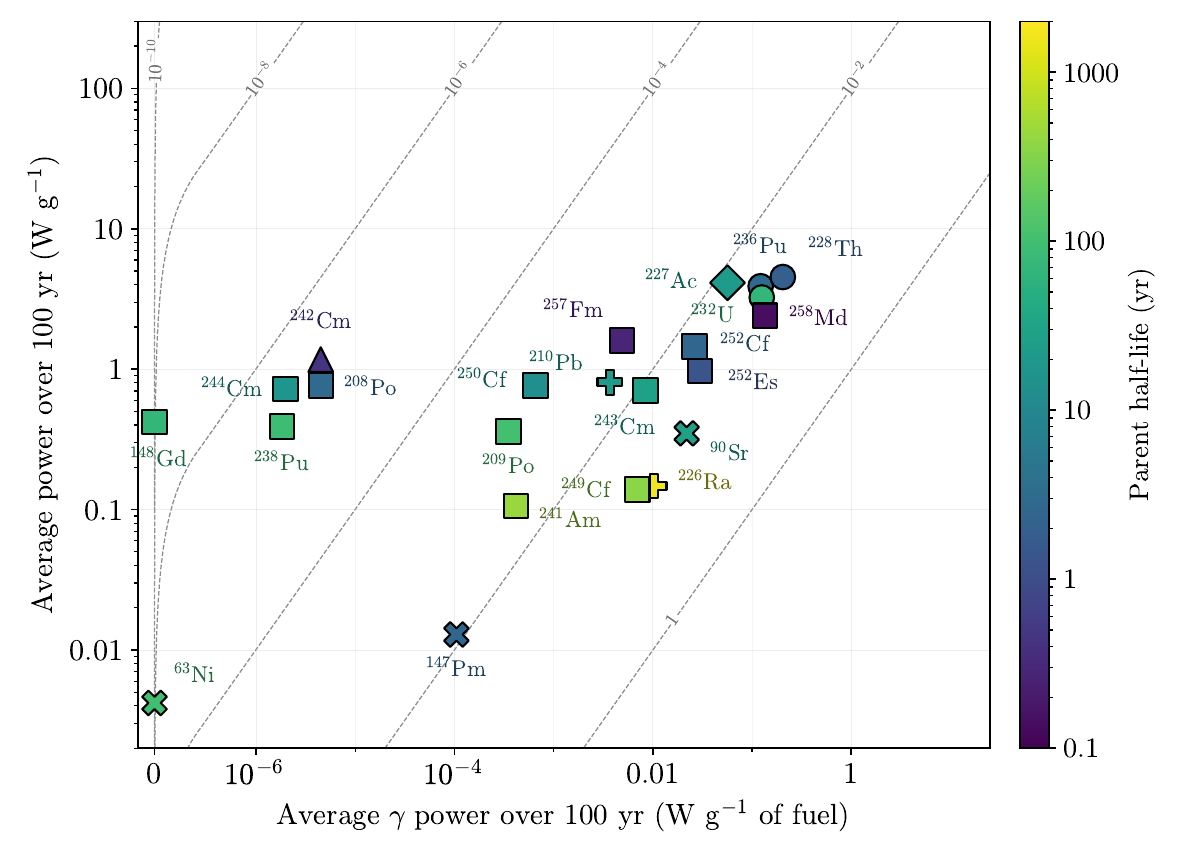}
\caption{}
\label{fig:cascade_search_a}
\end{subfigure}\\[2pt]
\begin{subfigure}{\columnwidth}
\centering
\includegraphics[width=\columnwidth]{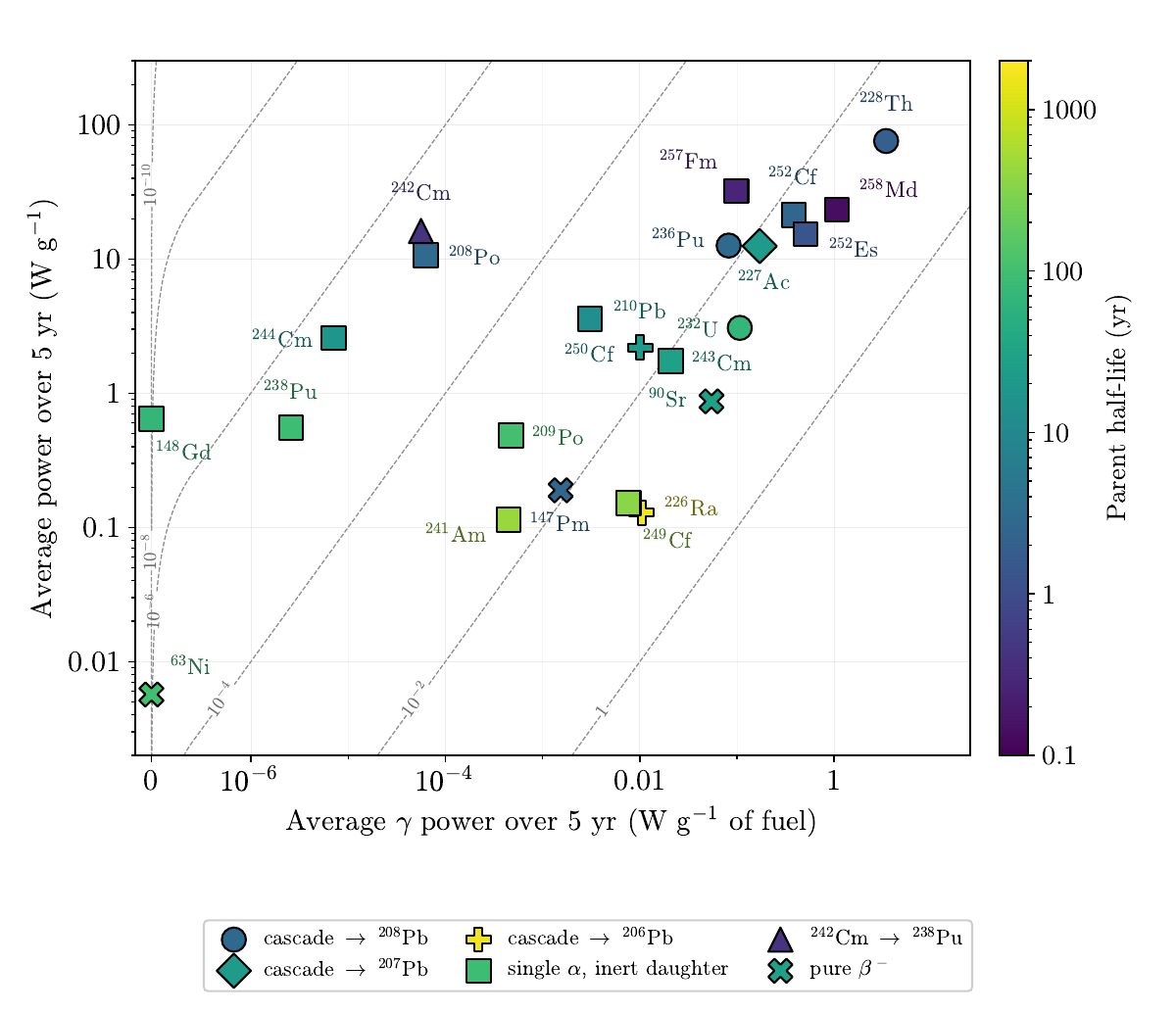}
\caption{}
\label{fig:cascade_search_b}
\end{subfigure}
\caption{Alpha-emitter parents across time-averaged specific thermal power and time-averaged gamma power per gram, over (a) 100 yr and (b) 5 yr periods, with color giving the parent half-life. Dashed lines mark contours of constant gamma power fraction. Photon power includes thick-target bremsstrahlung of the chain betas stopping in the fuel.}\label{fig:cascade_search}
\end{figure}

The chains considered here have significantly different energy release profiles, especially for those chains containing radioisotopes with strongly differing half-lives. \Cref{fig:alpha_release} shows the cumulative chain energy released per gram for fuels starting from each alpha emitter of \Cref{fig:decay} and from the long-lived alpha emitters including the full downstream chains. We estimate the useful energy release timescale in RTG fuels as between 1 and 100 years, although we extend the plotted time horizon for completeness, and because of potential future interest in fuels with much shorter or longer mission times. We also include several popular $\beta^-$ emitters for context. Of these $\beta^-$ emitters, only ${}^{90}$Sr exceeds a cumulative energy release of 0.1 GJ per gram within 10 years. 

The instantaneous power of a radionuclide chain is
\begin{equation}
P_\mrm{th}(t) = \sum_i \lambda_i\, N_i(t)\, Q_i^\mrm{th},
\label{eq:Pth}
\end{equation}
where $\lambda_i$ and $N_i$ are the decay constant and particle number of species $i$, $Q_i^\mrm{th}$ is the locally deposited energy per decay (alpha energy, beta kinetic heat, and absorbed gammas; antineutrinos escape), and the $N_i$ evolve according to \cite{Bateman1910},
\begin{equation}
\dot{N}_i = \Lambda_{ij} N_j, \;\;\;\;\; \dot{N}_i \equiv \frac{d N_i}{dt},
\label{eq:Ni}
\end{equation}
with $N_i(0)=\delta_{i,\mrm{Pu}}\,N_0$, where $\Lambda$ has components $\Lambda_{ij} = b_{j \to i}\,\lambda_j - \lambda_i\,\delta_{ij}$, with $b_{j \to i}$ the branching ratio of species $j$ into species $i$. For no branching, $b = 1$. An example of branching is $b_{{}^{212}\mrm{Bi} \to {}^{208}\mrm{Tl}} = 0.36$ and $b_{{}^{212}\mrm{Bi} \to {}^{212}\mrm{Po}} = 0.64$ for the ${}^{236}$Pu chain's one significant branching. For an unbranched chain ordered parent to daughter, the decay matrix is
\begin{equation}
\Lambda = \begin{pmatrix}
-\lambda_1 & 0 & \cdots & & 0 \\
\lambda_1 & -\lambda_2 & & & \vdots \\
0 & \lambda_2 & \ddots & & \\
\vdots & & \ddots & -\lambda_{n-1} & 0 \\
0 & \cdots & 0 & \lambda_{n-1} & 0
\end{pmatrix},
\label{eq:Lambda}
\end{equation}
where the final stable nuclide in the chain has $\lambda_n = 0$.

In \Cref{fig:specific_power} we solve $P_\mathrm{th} (t)$ (\Cref{eq:Ni}) for various chains of interest and show the results, normalized to initial fuel mass in grams. Notably, over the first decade since fuel preparation, the specific power varies by three orders of magnitude: at one year, $^{228}$Th has a specific power of over 100 W/g, one thousand times higher than $^{241}$Am at 0.1 W/g.

The ${}^{236}$Pu chain releases high specific power exceeding 2 W/g for about a century. Fresh ${}^{236}$Pu releases 18.1 W g${}^{-1}$. Since daughter buildup is fast for ${}^{228}$Th and below (since their half-lives are shorter) but slow for ${}^{232}$U, the specific power of the ${}^{236}$Pu chain stays high for decades: at $t = 1, 3, 10, 30$ yr the chain releases 14.5, 10.1, 5.5, 4.1 W g${}^{-1}$ respectively. At $t = 100$ yr the chain output exceeds ${}^{238}$Pu by a factor of 8. Integrated over 500 yr, the chain releases $1.8\times 10^{10}$ J g${}^{-1}$, 8.3 times the $2.2\times 10^{9}$ J g${}^{-1}$ of ${}^{238}$Pu. 

\Cref{fig:cascade_search} shows $\alpha$ and $\beta^-$ parents according to their average thermal power per gram of initial fuel over 100 yr and 5 yr battery time periods (chosen for representative mission times), and the average power of gamma photons the full chain emits per gram over the same time period, with color showing the parent half-life.

Only the ${}^{236}$Pu, ${}^{232}$U, and ${}^{228}$Th chain, and the one headed by ${}^{227}$Ac, release multi-W g${}^{-1}$ over 100 years. Every single-alpha parent and the ${}^{226}$Ra chain fall an order of magnitude lower; the nearest, ${}^{252}$Cf, reaches 1.5 W g${}^{-1}$ only by counting its 3.1\% spontaneous-fission heat. Over a 5 yr period the short-lived parents close the gap (${}^{228}$Th, ${}^{252}$Cf, and ${}^{242}$Cm lead), but ${}^{236}$Pu has a higher power density the standard RTG fuels by an order of magnitude. Within the chains, ${}^{228}$Th and ${}^{227}$Ac give slightly higher 100 yr averages than ${}^{236}$Pu because their short half-lives release the full chain within the time period, whereas much of the ${}^{236}$Pu chain remains at ${}^{232}$U.

In \Cref{fig:cascade_search} we demonstrate the tradeoff between shielding and power by plotting the average total power versus the average $\gamma$ power, with the caveat that radiation dosage is more complicated than $\gamma$ power alone. Averaged over their first century, the high-energy-density chains emit gamma power at the 0.1 W g${}^{-1}$ scale, over four orders of magnitude above ${}^{238}$Pu (something we address for the ${}^{236}$Pu chain in \Cref{sec:radon}), while the pure beta emitters and ${}^{148}$Gd emit essentially none. ${}^{148}$Gd follows the ${}^{238}$Pu power curve closely enough to serve as a radiation-free substitute, but it currently lacks a scalable production route. 

The ${}^{210}$Pb chain also offers significantly higher specific power than ${}^{238}$Pu. In equilibrium with its ${}^{210}$Po granddaughter, the ${}^{210}$Pb chain releases 2.7 W g${}^{-1}$, above ${}^{238}$Pu and the curiums over 100 years; its photon output is weak lines plus the bremsstrahlung from the ${}^{210}$Bi beta. With the neutron source removed by a light-element-free fuel and a shield mass comparable to ${}^{238}$Pu's at occupational thresholds (\Cref{sec:shieldmass}), ${}^{210}$Pb could function as a scalable ${}^{238}$Pu alternative for missions of up to a few decades, at nearly five times the specific power. In Appendix \ref{app:pb210} we show its fast production route from ${}^{231}$Pa, of $\sim$33 to 122 kg per GW$_\mrm{fus}$-yr.

\section{Production at fusion scale}
\label{sec:production}

\begin{figure}[!tb]
\centering
\begin{subfigure}{\columnwidth}
\centering
\includegraphics[width=\columnwidth]{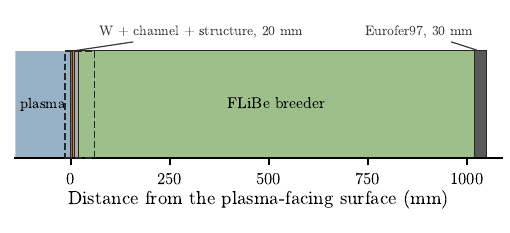}
\caption{}
\label{fig:geometry_a}
\end{subfigure}\\[2pt]
\begin{subfigure}{\columnwidth}
\centering
\includegraphics[width=\columnwidth]{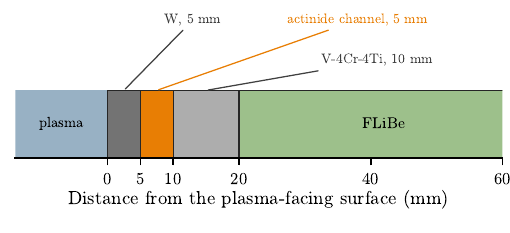}
\caption{}
\label{fig:geometry_b}
\end{subfigure}
\caption{Radial blanket build used in OpenMC scans. (a) The outboard build from the plasma-facing surface to the blanket back; the dashed box shows the region magnified in (b). (b) The first 60 mm of blanket: a 5 mm actinide channel of ${}^{232}$Th, ${}^{237}$Np, or ${}^{238}$U between the 5 mm tungsten first wall and the 10 mm structural layer. We vary the actinide channel thickness in \Cref{sec:scan}, using 5 mm as the standard value.}
\label{fig:geometry}
\end{figure}

\begin{figure}[!tb]
\centering
\begin{subfigure}{\columnwidth}
\centering
\includegraphics[width=\columnwidth]{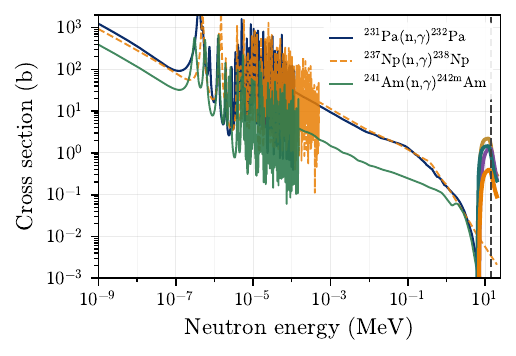}
\caption{}
\label{fig:production_xs_a}
\end{subfigure}\\[2pt]
\begin{subfigure}{\columnwidth}
\centering
\includegraphics[width=\columnwidth]{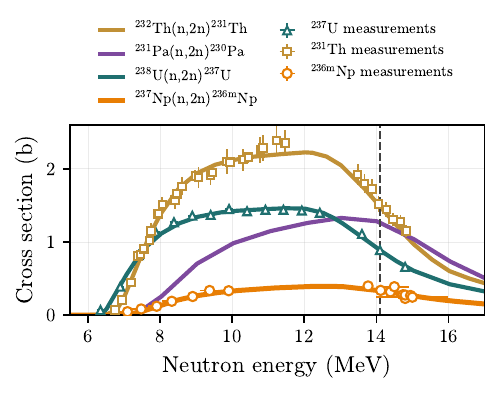}
\caption{}
\label{fig:production_xs_b}
\end{subfigure}\\[2pt]
\begin{subfigure}{\columnwidth}
\centering
\includegraphics[width=\columnwidth]{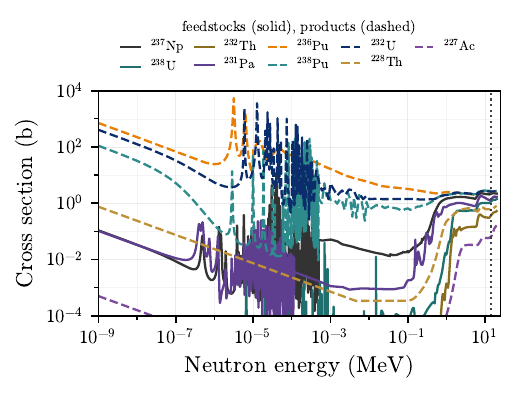}
\caption{}
\label{fig:production_xs_c}
\end{subfigure}
\caption{The main cross sections (ENDF/B-VIII.0). (a) Wide-energy range. (b) High-energy range. (c) Fission, for the feedstocks (solid) and bred products (dashed). The dashed vertical line marks 14.1 MeV.}
\label{fig:production_xs}
\end{figure}

\begin{figure*}[!tb]
\centering
\begin{subfigure}{0.32\textwidth}
\centering
\includegraphics[width=\textwidth]{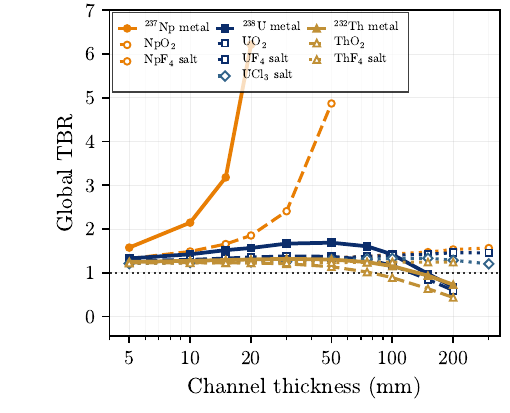}
\caption{}
\label{fig:blanket_scan_a}
\end{subfigure}
\hfill
\begin{subfigure}{0.32\textwidth}
\centering
\includegraphics[width=\textwidth]{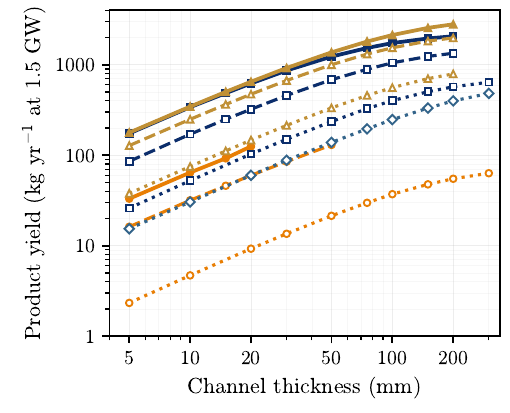}
\caption{}
\label{fig:blanket_scan_b}
\end{subfigure}
\hfill
\begin{subfigure}{0.32\textwidth}
\centering
\includegraphics[width=\textwidth]{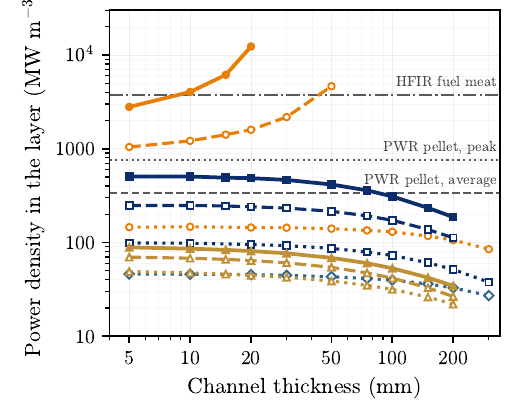}
\caption{}
\label{fig:blanket_scan_c}
\end{subfigure}\\[4pt]
\begin{subfigure}{0.32\textwidth}
\centering
\includegraphics[width=\textwidth]{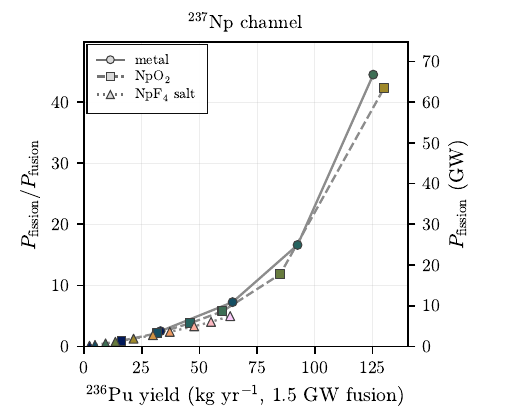}
\caption{}
\label{fig:blanket_scan_d}
\end{subfigure}
\hfill
\begin{subfigure}{0.32\textwidth}
\centering
\includegraphics[width=\textwidth]{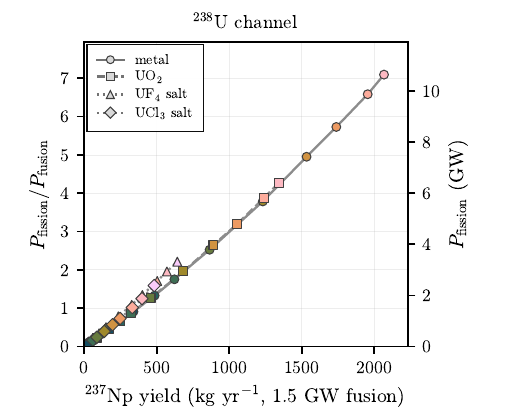}
\caption{}
\label{fig:blanket_scan_e}
\end{subfigure}
\hfill
\begin{subfigure}{0.32\textwidth}
\centering
\includegraphics[width=\textwidth]{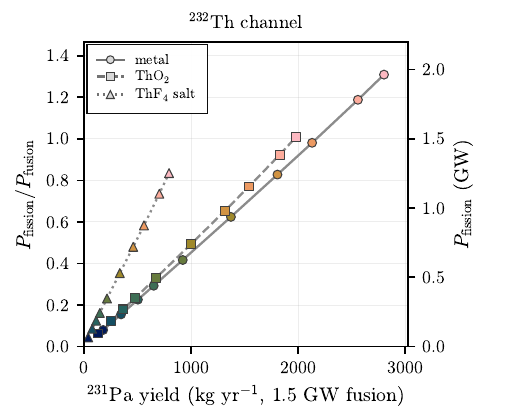}
\caption{}
\label{fig:blanket_scan_f}
\end{subfigure}\\[2pt]
\includegraphics[width=0.62\textwidth]{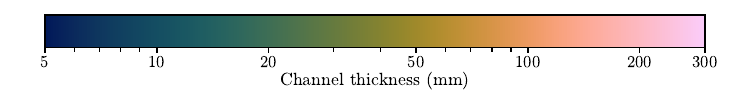}
\caption{OpenMC channel thickness scan for various actinides (blanket geometry in \Cref{fig:geometry}): metals (solid), oxides (dashed), molten salts (dotted). (a) Global tritium breeding ratio. (b) Annual product yield. (c) Volumetric power density. Legend in (a) applies to (a)-(c). (d)-(f) Fission-to-fusion power ratio and absolute fission power, versus product yield for ${}^{237}$Np, ${}^{238}$U, and ${}^{232}$Th channels. Salt materials are 15 mol\% NpF$_4$ in FLiBe, LiF-UF$_4$ 73/27, NaCl-UCl$_3$ 65.8/34.2 with 99at\% ${}^{37}$Cl, and LiF-ThF$_4$ 77.5/22.5. Materials with lithium are 90at\% ${}^{6}$Li.}
\label{fig:blanket_scan}
\end{figure*}

In this section we show production yields by irradiating feedstock with fast D-T neutrons. The feedstock is placed in the first layer of blanket, shown in \Cref{fig:geometry}.

We produce the battery fuels in three blanket channels, one per feedstock, each driven by a 14 MeV threshold \ntn\ reaction (cross sections in \Cref{fig:production_xs}) with the first-order fast neutron reactions
\begin{equation}
\begin{aligned}
&{}^{237}\mrm{Np}\ntn{}^{236\mrm{m}}\mrm{Np} \xrightarrow{\betam,\,22.5\,\mrm{h}} {}^{236}\mrm{Pu} \;\; (52.5\%),\\
&{}^{238}\mrm{U}\ntn{}^{237}\mrm{U} \xrightarrow{\betam,\,6.75\,\mrm{d}} {}^{237}\mrm{Np},\\
&{}^{232}\mrm{Th}\ntn{}^{231}\mrm{Th} \xrightarrow{\betam,\,25.5\,\mrm{h}} {}^{231}\mrm{Pa},
\end{aligned}
\label{eq:routes}
\end{equation}
with 14.1 MeV cross sections of 0.47, 0.89, and 1.49 b. 

\textit{${}^{237}$Np blanket}: The ${}^{237}$Np channel makes the ${}^{236}$Pu fuel directly, as well as co-producing ${}^{238}$Pu via neutron capture,
\begin{equation}
{}^{237}\mrm{Np}\mrm{(n,\gamma)}^{238}\mathrm{Np} \xrightarrow{\betam,\,2.10\,\mrm{d}} {}^{238}\mrm{Pu}.
\label{eq:Np237ng}
\end{equation}

\textit{${}^{238}$U blanket}: The ${}^{238}$U channel breeds the ${}^{237}$Np feedstock, which then undergoes further neutron reactions in \Cref{eq:routes,eq:Np237ng} to produce ${}^{236}$Pu and ${}^{238}$Pu. Another pathway to ${}^{238}$Pu in the ${}^{238}$U channel is
\begin{equation}
\begin{aligned}
& {}^{238}\mrm{U}\mrm{(n,\gamma)}^{239}\mathrm{U} \xrightarrow{\betam,\,23\,\mrm{m}} {}^{239}\mrm{Np} \\
& \xrightarrow{\betam,\,2.4\,\mrm{d}} {}^{239}\mrm{Pu}\mathrm{(n,2n)} {}^{238}\mrm{Pu}.
\end{aligned}
\label{eq:U238toPu238}
\end{equation}

\textit{${}^{232}$Th blanket}: The ${}^{232}$Th channel breeds ${}^{231}$Pa, the feedstock for ${}^{232}$U via
\begin{equation}
{}^{231}\mrm{Pa}\mrm{(n,\gamma)}{}^{232}\mrm{Pa} \xrightarrow{\betam,\,1.31\,\mrm{d}} {}^{232}\mrm{U}.
\label{eq:u232prod}
\end{equation}
${}^{231}$Pa is also the feedstock for ${}^{230}$U, which subsequently decays to ${}^{210}$Pb,
\begin{equation}
{}^{231}\mrm{Pa}\ntn{}^{230}\mrm{Pa} \xrightarrow{\betam,\,17.4\,\mrm{d}} {}^{230}\mrm{U} \;\; (8.4\%).
\label{eq:u230prod}
\end{equation}
Finally, ${}^{231}$Pa is a cow for the fuel ${}^{227}$Ac,
\begin{equation}
{}^{231}\mrm{Pa} \xrightarrow{\alpha,\,32.8\,\mrm{kyr}} {}^{227}\mrm{Ac},
\label{eq:pa231cowac227}
\end{equation}
where each ton of ${}^{231}$Pa produces $\sim$21 grams of ${}^{227}$Ac per year.

\subsection{Blanket scan: \texorpdfstring{${}^{237}$Np}{Np-237}, \texorpdfstring{${}^{238}$U}{U-238}, and \texorpdfstring{${}^{232}$Th}{Th-232} channels} \label{sec:scan}

\begin{table*}[!tb]
\centering
\caption{Overview of operating points and production rates in a blanket surrounding a 1.5 GW$_\mrm{fus}$ D-T fusion tokamak core (\Cref{fig:blanket_scan}). $t_\mrm{ch}$ is the channel thickness, $V_\mrm{ch}$ the channel volume, $M_\mrm{ch}$ the channel actinide inventory, $k_\mrm{eff}$ the neutron reactivity coefficient, TBR the full blanket tritium breeding ratio, $\dot{m}_\mrm{T}$ the tritium bred in excess of TBR = 1.1, $P_\mrm{tot}$ the total plant thermal power (fusion plus fission heat), and $q'''$ the volumetric heating averaged over the channel.}
\label{tab:hybrid}
\begin{tabular}{lcccccccccccccc}
\toprule
Channel & $t_\mrm{ch}$ & $V_\mrm{ch}$ & $M_\mrm{ch}$ & $k_\mrm{eff}$ & TBR & $\dot{m}_\mrm{T}$ & $P_\mrm{tot}$ & $q'''$ & \multicolumn{2}{c}{Product} & \multicolumn{2}{c}{Co-product} & ${}^{90}$Sr \\
        & (mm)         & (m${}^{3}$) & (t)          &               &     & (kg\,yr${}^{-1}$) & (GW)          & (MW\,m${}^{-3}$) & & (kg\,yr${}^{-1}$) & & (kg\,yr${}^{-1}$) & (kg\,yr${}^{-1}$) \\
\midrule
NpO$_2$ & 5 & 1.35 & 15 & 0.21 & 1.32 & 18 & 2.9 & 1044 & ${}^{236}$Pu & 16 & ${}^{238}$Pu & 673 & 9 \\
NpO$_2$ & 10 & 2.70 & 30 & 0.35 & 1.49 & 33 & 4.8 & 1213 & ${}^{236}$Pu & 32 & ${}^{238}$Pu & 1482 & 22 \\
NpO$_2$ & 15 & 4.06 & 45 & 0.45 & 1.66 & 47 & 7.2 & 1413 & ${}^{236}$Pu & 46 & ${}^{238}$Pu & 2453 & 38 \\
NpO$_2$ & 20 & 5.42 & 60 & 0.55 & 1.85 & 63 & 10 & 1594 & ${}^{236}$Pu & 60 & ${}^{238}$Pu & 3747 & 57 \\
NpO$_2$ & 30 & 8.15 & 90 & 0.68 & 2.41 & 110 & 19 & 2177 & ${}^{236}$Pu & 85 & ${}^{238}$Pu & 7792 & 117 \\

\midrule
${}^{238}$U & 5 & 1.35 & 26 & 0.05 & 1.33 & 19 & 2.2 & 507 & ${}^{237}$Np & 174 & --- & --- & 5 \\
${}^{238}$U & 50 & 13.7 & 260 & 0.20 & 1.69 & 50 & 7.2 & 416 & ${}^{237}$Np & 1230 & --- & --- & 38 \\
${}^{238}$U & 100 & 27.7 & 528 & 0.25 & 1.42 & 27 & 10 & 310 & ${}^{237}$Np & 1740 & --- & --- & 57 \\
${}^{238}$U$^{b}$, Np extraction & 50 & 13.7 & 260 & 0.20 & 1.69 & 50 & 7.2 & 416 & ${}^{237}$Np & 1270 & --- & --- & 47 \\
${}^{238}$U$^{c}$, Pu extraction & 50 & 13.7 & 260 & 0.20 & 1.69 & 50 & 7.2 & 416 & ${}^{238}$Pu & $\sim$500 & ${}^{236}$Pu & $\sim$5 & 52 \\
\midrule
${}^{232}$Th & 5 & 1.35 & 16 & 0.01 & 1.25 & 13 & 1.6 & 89 & ${}^{231}$Pa & 178 & ${}^{232}$U$^{f}$ & 9 & 0.8 \\
${}^{232}$Th & 50 & 13.7 & 160 & 0.03 & 1.30 & 17 & 2.4 & 68 & ${}^{231}$Pa & 1370 & ${}^{232}$U$^{f}$ & 80 & 6.5 \\
${}^{232}$Th & 100 & 27.7 & 325 & 0.04 & 1.15 & 4.2 & 3.0 & 53 & ${}^{231}$Pa & 2130 & ${}^{232}$U$^{f}$ & 130 & 10 \\
${}^{232}$Th$^{a}$, Pa retained & 100 & 27.7 & 325 & 0.04 & 1.15 & 4.2 & 3.0 & 53 & ${}^{232}$U & 804 & ${}^{210}$Pb & 6.3 & 28 \\
\midrule
${}^{231}$Pa/D$_2$O$^{d}$ & 5 & 1.35 & 0.33 & --- & 1.19 & 8.4 & 1.5 & 34 & ${}^{232}$U & 6.8 & ${}^{228}$Th & 1.0 & 0.2 \\
${}^{231}$Pa$^{e}$, solid & 20 & 5.4 & 83 & 0.37 & 1.43 & 28 & 7.1 & 1033 & ${}^{210}$Pb & 50 & ${}^{232}$U & 1490 & 39 \\
${}^{231}$Pa$^{e}$, solid & 200 & 56 & 877 & 0.83 & 1.14 & 3.4 & 78 & 1370 & ${}^{210}$Pb & 183 & ${}^{232}$U & 22500 & 535 \\
\bottomrule
\end{tabular}

\vspace{2pt}
\begin{minipage}{\textwidth}\footnotesize\raggedright
$^{a}$Thirty yr depletion averages with the protactinium retained (Appendix \ref{app:depletion}). The same channel also makes ${}^{228}$Th at 95 kg yr${}^{-1}$ and ${}^{227}$Ac at 0.4 kg yr${}^{-1}$.
$^{b}$Thirty yr depletion averages with ${}^{238}$U continuously replenished, neptunium and plutonium continuously extracted (Appendix \ref{app:depletion}).
$^{c}$As $^{b}$ but with plutonium-only extraction; the extracted plutonium is 12\% ${}^{238}$Pu by mass (see text).
$^{d}$ ${}^{231}$Pa at 244 g L${}^{-1}$ in D$_2$O, turning protactinium first made in thorium blanket into ${}^{232}$U and then used in this dedicated channel.\\
$^{e}$Solid ${}^{231}$Pa metal, no moderator, via ${}^{231}$Pa\ntn ${}^{230}$Pa $\to$ ${}^{230}$U $\to$ ${}^{210}$Pb of Appendix \ref{app:pb210} (\Cref{tab:pa231_blanket}), fed by Pa bred in a Th blanket.\\
$^{f}$Thirty yr average via the bred ${}^{233}$U's \ntn\ reaction (see text), additive to the ${}^{231}$Pa capture route of row $a$.
\end{minipage}
\end{table*}

\begin{figure}[b]
\centering
\includegraphics[width=\columnwidth]{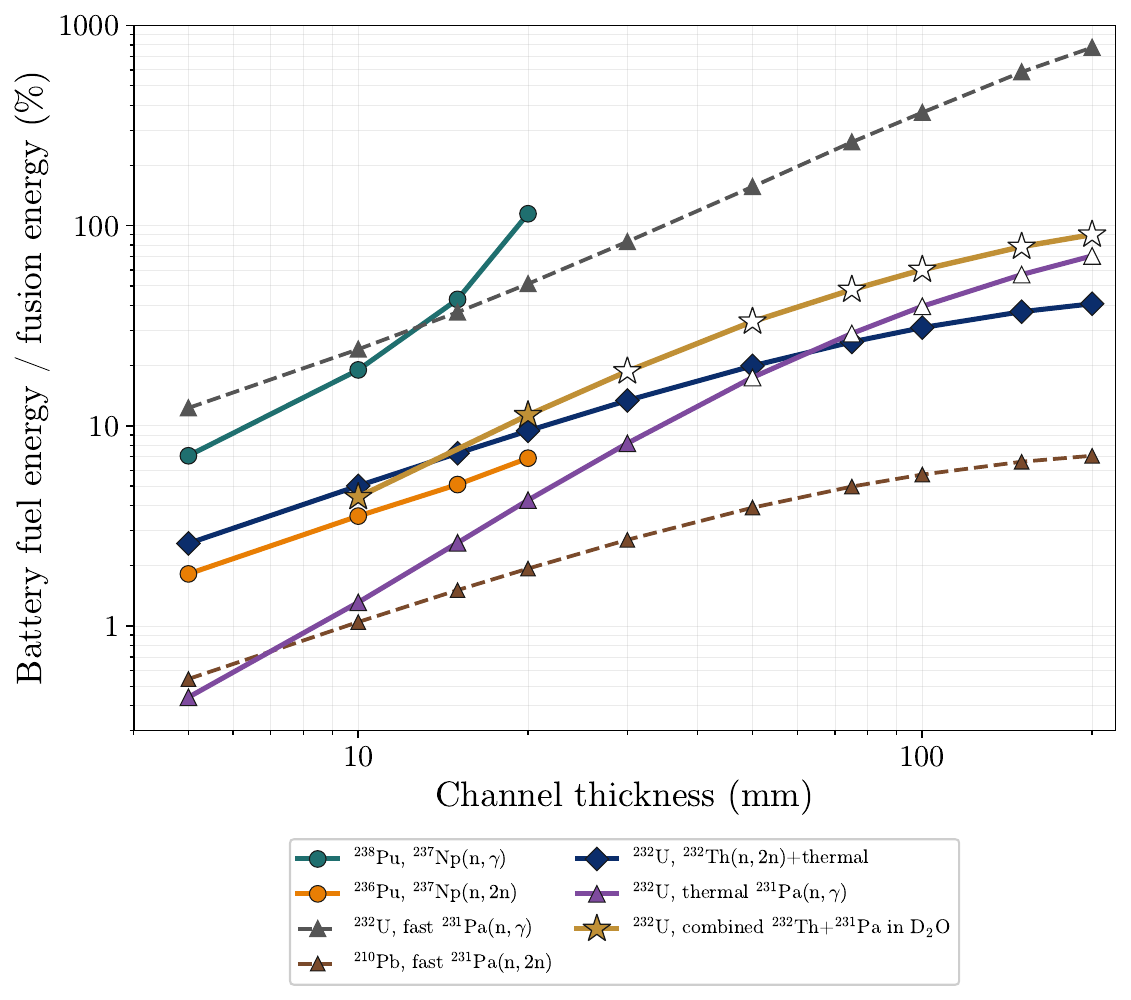}
\caption{Battery-fuel energy stored per unit of fusion energy versus channel thickness, from OpenMC scans. Ratios above 100\% (the ${}^{238}$Pu co-product, the fast ${}^{231}$Pa blanket) reflect fission multiplication rather than fusion productivity. Open symbols mark TBR $<$ 1; all ${}^{232}$U routes multiply the ${}^{231}$Pa capture rate by 0.42, the fraction recoverable as ${}^{232}$U (\Cref{sec:u232}). ${}^{231}$Pa-fed blankets presume availability of ${}^{231}$Pa, probably not available for a while (Appendix \ref{app:pb210}).}
\label{fig:fusion_battery_conversion}
\end{figure}

In our following modeling, we assume that a fusion power plant (FPP) must be tritium self-sufficient, requiring a tritium breeding ratio (TBR) above one~\cite{Abdou2021,Meschini2023}. TBR is the ratio of tritium bred in the blanket to tritium burned in the plasma.

We perform an OpenMC~\cite{openmc} tokamak scan with ENDF/B-VIII.0 cross sections~\cite{Brown2018ENDF}, varying the thickness of a single channel of pure ${}^{237}$Np metal, pure ${}^{238}$U, or pure ${}^{232}$Th, inserted between the tungsten first wall and an enriched Li (90\%at ${}^{6}$Li) FLiBe main breeder (\Cref{fig:geometry}). The standard channel is 5 mm thick (\Cref{fig:geometry_b}). Each thickness is a separate OpenMC run with the total radial thickness held fixed: every millimeter added to the channel is removed from the FLiBe layer on both the inboard (750 mm basecase) and outboard (1000 mm) sides, so a thick channel also displaces breeder. The plasma is the source of 14.1 MeV neutrons. The pure-FLiBe basecase (no actinide channel) gives TBR = 1.22 from ${}^9$Be\ntn\ followed mainly by ${}^6$Li capture. Each fission releases $\sim$200 MeV of recoverable heat. \Cref{fig:blanket_scan} shows the scans, and \Cref{tab:hybrid} gives representative operating points for a 1.5 GW$_\mrm{fus}$ D-T FPP. 

\begin{figure*}[!tb]
\centering
\begin{subfigure}{0.49\textwidth}
\centering
\includegraphics[width=\textwidth]{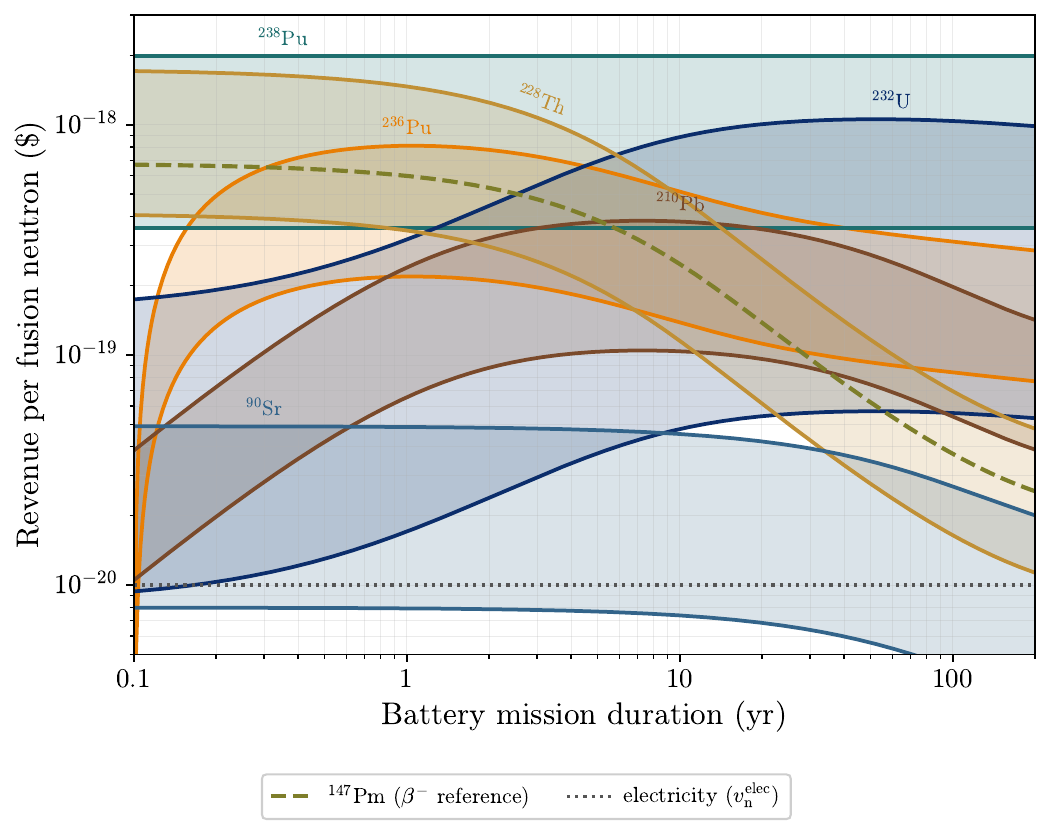}
\caption{}
\label{fig:vpn}
\end{subfigure}
\hfill
\begin{subfigure}{0.49\textwidth}
\centering
\includegraphics[width=\textwidth]{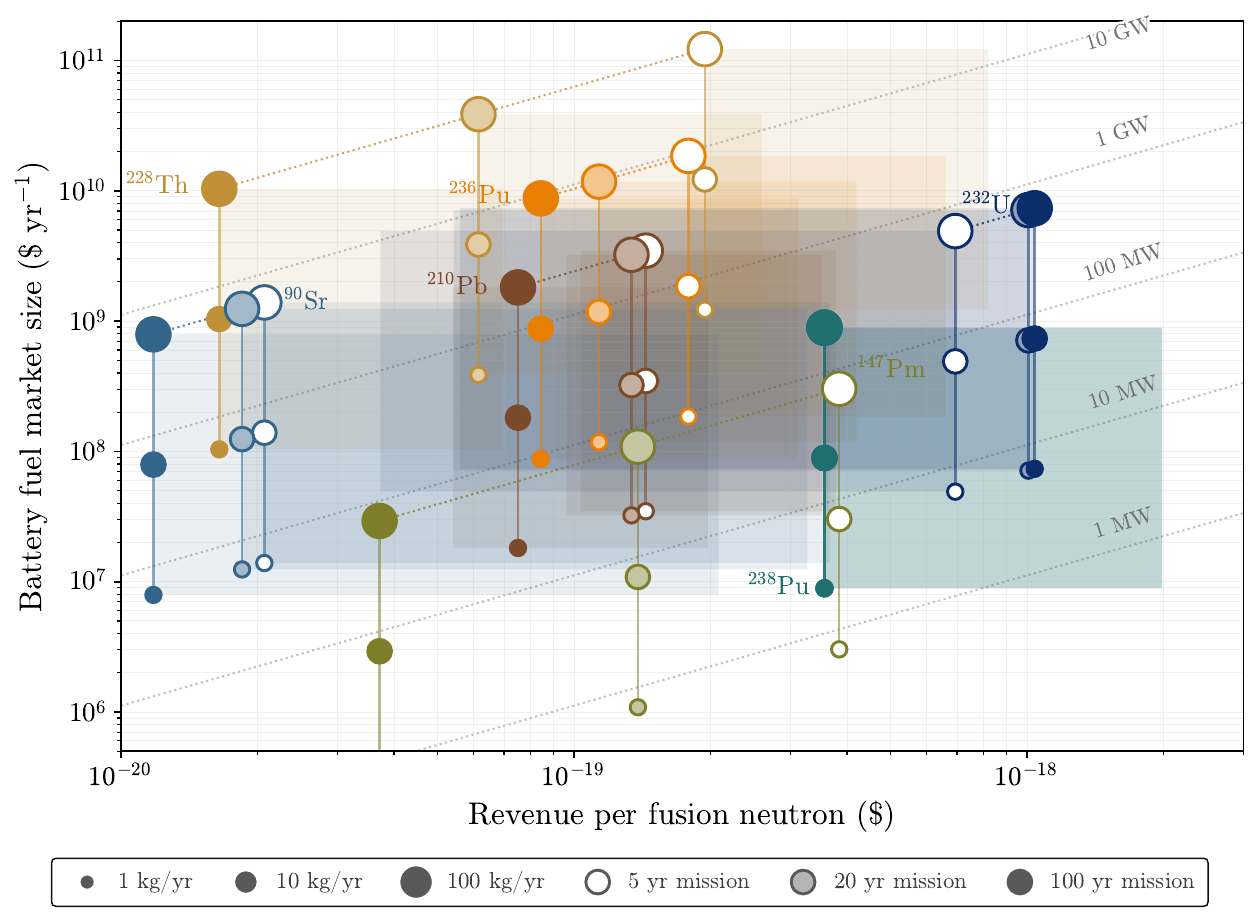}
\caption{}
\label{fig:market}
\end{subfigure}
\caption{Neutron revenue and market size. (a) Revenue per fusion neutron versus battery mission duration (\Cref{eq:vpn}): each fuel's revenue is scaled by the enhancement factor of \Cref{eq:enhancement}. Each band ranges from the thinnest to the thickest subcritical, tritium self-sufficient blanket from OpenMC scans. ${}^{236}$Pu, ${}^{238}$Pu and ${}^{90}$Sr come from ${}^{237}$Np\ntn\ in the 5 to 20 mm NpO$_2$ channel; ${}^{232}$U comes from a 5 to 30 mm ${}^{231}$Pa/D$_2$O channel; ${}^{210}$Pb comes from a 20 to 200 mm ${}^{231}$Pa channel; and ${}^{147}$Pm is the ${}^{148}$Nd\ntn\ pathway from the companion paper~\cite{parisi2026betaemitters}. (b) Battery fuel market size versus revenue per fusion neutron, at production scales of 1, 10, and 100 kg/yr (marker size) and mission durations of 5, 20, and 100 yr (open, tinted, and filled markers). Dotted diagonals are lines of constant D-T fusion power. Shaded rectangles cover the blanket thickness ranges in (a).}
\label{fig:vpn_market}
\end{figure*}

\textit{${}^{237}$Np blanket}: The ${}^{237}$Np channel is strongly fissile in the fast spectrum: fission above the $\sim$0.5 MeV threshold releases $\bar{\nu}\approx2.8$ neutrons at fission-spectrum energies, rising to 4.8 at 14.1 MeV~\cite{Brown2018ENDF}, and a channel of ${}^{237}$Np metal reaches criticality at $\sim$26.5 mm (an OpenMC scan over 5 to 30 mm gives $k_\mrm{eff} = 0.38, 0.60, 0.76, 0.88, 0.98, 1.05$ at 5/10/15/20/25/30 mm). Diluting the actinide as NpO$_2$ keeps the blanket subcritical even at 50 mm with $k_\mrm{eff} = 0.21, 0.35, 0.45, 0.55, 0.68, 0.79, 0.87$ at 5/10/15/20/30/40/50 mm. Instead, we set the thickness limit by volumetric heating in analogy to the highest heating limits tolerated in fission fuel: NpO$_2$ reaches the power density of HFIR fuel between 30 and 50 mm, so we take 30 mm as the thickest operating point. Fission events outnumber \ntn\ by up to a factor of 85: at 20 mm the channel produces 4.4 fissions, 0.052 \ntn, and 780 MeV of channel heat per source neutron, increasing TBR to 6.2. The system is a fusion-fission hybrid~\cite{Bethe1979} where fission heat is the main energy product and the \ntn\ reactions make a high-value byproduct: even for the thin 5 mm channel, the 1.5 GW$_\mrm{fus}$ plant produces 2.9 GW total thermal power, making $\sim$16 kg yr${}^{-1}$ of ${}^{236}$Pu, increasing to 85 kg yr${}^{-1}$ at 19 GW$_\mrm{th}$ for 30 mm (\Cref{fig:blanket_scan_a}, \Cref{tab:hybrid}). Operating a blanket with such high power would be challenging from a thermal perspective, but not necessarily beyond capabilities in the fission industry.

We compare the power density in the actinide channels to the fuel meat in fission reactors. A PWR pellet has $\sim$340 MW m$^{-3}$ at average rating and $\sim$760 MW m$^{-3}$ at peak~\cite{TodreasKazimi2011}, and HFIR, a power density reactor, $\sim$3750 MW m$^{-3}$ in its {U$_3$O$_8$}-Al fuel meat~\cite{ORNLHFIRGuide}. Our NpO$_2$ channel has 1044 MW m$^{-3}$ at 5 mm and 4643 MW m$^{-3}$ at 50 mm, so 3 to 14 times a PWR pellet, reaching HFIR near 50 mm; the ${}^{238}$U channel at 186 to 507 MW m$^{-3}$ and the ${}^{232}$Th channel at 34 to 89 MW m$^{-3}$ comparable to and below the pellet. Therefore heat removal, not criticality, limits the neptunium channel thickness (\Cref{fig:blanket_scan_c}), although the two are closely related.

\textit{${}^{238}$U blanket}: The ${}^{238}$U channel has much weaker neutron multiplication, with fission events outnumbering \ntn\ by only $\sim$2: a 50 mm channel has 0.38 fissions, 0.19 \ntn, and 67 MeV of channel heat per source neutron, and it stays tritium self-sufficient out to $\sim$145 mm, beyond which channel absorption decreases TBR below one. We scanned the total blanket thickness from 5 to 200 mm. It also adds a \nthn\ channel that opens around 12 MeV, with a \nthn:\ntn\ ratio of 0.42 across all thicknesses (against 0.05 for ${}^{237}$Np). The channel has much reduced ${}^{236}$Pu output for a much larger ${}^{237}$Np breeding rate, from 174 kg yr${}^{-1}$ of ${}^{237}$Np at 5 mm to 1.74 t yr${}^{-1}$ at 100 mm, with the 50 mm point producing $\sim$1.2 t yr${}^{-1}$ at 7.2 GW thermal (\Cref{tab:hybrid}); the bred ${}^{237}$Np can be chemically separated and either sold as feedstock or re-irradiated for ${}^{236}$Pu in a second cycle, or left in place as feedstock for the ${}^{236}$Pu chain. Neutron capture on ${}^{238}$U also breeds ${}^{239}$Pu, which under continuous elemental extraction leaves with the stream at 123 kg yr${}^{-1}$ for the 5 mm channel and 3.5 t yr${}^{-1}$ at 50 mm with extremely high isotopic purity (Appendix \ref{app:depletion}). Continuous extraction of Pu and Np presents the largest safeguards concern in a ${}^{238}$U blanket (Appendix \ref{app:u233}). However, if just Pu were extracted and neutron reactions were allowed to take place on ${}^{237}$Np, this would likely safeguard the blanket material through the presence of significant quantities of ${}^{236}$Pu. Overall, we judge the ${}^{238}$U channel to be relatively uninteresting compared with the thorium, neptunium, and protactinium channels from the perspective of making nuclear batteries, and of much higher proliferation concern.

\textit{${}^{232}$Th blanket}: The ${}^{232}$Th channel is highly fertile, with little fission multiplication except for high neutron energies, and was also scanned from 5 to 200 mm. Its TBR crosses under one near 135 mm. Its \ntn\ breeds ${}^{231}$Pa through 25.5 h ${}^{231}$Th, at up to $\sim$1.9 tonnes per GW-yr at 200 mm or $\sim$1.6 tonnes per GW-yr at the tritium self-sufficiency limit (\Cref{fig:blanket_scan_a}). Beyond a thin blanket with $\sim$10 mm thickness, the softening neutron spectrum makes the ${}^{232}$Th\ngamma\ rate overtake its \ntn\ (\Cref{fig:blanket_scan_b}), breeding fissile ${}^{233}$U alongside the ${}^{231}$Pa. The bred ${}^{233}$U is also a target, its \ntn\ ($\sim$0.4 b at 14.1 MeV, against 1.49 b for ${}^{232}$Th) converts the growing inventory to ${}^{232}$U, a heuristic $\sim$9, 80, and 130 kg yr${}^{-1}$ at 5, 50, and 100 mm averaged over 30 yr (the ${}^{232}$U co-product column of \Cref{tab:hybrid}), and additive to the ${}^{231}$Pa capture route of \Cref{tab:hybrid} row $a$; the safeguards consequences are described with depletion simulations in Appendix \ref{app:u233}, and the conversion of the bred ${}^{231}$Pa to ${}^{232}$U is discussed in \Cref{sec:u232}.

\Cref{fig:blanket_scan_d,fig:blanket_scan_e,fig:blanket_scan_f} show the fission-to-fusion power ratio against product yield for the three blankets: every ${}^{237}$Np operating point lies at $P_\mrm{fission}/P_\mrm{fusion} = 2.5$ to 45, deep in fission-fusion hybrid territory, while the ${}^{238}$U blanket crosses power parity near 11 mm and the ${}^{232}$Th blanket reaches it only near 100 mm.

The yields above are per-source-neutron rates for a fresh channel. To follow a channel through a full plant irradiation, and to determine the long-lived radioisotope inventory it leaves behind, we also perform an OpenMC depletion simulation of the 5 mm ${}^{237}$Np channel over 30 years of continuous irradiation followed by 100 years of cooling (Appendix \ref{app:depletion}). Even the thin 5 mm channel converts roughly half of all source neutrons, so it burns quickly: 87\% of the ${}^{237}$Np is consumed by year 30, and the remaining ${}^{236}$Pu inventory peaks at only $\sim$61 kg near year five because the 2.86 yr product decays and transmutes as fast as it forms. Scalable ${}^{236}$Pu production therefore requires extraction on a cycle short compared to the fuel's 2.86 yr half-life, its $\sim$8 yr in-flux fission burnup, and the $\sim$13 yr feedstock burnup time. More details are discussed in Appendix \ref{app:depletion}. The plutonium these channels make is mostly ${}^{238}$Pu, the ${}^{236}$Pu fraction being 2.4 at\% at 5 mm. Appendix \ref{app:purity} shows that a neutron absorber behind the channel raises it to 6.2 at\% without a tradeoff in TBR.

\subsection{Flight-grade ${}^{238}$Pu by decay} \label{sec:pu238flight}

The NpO$_2$ blanket also co-produces significant quantities of ${}^{238}$Pu, at 0.7 to 7.8 t yr${}^{-1}$, three orders of magnitude above current kilogram-scale production~\cite{dustin2021assessment}. However, due to the largely fast neutron spectrum, the bred plutonium starts far from flight grade. The fresh ${}^{236}$Pu fraction of the plutonium falls from 2.4\% in the 5 mm channel to 1.1\% at 30 mm, with thicker channels moderating further and favoring capture over (n,2n), while ${}^{238}$Pu fuel has historically had ${}^{236}$Pu near 2 ppm~\cite{Nelson2023}.

However, freshly made Pu in the Np blanket is up to $\sim$2.4\% ${}^{236}$Pu, meaning that we must wait for the ${}^{236}$Pu to decay. ${}^{236}$Pu decays 31 times faster than ${}^{238}$Pu, so the atom ratio falls as $e^{-(\lambda_{236}-\lambda_{238})t}$, a factor of ten every 9.8 years. Plutonium from the 30 mm channel falls below 100 ppm at 20 yr, 10 ppm at 30 yr, and 2 ppm at 37 yr with 75\% of its ${}^{238}$Pu remaining. In steady state, the 7.8 t yr${}^{-1}$ of fresh ${}^{238}$Pu production becomes 5.8 t yr${}^{-1}$ of flight-grade material. Reaching 2 ppm within ten years would instead require breeding at 21 ppm, five hundred times below the 30 mm channel, beyond what channel thickness can do ($k_\mrm{eff}$ is already 0.68 at 30 mm), though a capture-only ${}^{237}$Np station shielded from the uncollided 14 MeV flux could start far lower. Continuous extraction also shortens the wait: material recovered early in a 30 yr irradiation ages in the stream, and the 5 mm depletion run of Appendix \ref{app:u233} ends holding 0.9\% versus its 2.4\% fresh ratio. Since ${}^{232}$U and its daughters accumulate during storage, the aged plutonium product requires further chemical purification to obtain isotopically pure ${}^{238}$Pu.

\subsection{Fusion energy converted to battery fuel} \label{sec:fusion_to_battery}

A useful metric for a fusion-driven battery economy is the decay energy stored in battery fuel per unit of fusion energy. This is the answer to the question: how much deployed nuclear battery power can we support per unit of deployed fusion power?

If a fraction $f$ of source neutrons breeds a chain with total useful heat $Q_\mrm{chain}$, their ratio is
\begin{equation}
\frac{P_\mrm{battery}}{P_\mrm{fusion}} \;\approx\; \frac{f\,Q_\mrm{chain}}{Q_\mrm{DT}},
\label{eq:fbatt}
\end{equation}
which we show in \Cref{fig:fusion_battery_conversion}, where $Q_\mrm{DT}=17.6$ MeV and $f$ obtained from OpenMC simulations  (\Cref{fig:blanket_scan}). Ratios above 100\% typically indicate production from neutrons obtained via fission multiplication rather than production directly from the fast neutrons from fusion: the ${}^{238}$Pu co-product reaches 115\% at the 20 mm ${}^{237}$Np blanket and a fast ${}^{231}$Pa blanket exceeds 700\% (\Cref{fig:blanket_scan_d,fig:blanket_scan_e,fig:blanket_scan_f}), while direct ${}^{236}$Pu stores 2 to 7\%.

\subsection{Revenue per fusion neutron} \label{sec:vpn}

Following~\cite{parisi2025isotope,parisi2026neutronvalue,parisi2026betaemitters}, the revenue a fusion plant earns per source neutron diverted to a battery isotope is
\begin{equation}
v = v_\mrm{n}^\mrm{elec}\, f\, \frac{\eta\,Q_\mrm{chain}}{\eta_\mrm{th}\, E_\mrm{n}}\, \frac{\mathcal{P}}{\mathcal{M}},
\label{eq:vpn}
\end{equation}
where $v_\mrm{n}^\mrm{elec}\approx\$10^{-20}$ is the revenue from a neutron sold as electricity~\cite{parisi2026neutronvalue}, $f$ is the conversion fraction, $\eta\,Q_\mrm{chain}$ is the electric energy drawn from one bred atom ($\eta$ the RTG efficiency), $\eta_\mrm{th}=0.3$ and $E_\mrm{n}=14.1$ MeV convert to plant electricity, $\mathcal{P}$ is the price premium of an actinide RTG over grid power, and $\mathcal{M}=100$ is the battery structure-to-fuel cost ratio.

\begin{figure}[bt]
\centering
\includegraphics[width=\columnwidth]{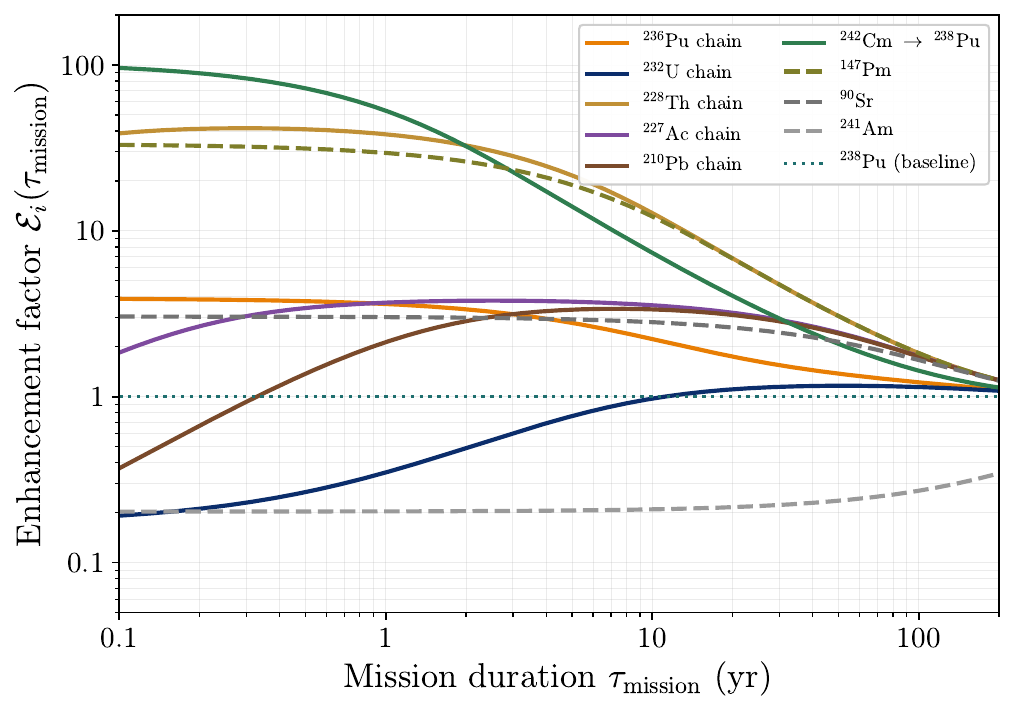}
\caption{Enhancement factor $\mathcal{E}_i$ (\Cref{eq:enhancement}) for different chains relative to ${}^{238}$Pu.}
\label{fig:enhancement_factor}
\end{figure}

A caveat of~\cite{parisi2026betaemitters}, that $\mathcal{P}$ was measured for an $\alpha$-emitting ${}^{238}$Pu RTG but applied to a $\beta^-$ emitter, does not apply here: ${}^{236}$Pu is a candidate $\alpha$ RTG fuel, so the comparison to ${}^{238}$Pu is more appropriate. With $Q_\mrm{chain}=45$ MeV, $\eta\approx0.07$, $f\approx0.5\%$ at the 5 mm channel (\Cref{fig:blanket_scan}), and the ${}^{238}$Pu-derived $\mathcal{P}\approx3.8\times10^{5}$~\cite{parisi2026betaemitters}, the full-chain revenue is $v_{{}^{236}\mrm{Pu}}\approx1.4\times10^{-19}$ per source neutron, an order of magnitude above a neutron spent making ${}^{147}$Pm~\cite{broderick2019reactor}, about 14 times a neutron spent on electricity. \Cref{fig:vpn} shows the value against mission duration using the convention of~\cite{parisi2026betaemitters}. Because $\mathcal{P}$ is measured for ${}^{238}$Pu, a fuel is credited only for the decay heat it releases within the mission, relative to ${}^{238}$Pu on the same mission time, through the enhancement factor
\begin{equation}
\mathcal{E}_i(\tau_\mrm{mission}) = \frac{E_i(\tau_\mrm{mission})/Q_i}{E_{238}(\tau_\mrm{mission})/Q_{238}},
\label{eq:enhancement}
\end{equation}
the ratio of decay-heat fractions released by $\tau_\mrm{mission}$, with $E_i(\tau)$ the cumulative chain energy deposited up to $\tau$ and $Q_i$ the full-chain energy. We plot $\mathcal{E}_i$ in \Cref{fig:enhancement_factor}. Referencing long-lived ${}^{238}$Pu makes power released earlier than ${}^{238}$Pu valuable: short-lived fuels are strongly enhanced ($\mathcal{E}\approx19$ for ${}^{147}$Pm at $\tau_\mrm{mission}=5$ yr; $\mathcal{E}\approx3$ for the ${}^{236}$Pu chain at year-scale missions), while for all fuels $\mathcal{E}_i\to1$ as $\tau_\mrm{mission}\to\infty$.

In \Cref{fig:market} we convert the revenue per neutron into market size and required plant capacity. The annual market is the revenue per neutron times the neutrons spent per year, and the required fusion power scales with the same neutron rate, so lines of constant fusion power are diagonals: mission duration moves a fuel along its diagonal. Producing 1 kg/yr takes 45 MW$_\mrm{fus}$ for ${}^{236}$Pu, 22 MW$_\mrm{fus}$ for ${}^{210}$Pb, 6.3 MW$_\mrm{fus}$ for ${}^{232}$U, and 1.0 MW$_\mrm{fus}$ for the ${}^{238}$Pu co-product; these differences are largely driven by differing cross sections and feedstock densities. At 100 kg/yr, the hypothetical ${}^{236}$Pu fuel market is roughly \$10 to \$20 billion per year on $\sim$4.5 GW$_\mrm{fus}$ of capacity, while ${}^{232}$U has a hypothetical \$5 to \$7 billion market on $\sim$one-fifth of the power.

\begin{figure*}[!tb]
\centering
\includegraphics[width=\textwidth]{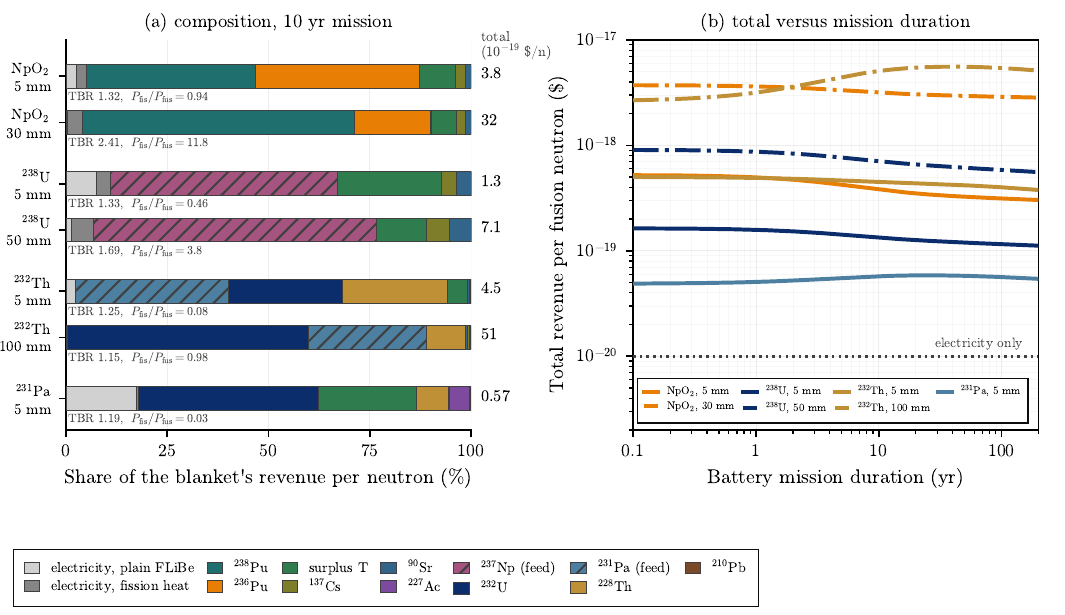}
\caption{Total revenue per fusion neutron of a whole blanket at 1.5 GW$_\mrm{fus}$ of D-T fusion, summing every saleable isotope and the electricity. (a) Composition for a 10 yr battery mission (b) Total against mission duration. Production rates are 30 yr averages from the depletion runs of \Cref{app:depletion}. Pricing in \Cref{tab:prices}.}
\label{fig:blanket_value}
\end{figure*}

\begin{table}[tb]
\centering
\caption{Value of each product (using \Cref{eq:vpn}) for a 10 yr mission. Prices are calculated from ${}^{238}$Pu premium rather than being assumed independently or using external market data. Feedstocks are priced by what a second stage converts them into. Electricity is \$43 per MWh.}
\label{tab:prices}
\begin{tabular}{lrrrr}
\toprule
Product & \thalf & \$M/kg & Ci/g & \$/Ci \\
\midrule
${}^{228}$Th & 1.9 yr & 725 & 820 & 884 \\
${}^{227}$Ac & 21.8 yr & 219 & 72 & 3030 \\
${}^{236}$Pu & 2.9 yr & 146 & 530 & 276 \\
${}^{232}$U & 68.9 yr & 63 & 22 & 2800 \\
${}^{210}$Pb & 22.2 yr & 35 & 77 & 455 \\
${}^{90}$Sr & 28.8 yr & 13 & 138 & 97 \\
${}^{238}$Pu & 87.7 yr & 8.9 & 17 & 519 \\
${}^{137}$Cs & 30.1 yr & 8.8 & 87 & 101 \\
{${}^{147}$Pm} & {2.6 yr} & {2.0} & {927} & {2.1} \\
\midrule
${}^{3}$H & 12.3 yr & 30 & 9620 & 3.1 \\
${}^{231}$Pa$^{a}$ & 32.8 kyr & 26 & 0.047 & --- \\
${}^{237}$Np$^{a,b}$ & 2.1 Myr & 6.8 & 0.0007 & --- \\
\bottomrule
\end{tabular}

\vspace{2pt}
{\footnotesize\raggedright
$^{a}$Feedstock priced through the products made from it in the blanket.\\
$^{b}$${}^{237}$Np has been reported for sale at \$0.66M/kg in 2003~\cite{pomona_neptunium}.\par}
\end{table}

\Cref{eq:vpn} shows the revenue per neutron for each radioisotope, and \Cref{tab:prices} the resulting per-kilogram and per-curie values. Because the actinide blanket materials co-produce multiple products, it is useful to calculate the revenue per source neutron for the whole blanket. \Cref{fig:blanket_value} shows what a whole blanket earns per source neutron: every saleable decay-heat isotope, the feedstock it breeds, the surplus tritium it breeds above its own needs, and the electricity sold at the fission-raised thermal power. Production rates are 30 yr averages from the depletion runs of \Cref{app:depletion}, with feedstock inventories held near their initial values by constant replenishment. All tritium bred above TBR = 1.1 is surplus tritium, priced at \$30M kg$^{-1}$. We assume that holding tritium costs nothing, since tritium decays atom for atom to ${}^{3}$He, which is comparably valuable. The price of \$30M kg$^{-1}$ is likely far too high for tritium and ${}^{3}$He once many fusion power plants are deployed. We exclude the value of fissile ${}^{233}$U and plutonium isotopes heavier than ${}^{238}$Pu. The ${}^{238}$Pu co-product is priced at the flight-grade \$8.9M kg${}^{-1}$ of \Cref{tab:prices}. 

There are several caveats to \Cref{fig:blanket_value}. We do not include the effects of shield mass, which could be particularly misleading at lower thermal powers where ${}^{238}$Pu RTGs are flown unshielded, although as shown in \Cref{sec:shieldmass}, shielding is likely required for human spaceflight and higher thermal power. No costs are included; it is only revenue. These costs will likely be substantial, including extraction from an actinide stream, isotope separation, fabrication, and safeguards. Pricing ${}^{236}$Pu and ${}^{238}$Pu as two products assumes they can be separated, which is highly challenging for ${}^{236}$Pu using the fast neutron spectrum in a blanket, and ${}^{238}$Pu can only be sold decades later when the ${}^{236}$Pu content has decayed sufficiently. The same argument applies to ${}^{232}$U and ${}^{234}$U, although the isotopic purity of ${}^{232}$U is much higher than that of ${}^{236}$Pu (Appendix~\ref{app:purity}). Finally, we assume the market absorbs all products at the fixed price, though one 30 mm blanket makes ${}^{238}$Pu at up to thousands of times the 1.5 kg\,yr${}^{-1}$ NASA has requested~\cite{SpaceNewsPu238,ORNLPu238}.

\subsection{\texorpdfstring{${}^{227}$Ac}{Ac-227} and \texorpdfstring{${}^{232}$U}{U-232} from \texorpdfstring{${}^{231}$Pa}{Pa-231}} \label{sec:u232}
The ${}^{231}$Pa bred in the ${}^{232}$Th blanket of \Cref{sec:scan} is a path to two further alpha fuels: neutron capture converts it to ${}^{232}$U, while its 32,760 yr $\alpha$ decay feeds the ${}^{227}$Ac chain. For scale, the world's entire separated (and declared) ${}^{231}$Pa stock is the 125 g the UK Atomic Energy Authority recovered in 1961~\cite{CEN1961Pa} --- a 1.5 GW$_\mrm{fus}$ plant at the tritium self-sufficiency limit breeds that much in under half an hour (\Cref{fig:blanket_scan_a}).

\textit{${}^{232}$U}: ${}^{232}$U has been studied as an RTG fuel since 1960~\cite{Rohrmann1960,Kulikov2020}, releasing $\sim$39 MeV per atom on the 69 yr ${}^{232}$U half-life. The production route in a Th blanket is
\begin{equation}
\begin{aligned}
{}^{232}\mrm{Th}\ntn{}^{231}\mrm{Th} &\xrightarrow{\betam,\,25.5\,\mrm{h}} {}^{231}\mrm{Pa}\,\ngamma\,{}^{232}\mrm{Pa} \\
&\xrightarrow{\betam,\,1.31\,\mrm{d}} {}^{232}\mrm{U}.
\end{aligned}
\label{eq:u232route}
\end{equation}
The first step has a relatively high cross section of $\sigma_\mrm{n,2n}\approx1.5$ b at 14 MeV~\cite{Brown2018ENDF,Perkin1961}; the second step has a large ${}^{231}$Pa resonance integral of $\sim$500 b~\cite{Brown2018ENDF}. Therefore, the ${}^{231}$Pa(n,$\gamma$) reaction rate can be increased significantly by putting ${}^{231}$Pa in a D$_2$O layer in order to moderate the fast neutrons.

\begin{figure}[tb]
\centering
\begin{subfigure}{\columnwidth}
\centering
\includegraphics[width=\columnwidth]{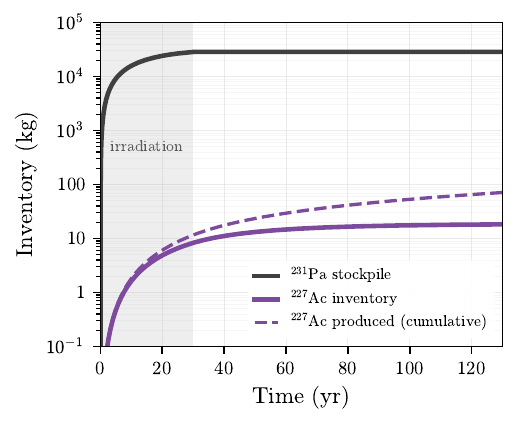}
\caption{}
\label{fig:ac227_growth}
\end{subfigure}\\[4pt]
\begin{subfigure}{\columnwidth}
\centering
\includegraphics[width=\columnwidth]{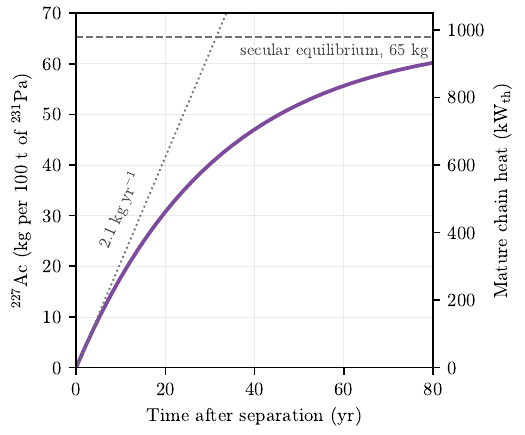}
\caption{}
\label{fig:ac227_per_tonne}
\end{subfigure}
\caption{${}^{227}$Ac from stockpiled ${}^{231}$Pa. (a) Output of the 28.6 t stockpile of the ${}^{232}$Th blanket that retains protactinium (\Cref{fig:deptraj_d}), extracted after a 30 yr irradiation. (b) Ingrowth in 100 t of freshly separated ${}^{231}$Pa, with the thermal power of the mature chain at 15 W/g on the right axis. Dashed: equilibrium at 65 kg; dotted: the 2.1 kg yr${}^{-1}$ initial growth rate.}
\label{fig:ac227}
\end{figure}

\textit{${}^{227}$Ac}: One kilogram of ${}^{231}$Pa yields $\approx 21$ mg yr${}^{-1}$ of ${}^{227}$Ac, so kilogram-per-year ${}^{227}$Ac recovery requires the tonne-scale ${}^{231}$Pa stockpile that accumulates in a $^{232}$Th blanket. A fresh tonne of ${}^{231}$Pa grows ${}^{227}$Ac at 21 g yr${}^{-1}$ from the moment of separation, and its inventory saturates at 0.65 kg with $\sim$10 kW$_\mrm{th}$ of heat. See \Cref{fig:ac227_per_tonne} for a 100 t ${}^{231}$Pa inventory. With ${}^{231}$Pa produced at 3 t yr${}^{-1}$ by a 1.5 GW$_\mrm{fus}$ plant, ${}^{227}$Ac output exceeds 1 kg yr${}^{-1}$ sixteen years after startup, and the same inventory produces the alpha therapy isotope ${}^{223}$Ra~\cite{marques2018targeted}. \Cref{fig:ac227_growth} shows the ${}^{232}$Th blanket (that retains Pa) of \Cref{fig:deptraj_d}; its 28.6 t ${}^{231}$Pa stockpile at 30 yr, extracted and left to decay, provides $\sim$0.6 kg yr${}^{-1}$ of ${}^{227}$Ac. Since neutron burning to ${}^{232}$U outpaces decay to ${}^{227}$Ac by four orders of magnitude, in order to make significant quantities of ${}^{227}$Ac, the ${}^{231}$Pa stockpile for making ${}^{227}$Ac must sit outside the blanket. ${}^{227}$Ac heat sources have precedent: a 1950 study sized a $\sim$70 g yr${}^{-1}$ radium-irradiation route~\cite{ANL1950}, and SCK/CEN produced ${}^{227}$Ac from irradiated radium and developed an Ac$_2$O$_3$ heat source~\cite{Baetsle1971}; the stockpile route replaces the scarce ${}^{226}$Ra feedstock with bred ${}^{231}$Pa. For much longer timescales, Appendix \ref{app:pa231} shows the ${}^{231}$Pa decay chain, or ${}^{230}$Th from a second \ntn\ reaction, as milliwatt per gram heat sources over tens of millennia.

\section{Shield mass comparison of battery fuels} \label{sec:shieldmass}

In this section we discuss the shielding requirements of these different alpha emitters. Some RTG applications are mass-sensitive, spaceflight above all, where every kilogram of shield displaces payload. Others, such as seabed or remote terrestrial power, tolerate tonne-class enclosures, as the shielded ${}^{90}$Sr generators did. The comparisons below matter most for mass-sensitive applications. Most of the work in this section is an extension of the ideas in~\cite{Nelson2023}.

A radioisotope source must meet a specified dose limit at its surface, and if required, the shield can outweigh the fuel and energy converter combined, so shield mass is an important figure of merit for a nuclear battery. Ref.~\cite{Nelson2023} calculated the required shield thicknesses for seven candidate fuels in a spherical geometry, a 100 W$_\mrm{th}$ fuel sphere with a 1 mm Ir clad, using graphite where neutrons set the dose and depleted uranium (DU) where photons do (see also the radionuclide comparison of~\cite{dustin2021assessment}). Thicknesses of different materials are hard to compare, graphite being roughly ten times less dense than DU, so in \Cref{tab:shieldmass} we convert the comparison to total shield mass in the same geometry, which reproduces the totals of~\cite{Nelson2023} for their seven fuels. We also add the ${}^{210}$Pb and ${}^{236}$Pu chains of this work. For ${}^{210}$Pb we computed the DU thickness with OpenMC photon transport of its bremsstrahlung source, converting the photon flux at the surface to an effective dose rate with the standard ICRP-116 coefficients~\cite{ICRP116}, benchmarked on the ${}^{90}$Sr case of~\cite{Nelson2023} (8.8 cm against their 7.7), and rounded the computed 5.5 cm (for ${}^{210}$Pb) up to 6.0 cm. For the ${}^{236}$Pu chain we use the ${}^{232}$U thickness. Following~\cite{Nelson2023}, we evaluate each dose at the fuel's radiological peak: fresh for ${}^{238}$Pu, ${}^{241}$Am, and ${}^{244}$Cm; one year for ${}^{227}$Ac and ${}^{228}$Th; ten years for ${}^{232}$U; eighteen years for the ${}^{236}$Pu chain; one month for ${}^{210}$Pb. These are worst-case values over the source life, as freshly separated chain fuels start with much lower doses (Appendix \ref{app:mars}). Note that \Cref{tab:shieldmass} is for a contact dose of 10 mrem/h and a 100 W$_\mrm{th}$ source. A more lenient dose limit of 100 mrem/h requires no ${}^{238}$Pu or ${}^{241}$Am shielding for a 100 W$_\mrm{th}$ source~\cite{Nelson2023}.

\begin{figure*}[tb]
\centering
\begin{subfigure}{0.49\textwidth}
\centering
\includegraphics[width=\textwidth]{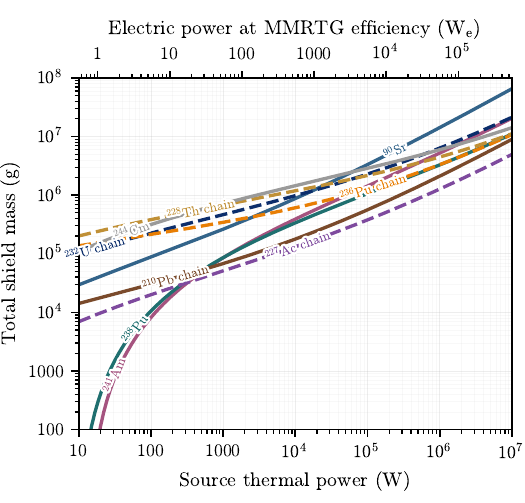}
\caption{10 mrem/h contact dose}
\label{fig:shieldmass_a}
\end{subfigure}
\hfill
\begin{subfigure}{0.49\textwidth}
\centering
\includegraphics[width=\textwidth]{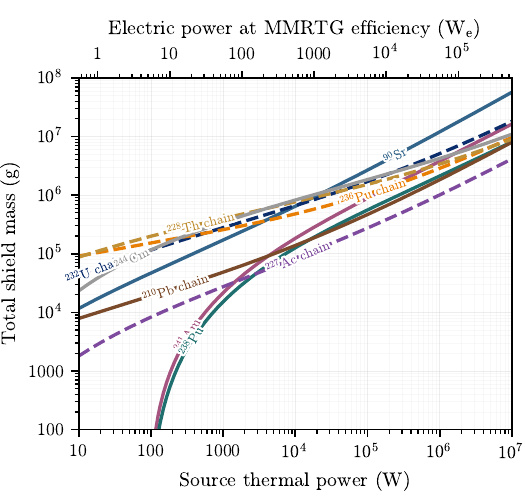}
\caption{100 mrem/h contact dose}
\label{fig:shieldmass_b}
\end{subfigure}
\begin{subfigure}{0.49\textwidth}
\centering
\includegraphics[width=\textwidth]{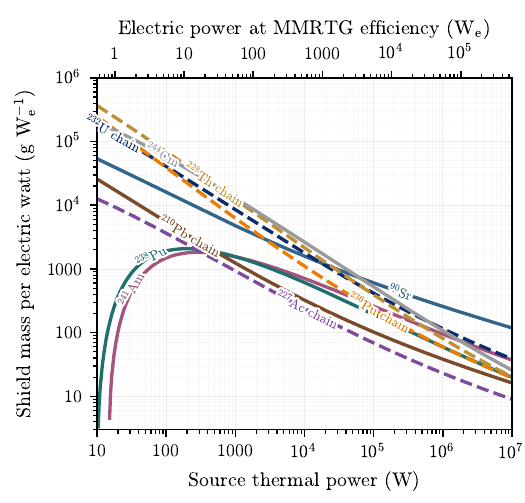}
\caption{10 mrem/h, per electric watt}
\label{fig:shieldpw_a}
\end{subfigure}
\hfill
\begin{subfigure}{0.49\textwidth}
\centering
\includegraphics[width=\textwidth]{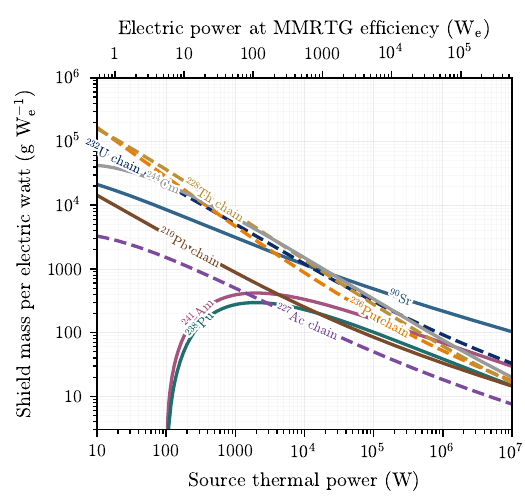}
\caption{100 mrem/h, per electric watt}
\label{fig:shieldpw_b}
\end{subfigure}
\caption{Total shield mass versus source thermal power at contact dose thresholds of (a) 10 and (b) 100 mrem/h, in minimum-mass configurations: DU for the photon-limited fuels, with thicknesses and slopes from~\cite{Nelson2023}, and LiH for the neutron-limited ${}^{238}$Pu, ${}^{241}$Am, and ${}^{244}$Cm. Panels (c) and (d) give the same shields per watt of electric output at the 5.5\% conversion efficiency of the MMRTG~\cite{lee2015rps}, which also gives the top x-axes. Curves end where the bare source meets the threshold. Each fuel is evaluated at its radiological peak: fresh for ${}^{238}$Pu, ${}^{241}$Am, and ${}^{244}$Cm, one year for ${}^{227}$Ac and ${}^{228}$Th, ten years for ${}^{232}$U, eighteen for the ${}^{236}$Pu chain, one month for ${}^{210}$Pb.}
\label{fig:shieldmass}
\end{figure*}

We make three main conclusions from \Cref{tab:shieldmass} for 100 W$_\mrm{th}$-scale sources. First, at 100 W$_\mrm{th}$ the ${}^{210}$Pb chain needs 30 kg of shield, less than the 65 kg of graphite-shielded ${}^{238}$Pu but more than its 11 kg LiH minimum. The ${}^{210}$Pb shield grows much more slowly with power, 1 cm of DU per decade against 13 cm of LiH, so above roughly a kilowatt ${}^{210}$Pb needs less shield than ${}^{238}$Pu under either configuration (\Cref{fig:shieldmass}). Lead metal also gives the chain's alphas no ($\alpha$,n) target, so it emits no neutrons at all, where PuO$_2$ produces $\sim$$2{\times}10^{4}$ n/s per gram of ${}^{238}$Pu from its oxygen. PbTe keeps this neutron-free property and melts at 1197 K instead of lead's 601. Second, the ${}^{227}$Ac chain has the lightest shield of the gamma chains at 20 kg: 36 MeV per decay makes its source tiny and its photon output per watt only about twice ${}^{210}$Pb's, absorbed in half a centimeter of DU. However, in the LiH minimum column, ${}^{241}$Am and ${}^{238}$Pu are lighter at this power, 8 and 11 kg. Third, the ${}^{208}$Tl chains (${}^{232}$U, ${}^{228}$Th, and mature ${}^{236}$Pu without the radon separation of \Cref{sec:radon}) need $\sim$200 to 400 kg. One assumption behind the DU-only rows is that the gamma chains do produce ($\alpha$,n) neutrons in their oxides, but behind the DU their neutron dose is below 1 mrem/h at this power, so no neutron shield is needed at the 10 mrem/h threshold. A requirement below 1 mrem/h would also need neutron moderation, and the layered graphite and DU shield designs of~\cite{Nelson2023} are then necessary and an order of magnitude heavier.

\begin{table}[tb]
\centering
\caption{Total shield mass for 100 W$_\mrm{th}$ sources at a 10 mrem/h contact dose, in the mass-optimal configuration of~\cite{Nelson2023}: graphite where neutrons set the dose, DU alone for the gamma chains. Only ${}^{238}$Pu, ${}^{241}$Am, and ${}^{148}$Gd can operate unshielded; every other fuel is Sv/h-class bare. Fuel is sized to hold 100 W$_\mrm{th}$ for 20 years (5 for ${}^{228}$Th), and each dose is evaluated at the fuel's radiological peak: fresh for ${}^{238}$Pu, ${}^{241}$Am, and ${}^{244}$Cm; one year for ${}^{227}$Ac and ${}^{228}$Th; ten years for ${}^{232}$U; eighteen for ${}^{236}$Pu; one month for ${}^{210}$Pb. Masses for the seven fuels of~\cite{Nelson2023} are from its Table A-1; the rest are this work. The last two columns give the minimum-mass configuration, LiH in place of graphite. Rows are ordered by minimum mass.}
\label{tab:shieldmass}
\setlength{\tabcolsep}{3.5pt}
\footnotesize
\begin{tabular}{lrrrrr}
\toprule
Fuel & Graphite & DU & Shield & LiH & Min. \\
 & (cm) & (cm) & (kg) & (cm) & (kg) \\
\midrule
${}^{148}$Gd$^{d}$ & 0 & 0 & 0 & 0 & 0 \\
${}^{241}$Am & 12.0 & 0 & 33 & 10.8 & 8 \\
${}^{238}$Pu & 17.0 & 0 & 65 & 13.0 & 11 \\
${}^{227}$Ac chain & 0 & 5.5 & 20 & --- & 20 \\
${}^{210}$Pb chain$^{b}$ & 0 & 6.0 & 30 & --- & 30 \\
${}^{90}$Sr & 0 & 7.7 & 76 & --- & 76 \\
${}^{236}$Pu chain$^{b,c}$ & 0 & 13.0 & 213 & --- & 213 \\
${}^{232}$U chain & 0 & 13.1 & 223 & --- & 223 \\
${}^{244}$Cm$^{a}$ & 67.1 & 0 & 3250 & 44.6 & 320 \\
${}^{228}$Th chain & 0 & 16.3 & 385 & --- & 385 \\
\bottomrule
\end{tabular}

\vspace{2pt}
{\footnotesize\raggedright
$^{a}$The Cd/Gd layers of the mixed ${}^{244}$Cm shield are counted as graphite.\\
$^{b}$This work: the ${}^{210}$Pb DU from OpenMC photon transport of its bremsstrahlung source, benchmarked on the ${}^{90}$Sr case of~\cite{Nelson2023} and rounded up; the ${}^{236}$Pu DU is sized the same as for ${}^{232}$U.\\
$^{c}$Mature chain, no radon separation; DU sized as for the ${}^{232}$U chain, whose emissions it shares.\\
$^{d}$A truly pure alpha emitter: a single alpha to stable ${}^{144}$Sm with no photon, electron, or neutron emission. It is exceptionally hard to make at any meaningful quantity.\par}
\end{table}

\begin{figure*}[tb]
\centering
\begin{subfigure}{0.49\textwidth}
\centering
\includegraphics[width=\textwidth]{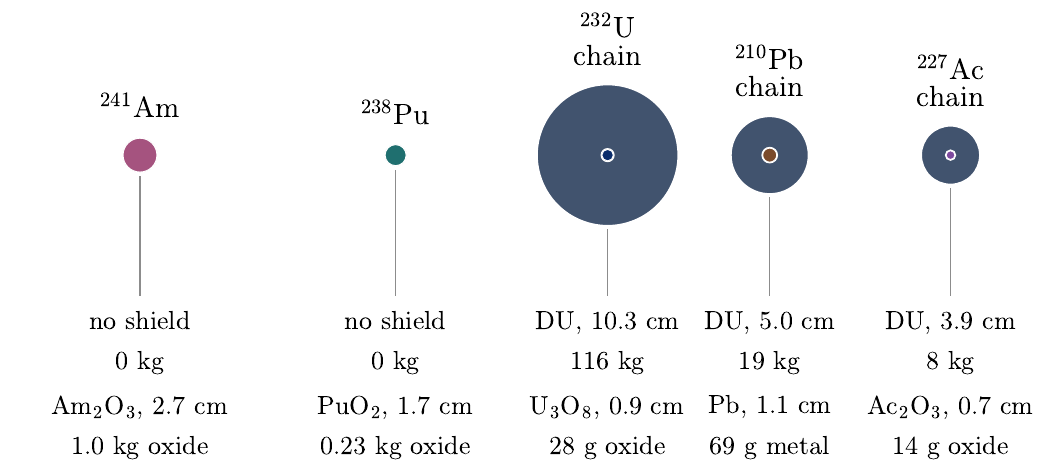}
\caption{100 W$_\mrm{th}$}
\label{fig:shield_schematic_a}
\end{subfigure}
\hfill
\begin{subfigure}{0.49\textwidth}
\centering
\includegraphics[width=\textwidth]{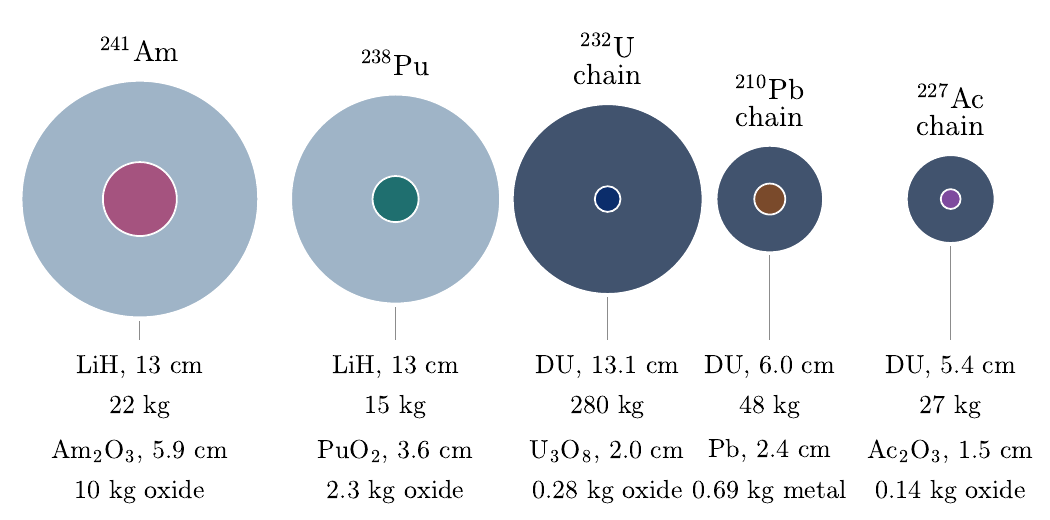}
\caption{1 kW$_\mrm{th}$}
\label{fig:shield_schematic_b}
\end{subfigure}\\[6pt]
\begin{subfigure}{0.49\textwidth}
\centering
\includegraphics[width=\textwidth]{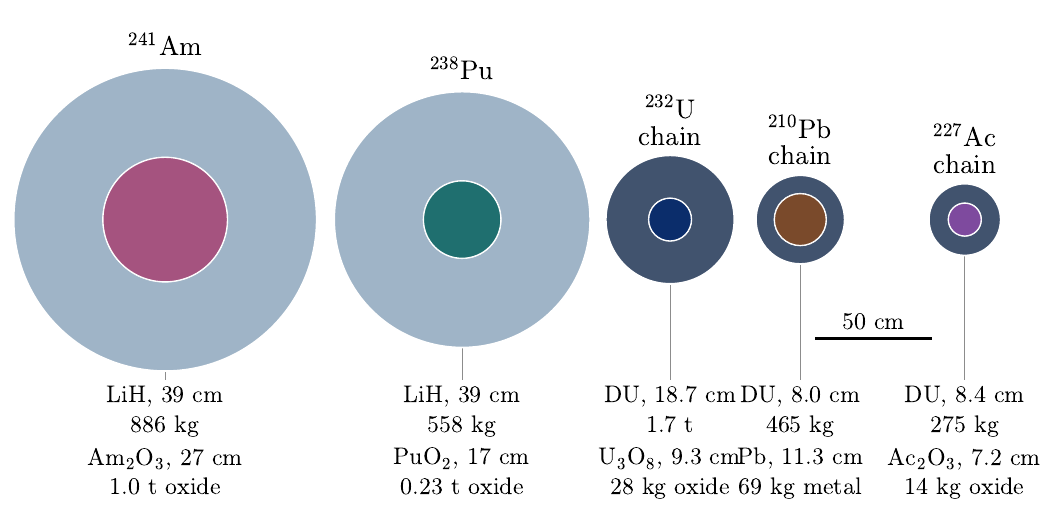}
\caption{100 kW$_\mrm{th}$, reduced scale}
\label{fig:shield_schematic_c}
\end{subfigure}
\hfill
\begin{subfigure}{0.49\textwidth}
\centering
\includegraphics[width=\textwidth]{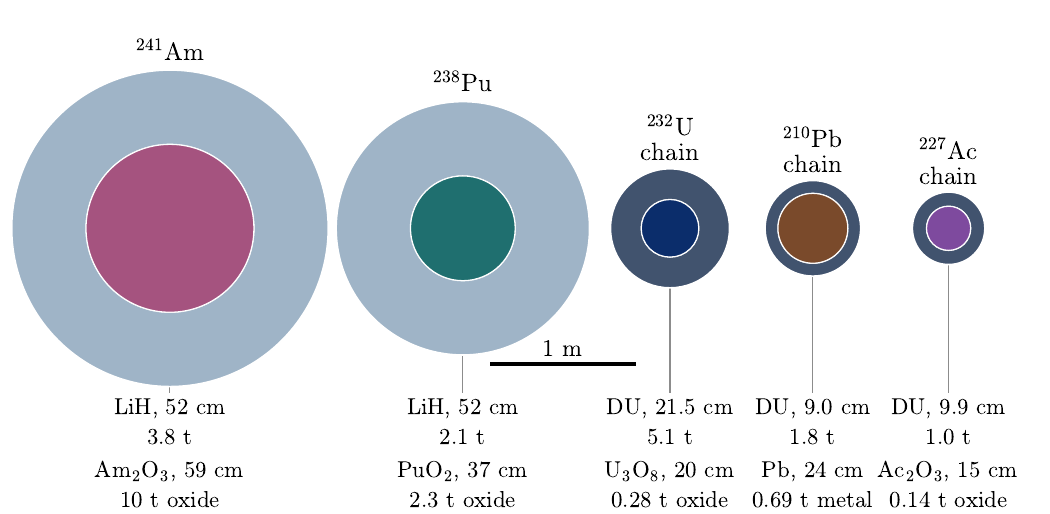}
\caption{1 MW$_\mrm{th}$, reduced scale}
\label{fig:shield_schematic_d}
\end{subfigure}
\caption{Shield configurations at 100 mrem/h on contact for (a) 100 W$_\mrm{th}$, (b) 1 kW$_\mrm{th}$, (c) 100 kW$_\mrm{th}$, and (d) 1 MW$_\mrm{th}$ sources, drawn to scale on one length scale in the spherical fuel, iridium clad, shield geometry of Fig.~7 of Ref.~\cite{Nelson2023}. At 100 W$_\mrm{th}$ the ${}^{238}$Pu and ${}^{241}$Am pellets need no shield at all; at 1 kW$_\mrm{th}$ they take the minimum-mass LiH of \Cref{fig:shieldmass}, while the chains take depleted uranium for bremsstrahlung and line gammas at all powers. By 100 kW$_\mrm{th}$ the shield mass ordering is reversed, and at 1 MW$_\mrm{th}$ the ${}^{227}$Ac chain needs a quarter of the ${}^{241}$Am shield mass. Panels (a) and (b) share one length scale; (c) and (d) are reduced, with their own bars. Near 1 MW$_\mrm{th}$ the ($\alpha$,n) neutron dose of the oxide chains reaches the threshold and a thin LiH liner, tens of kg, would be added. The fuel spheres of (c) and (d) are illustrative; heat removal would subdivide the fuel inside the one shield. At the lower powers shown, the ($\alpha$,n) neutrons of the DU-shielded oxides stay below the threshold unshielded~\cite{Nelson2023}.}
\label{fig:shield_schematic}
\end{figure*}

In \Cref{fig:shieldmass} we compare the shielding requirements versus source power, and in \Cref{fig:shield_schematic} we show four configurations at four thermal powers. We find two shielding regimes. At low power the neutron-limited fuels are the light ones: below a few tens of watts at 10 mrem/h, and a few hundred at 100 mrem/h, ${}^{238}$Pu and ${}^{241}$Am need no shield at all, and their LiH shields stay lighter than any chain fuel's DU up to a few hundred watts, while the chain fuels make Sv/h bare at any power and always need their DU, 20 kg for ${}^{227}$Ac and hundreds for the ${}^{208}$Tl chains at 100 W$_\mrm{th}$. At high power the ordering inverts and the DU-shielded chains are the light ones: above a few hundred watts the ${}^{227}$Ac chain has the lightest shield of any fuel, ${}^{210}$Pb follows, and the LiH-shielded ${}^{238}$Pu, ${}^{241}$Am, and ${}^{244}$Cm are the heaviest, ${}^{227}$Ac needing a quarter of the ${}^{241}$Am mass at 1 MW$_\mrm{th}$.

${}^{238}$Pu RTG practice to date has never flown a dedicated radiation shield, since the iridium clad, graphite impact shells, and aeroshell around their fuel are for reentry and impact containment. The 2 kW$_\mrm{th}$ MMRTG~\cite{lee2015rps}, an assembly of eight 250 W$_\mrm{th}$ general purpose heat source (GPHS) modules, runs a few hundred mrem/h at its housing and a few mrem/h at 1 m, accepted and managed by distance and procedure, and Apollo crews hand-fueled SNAP-27 generators on the lunar surface~\cite{bennett2006space}. The terrestrial ${}^{90}$Sr generators, facing significant bremsstrahlung, were always shielded. For scale, a 10 mrem/h contact dose roughly doubles the 7.7 mrem/h galactic cosmic ray background a crew in transit to the Moon or Mars already receives~\cite{zeitlin2013measurements}, and on the Martian surface the background is 2.7 mrem/h~\cite{hassler2014mars}.

This inversion is caused by the different attenuation of neutrons and photons in materials. Each factor of ten in source power requires $\sim$one more factor of ten of attenuation, and that factor takes very different thicknesses for the two radiation types: 1.5 to 3 cm of DU for the chains' photons~\cite{Nelson2023}, compared with 13 cm of LiH~\cite{lahti1968fission} or 16 to 19 cm of graphite for the fast neutrons of ${}^{238}$Pu, ${}^{241}$Am, and ${}^{244}$Cm, which no dense metal stops and only light nuclei moderate. The shell mass grows approximately as thickness times radius squared, so the thick neutron shields also increase their own radius as power rises, and their mass increases roughly ten times faster per decade of power than the photon shields'. The two classes reverse shielding requirements at a few hundred watts. This holds even though \Cref{fig:shieldmass} gives the neutron-limited fuels their best, lowest shielding mass, LiH, flight-proven on SNAP-10A and breeding only tens of microcuries of tritium per 100 W$_\mrm{th}$ over twenty years. Along every curve of \Cref{fig:shieldmass} the total shield mass grows sublinearly with power, $m \propto P^{\alpha}$ with local exponents $\alpha \approx 0.4$ to 0.9, so per electric watt (\Cref{fig:shieldmass}c,d) the specific shield mass $m/P_\mrm{e} \propto P^{\alpha-1}$ falls with power: at 55 kW$_\mrm{e}$ the ${}^{227}$Ac shield costs 0.02 kg per W$_\mrm{e}$, thirty times less than at 100 W$_\mrm{th}$. Near 1 MW$_\mrm{th}$ the oxide chains' ($\alpha$,n) neutron dose grows to the order of the 100 mrem/h threshold itself, so a thin LiH layer, of order 10 cm and tens of kg versus their tonne of DU, must be added to keep the total dose below the threshold. ${}^{210}$Pb metal, with no oxygen target, needs no such layer at any power.

Subdividing the source into smaller units helps only where shield mass rises faster than linearly with power, which happens only at the onset of the ${}^{238}$Pu and ${}^{241}$Am shielding requirements. Here, modules each stay below the bare threshold and are spatially dispersed so that no accessible surface sees more than one, and then they need no shield at all. Everywhere else the growth is sublinear ($\alpha < 1$), so $N$ modules weigh $N^{1-\alpha}$ times the single shield when dispersed and $N$ times when co-located. For the chain fuels one shield is always the minimum, with fuel subdivided inside it for heat removal, as in the MMRTG's eight GPHS modules.

There may be creative ways to reduce the shield mass further. On a robotic vehicle ${}^{210}$Pb may only need a shadow shield, a narrow cone toward the payload covering a few percent of the full sphere at a few kilograms for 100 W$_\mrm{th}$, with the full enclosure staying on the ground as a handling cask. Where shield mass is cheap, fusion breeding supplies the high-power chains in quantity, and where it is costly, ${}^{210}$Pb is the light-shield fuel. For very small sources without shielding, ${}^{238}$Pu and ${}^{241}$Am are best.

There may also be ways to significantly reduce the relatively heavy shielding of the ${}^{208}$Tl chains. The 2.6 MeV line that sets the ${}^{236}$Pu chain's shield comes from ${}^{208}$Tl, and every path to ${}^{208}$Tl passes through 55.6 s ${}^{220}$Rn gas. Separating that radon as it appears removes the gamma at its source, and freshly separated fuel emits only a few percent of its mature output in its first months anyway (\Cref{app:mars}). \Cref{sec:radon} develops four implementations, from a shielded plate-out chamber that keeps the heat to venting the gas.

\section{Separating the ${}^{236}$Pu chain at ${}^{220}$Rn} \label{sec:radon}

The main challenge of using the ${}^{236}$Pu chain in a nuclear battery is the 2.6 MeV ${}^{208}$Tl gamma (Appendix \ref{app:decaydata}), which has historically disqualified ${}^{236}$Pu as RTG fuel: an ORNL study~\cite{Nelson2023} finds that 100 W$_\mrm{th}$ oxide sources of ${}^{232}$U, ${}^{228}$Th, or ${}^{228}$Ra at peak activity require 10 to 15 cm of depleted uranium for occupational contact dose. We describe possible approaches that reduce this shielding while keeping the dose below occupational limits.

Only one of the chain's eleven isotopes is a noble gas, ${}^{220}$Rn, so we propose separating the chain there: ${}^{220}$Rn diffuses out of a porous fuel, taking the gamma-emitting lower segment away while the four-alpha upstream chain (${}^{236}$Pu $\to$ ${}^{232}$U $\to$ ${}^{228}$Th $\to$ ${}^{224}$Ra, 22 MeV) stays behind as heat. Radon emanation is well characterized in geophysics~\cite{Sakoda2011}, in ${}^{224}$Ra powders~\cite{Danylec2018}, and in the ${}^{212}$Pb generators of targeted alpha therapy~\cite{Yong2015,Kokov2022,Pretze2025}, reaching fractions of 0.5 to 0.7 when the grain size of the fuel material approaches the $\sim$50 nm ${}^{220}$Rn recoil range~\cite{Sakoda2011,Danylec2018}: the fuel must therefore be porous enough for ${}^{220}$Rn to escape within its 55.6 s half-life.

\begin{figure*}[!tb]
\centering
\begin{subfigure}{0.49\textwidth}
\centering
\includegraphics[width=\textwidth]{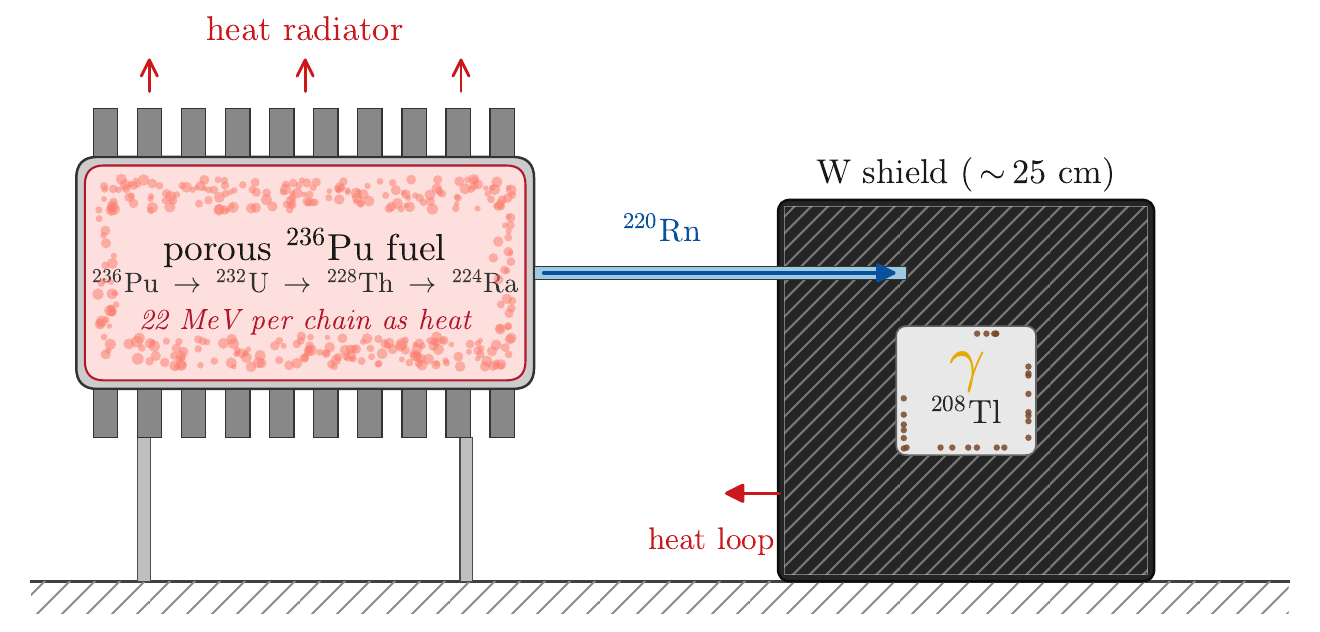}
\caption{}
\label{fig:radon_chamber}
\end{subfigure}
\hfill
\begin{subfigure}{0.49\textwidth}
\centering
\includegraphics[width=\textwidth]{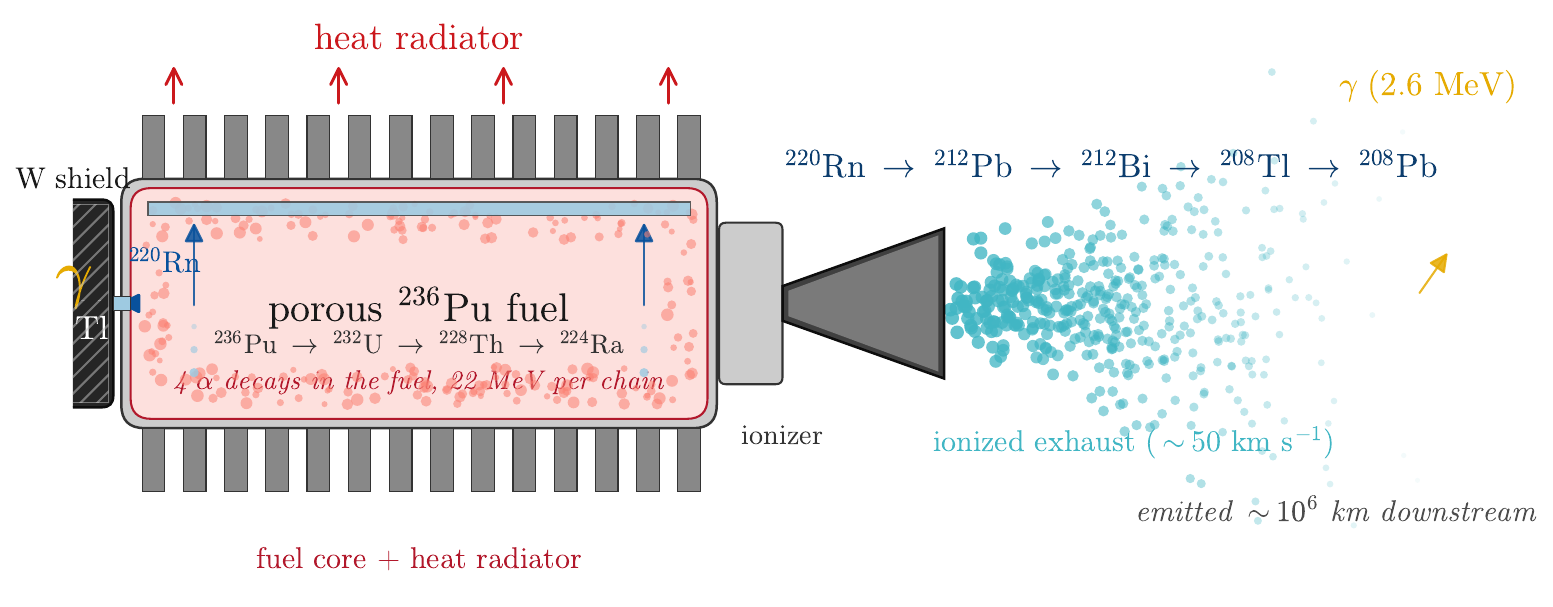}
\caption{}
\label{fig:radon}
\end{subfigure}\\[2pt]
\begin{subfigure}{0.49\textwidth}
\centering
\includegraphics[width=\textwidth]{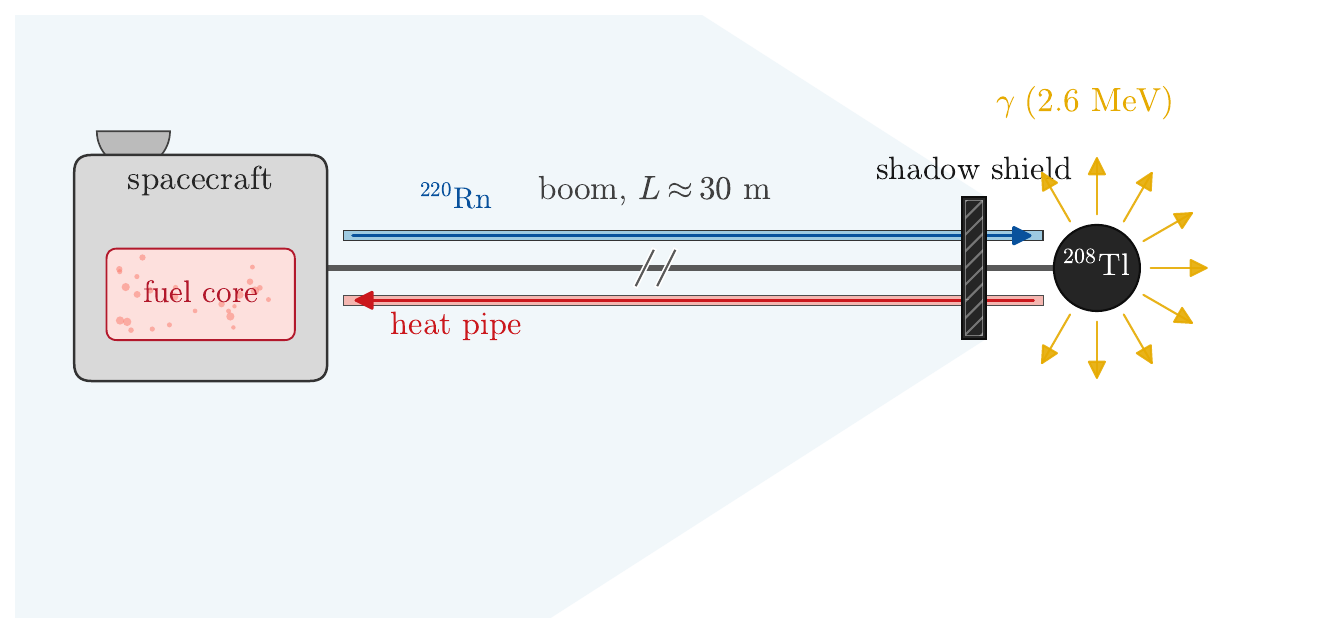}
\caption{}
\label{fig:radon_boom}
\end{subfigure}
\hfill
\begin{subfigure}{0.49\textwidth}
\centering
\includegraphics[width=\textwidth]{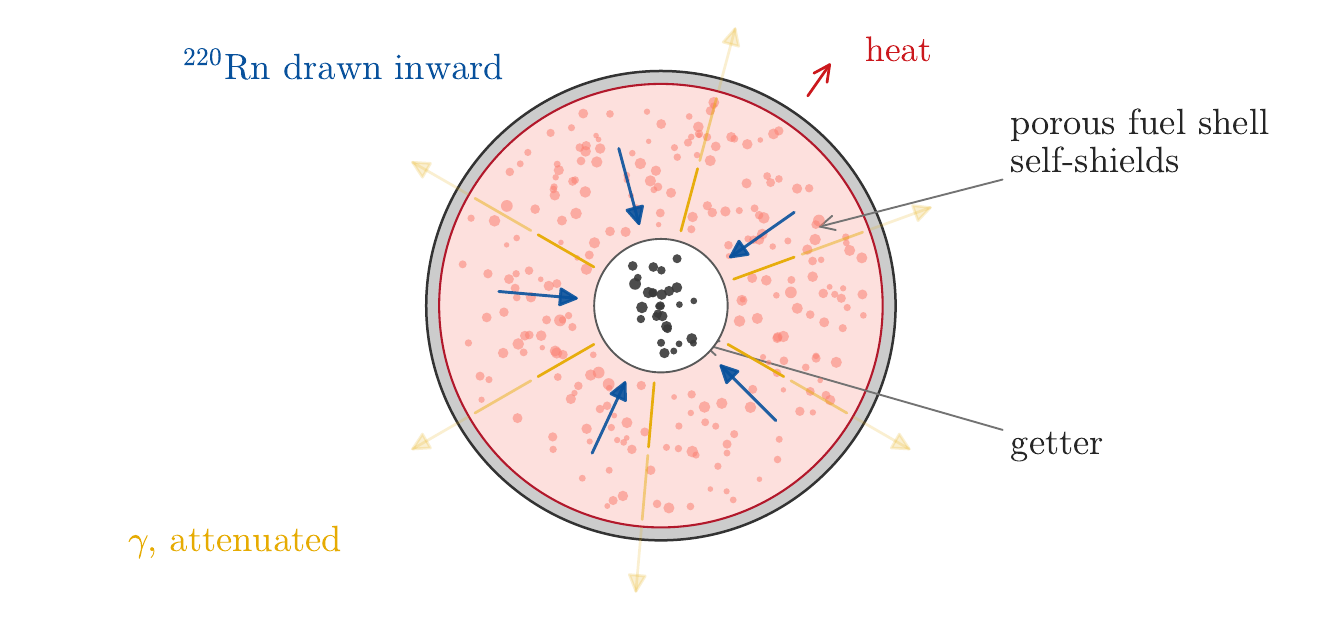}
\caption{}
\label{fig:radon_selfshield}
\end{subfigure}
\caption{Four ways to manage the ${}^{208}$Tl gamma. In each, a porous fuel core contains the ${}^{236}$Pu $\to$ ${}^{232}$U $\to$ ${}^{228}$Th $\to$ ${}^{224}$Ra chain (22 MeV per chain as heat) and emanated ${}^{220}$Rn is collected within its 55.6 s half-life. (a) The gas finishes its chain in a thick chamber, the daughters plate out on the chamber walls and the chain heat returns through a secondary loop. (b) In space: an ionizer and bell nozzle exhaust the gas-phase segment as a $\sim$50 km s${}^{-1}$ ion stream, the 10.6 h ${}^{212}$Pb half-life delays the ${}^{208}$Tl ingrowth, placing the 2.6 MeV source $\sim 10^{6}$ km behind the spacecraft. (c) Propellant-free: the gas-phase segment decays in a chamber on a $\sim$30 m boom, the spacecraft intercepts $A/(4\pi L^{2}) \approx 2\times10^{-3}$ of flux, a $\sim$10 kg shadow shield at the chamber supplies the rest. (d) Fully passive: ${}^{220}$Rn is drawn to a central getter and the fuel shell self-shields the resulting source.}
\label{fig:radon_options}
\end{figure*}

Here we describe four potentials ways to exploit this, each sketched in \Cref{fig:radon_options}. The simplest, for a stationary battery, pipes ${}^{220}$Rn to a shielded back-end chamber (\Cref{fig:radon_chamber}), where the ${}^{216}$Po (0.15 s) and its successors plate out on the walls, localizing the ${}^{208}$Tl gamma. At the 5 W g${}^{-1}$ quasi-equilibrium point (after $\sim$10 years of ${}^{236}$Pu decay, \Cref{fig:specific_power}) each kilogram of fuel supplies about 100 W of 2.6 MeV photons ($\sim$100 Sv/h unshielded at 1 m), and the $10^{7}$ reduction to a 10 $\mu$Sv/h surface takes $\sim$25 cm of tungsten or $\sim$40 cm of lead (tenth-value layers 2.8 and 4.8 cm), roughly 2 or 5 t for a compact chamber. Because attenuation is exponential, the shield grows by only one tenth-value layer per factor of ten in fuel inventory, so this is a fixed cost that acceptable for a ground installation with a secondary thermal loop, but likely prohibitive in spaceflight.

The second option, for a spacecraft in vacuum, vents ${}^{220}$Rn through an ionizer-accelerator and nozzle (\Cref{fig:radon}): the gamma appears only after ${}^{212}$Pb (10.6 h) and ${}^{212}$Bi (61 min) decay, so at the $\sim$40 km/s of a gridded ion thruster running on the radon~\cite{herman2008next} the ${}^{208}$Tl source forms a safe distance $\sim$$10^{6}$ km behind the vehicle. Radon that decays before ejection plates ${}^{212}$Pb onto the walls of the feed line and releases its 36\% ${}^{208}$Tl branch aboard, so the transit time from fuel to nozzle must be short compared with radon's 55.6 s half-life. The exhaust rate is about 20 mg per day per kilogram of fresh ${}^{236}$Pu at full emanation.

The third option, when propellant loss is unacceptable, decays the gas-phase segment in a chamber on a deployable boom (\Cref{fig:radon_boom}): a 25 m${}^{2}$ spacecraft at $L=30$ m intercepts only $A/(4\pi L^2) \approx 2\times 10^{-3}$ of the gamma flux. A shadow shield mounted at the source need only cover the spacecraft's solid angle, so 5 cm of tungsten ($\sim$10 kg) pushes the total reduction factor past $10^4$. A tether could extend the separation, for example, to $L=300$ m ($\sim$$10^{-5}$ intercepted). Alternatively, a long thin fuel rod viewed end-on is largely self-shielding, so inventory can grow at minimal on-axis dose cost, protecting astronauts and equipment along the axis.

The fourth option is fully passive: drive ${}^{220}$Rn inward to a central getter cavity and let the actinide shell self-shield the source (\Cref{fig:radon_selfshield}). With a linear photon attentuation of $\mu \approx 0.9$ cm${}^{-1}$ in Pu metal at 2.6 MeV, spheres of 1, 8, and 100 kg attenuate by about one, three, and nine e-foldings net of buildup, while the required attenuation grows by one tenth-value layer per decade of inventory, so self-shielding alone would need a tonne-scale sphere. Shown in \Cref{tab:configs}, it instead reduces the external shield by 30 to 55\%. Only radon born within a diffusion length $\sqrt{D\,\thalf} \sim 7$ cm of the cavity reaches it before decaying (gas-phase $D \approx 0.9$ cm$^2$ s$^{-1}$ at 900 K, scaling the measured 0.12 cm$^2$ s$^{-1}$ of radon in air at room temperature~\cite{hirst1939diffusion} as $T^{1.75}$~\cite{fuller1966new}), so a thicker shell needs forced circulation to draw the radon inward.

An important caveat is that venting is only safe in vacuum, far away from life: a terrestrial battery must contain ${}^{220}$Rn and its daughters. Assembly can also leave its shielding behind for certain space applications: freshly separated ${}^{236}$Pu emits 0.04\% of its mature gamma output one month after separation (Appendix \ref{app:mars}), so a battery docked in Earth orbit is handled at its quietest by relatively massive depot infrastructure that never leaves orbit, and the vehicle can depart with only its few-kg shadow shield while the source matures en-route.

\section{Discussion} \label{sec:discussion}

We have shown that the coming availability of fusion neutrons will enable new capabilities for nuclear batteries, producing new alpha emitter fuels with over ten times the energy density of existing alpha emitters. These fuels are of two kinds: chains whose production pathways are proposed here for the first time (${}^{236}$Pu, ${}^{210}$Pb, and ${}^{227}$Ac from stockpiled ${}^{231}$Pa), and chains identified as heat-source candidates in the 1960s whose routes were blocked by scarce natural ${}^{230}$Th and ${}^{231}$Pa feedstock, which this work breeds at tonne scale from ${}^{232}$Th (${}^{232}$U, ${}^{228}$Th)~\cite{Rohrmann1960,Rohrmann1963}. Beyond enabling these new fuels, fusion neutrons will significantly expand the supply of established ones, ${}^{238}$Pu above all, and allow nuclear batteries at powers far beyond what RTGs have achieved, the range NASA's Kilopower fission reactor program targeted and beyond~\cite{Gibson2017,Poston2020KRUSTY}: at hundreds of kilowatts to megawatts thermal. These batteries could complement the 100 kW$_\mrm{e}$-class reactors of NASA's ongoing Fission Surface Power project~\cite{NASAFSP}.

\begin{figure*}[!tb]
\centering
\includegraphics[width=\textwidth]{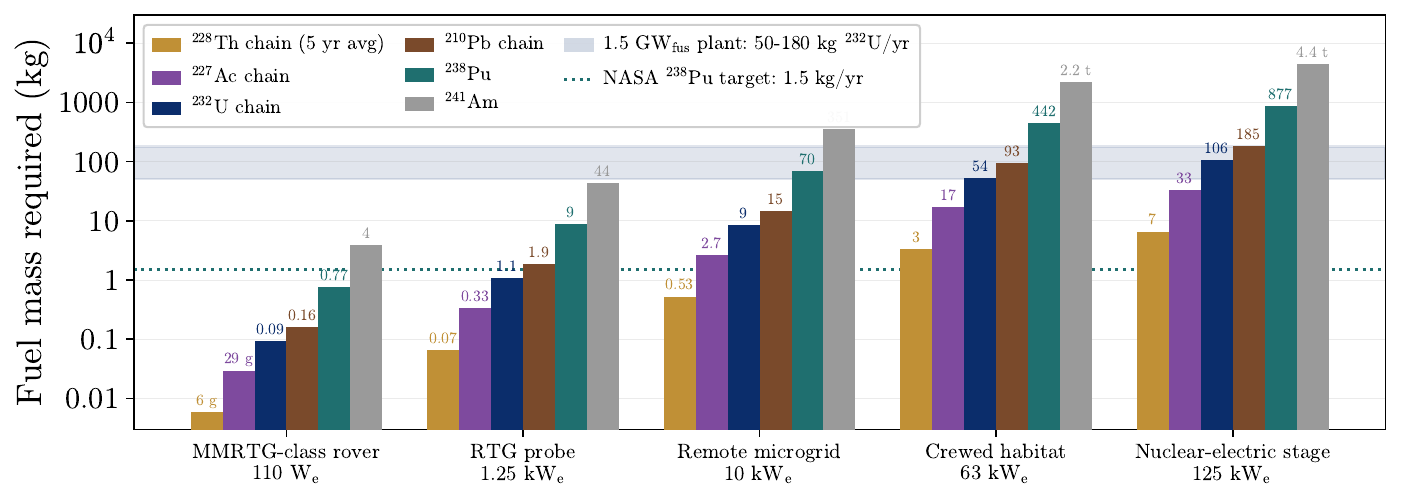}
\caption{Fuel mass to supply the electric power for various missions (\Cref{tab:configs}) with the ${}^{228}$Th chain (76 W g${}^{-1}$), the ${}^{227}$Ac chain (15), the ${}^{232}$U chain (4.7), the ${}^{210}$Pb chain (2.7), ${}^{238}$Pu (0.57), and ${}^{241}$Am (0.114) all through the same 25\% converter. Shield mass is excluded. The dotted line is NASA's annual ${}^{238}$Pu target.}
\label{fig:fuel_mass}
\end{figure*}

\begin{table}[bt]
\centering
\setlength{\tabcolsep}{1.5pt}
\footnotesize
\begin{tabular}{lccll}
\toprule
Application            & Fuel    & $P_\mrm{e}$ & Gamma management & Shield \\
                      & (kg)    & (kW)        & & (t) \\
\midrule
Undersea glider       & 0.1     & 0.13        & self-shielded getter & 0.3 \\
Remote microgrid      & 8       & 10          & shielded chamber & 1.3 \\
\midrule
RTG probe             & 1       & 1.25        & vent or boom & 0.04 \\
Crewed habitat        & 50      & 63          & fully contained & 3.2 \\
Nuclear-electric stage & 100    & 125         & vent overboard & 0.3 \\
NEP freighter (${}^{228}$Th) & 530 & 10000 & vent overboard & 0.9 \\
\midrule
Rover (${}^{227}$Ac) & 0.07 & 0.25 & DU shield & 0.05 \\
Seabed sensor (${}^{210}$Pb) & 0.007 & 0.005 & DU shield & 0.02 \\
\bottomrule
\end{tabular}
\caption{Example ${}^{232}$U chain battery configurations, with ${}^{227}$Ac and ${}^{210}$Pb comparisons at low power, where a sealed source behind tens of kg of DU needs no radon management at all. Electric powers assume 25\% heat-to-electricity conversion~\cite{Lange2008}. Shields are DU at a 10 mrem/h contact dose from the model of \Cref{fig:shieldmass}; in the vented and chamber rows, radon separation at 90\% emanation removes one tenth-value layer of DU; the ${}^{228}$Th freighter is sized on its 76 W/g five-year average; the crewed habitat keeps its full chain contained and takes the unreduced shield; the robotic space vehicles use a 7\% shadow cone. NEP: nuclear-electric propulsion.}
\label{tab:configs}
\end{table}

\Cref{fig:fuel_mass} shows the capabilities enabled by fusion-neutron production of alpha emitters for nuclear batteries. Assuming 25\% heat-to-electricity conversion efficiency, considered achievable for dynamic Stirling converters~\cite{Lange2008}, each electric watt of the ${}^{232}$U chain needs 9 times less fuel than ${}^{238}$Pu and 46 times less than ${}^{241}$Am: 1 kg provides 1.3 kW$_\mrm{e}$; through the same converter, a Voyager RTG's 4.5 kg of ${}^{238}$Pu would provide 0.6 kW$_\mrm{e}$~\cite{bennett2006space}. Future converter technology could raise the efficiency further~\cite{bennett2006space,Lange2008,Ambrosi2019}.

The largest uncertainty for the ${}^{236}$Pu chain is to demonstrate its safety in manned spaceflight and other use-cases where humans will be close to the RTG. The nuclear battery must demonstrate ${}^{220}$Rn emanation fractions near 0.9 at $\sim$100 W cm${}^{-3}$ ~\cite{Nelson2023} in the gamma chamber. A ${}^{237}$Np blanket with $k_\mrm{eff}$ up to 0.68 at its thickest NpO$_2$ operating point (pure ${}^{237}$Np metal would go critical near 26.5 mm) needs engineered subcritical margin and safeguards. Continuous molten-salt extraction of these blankets must be demonstrated at scale, building on FLiBe blanket technology under active development~\cite{Forsberg2020}. These channels also breed fissile material considerably faster than the breakout configurations previously analyzed for fusion plants~\cite{Ball2025}, at one significant quantity of ${}^{239}$Pu every 20 hours for the ${}^{238}$U channel, which makes safeguards a significant design requirement. We work this through in Appendix \ref{app:u233}. 

\begin{table}[!tb]
\centering
\caption{Fraction of the fuel rest energy $mc^{2}$ released as usable heat. The lithium-ion row is per kilogram of pack at 250 Wh/kg.}
\label{tab:mc2}
\begin{tabular}{lcc}
\toprule
Source & $E/mc^{2}$ & $E$ (GJ g$^{-1}$) \\
\midrule
Lithium-ion battery & $1.0\times10^{-11}$ & $9\times10^{-7}$ \\
${}^{147}$Pm ($\betam$, 62 keV mean) & $4.5\times10^{-7}$ & 0.041 \\
${}^{90}$Sr/${}^{90}$Y ($\betam$, 1.13 MeV) & $1.4\times10^{-5}$ & 1.2 \\
${}^{238}$Pu ($\alpha$, 5.59 MeV) & $2.5\times10^{-5}$ & 2.3 \\
${}^{210}$Pb chain ($\betam\betam\alpha$, 5.8 MeV) & $3.0\times10^{-5}$ & 2.7 \\
${}^{236}$Pu chain (45 MeV) & $2.0\times10^{-4}$ & 18 \\
${}^{250}$Cm (74\% SF, 149 MeV) & $6.4\times10^{-4}$ & 57 \\
${}^{235}$U fission (200 MeV) & $9.1\times10^{-4}$ & 82 \\
D-T fusion (17.6 MeV) & $3.8\times10^{-3}$ & 340 \\
\bottomrule
\end{tabular}
\end{table}

These results motivate dedicated feasibility studies. Priorities include extraction chemistry on hot actinide streams~\cite{AMPPEX,CEN1961Pa,McAlister2018} and remote fuel fabrication at kilogram scale, demonstration of ${}^{220}$Rn separation at full activity~\cite{Yong2015,Kokov2022,Pretze2025}, shield and heat-rejection engineering beyond the spherical estimates of \Cref{sec:shieldmass}~\cite{Nelson2023,dustin2021assessment}, launch safety and licensing for fuels without flight heritage~\cite{lee2015rps,bennett2006space}, integration of actinide channels into tritium-breeding blankets over a plant lifetime~\cite{Sorbom2015,Forsberg2020,Meschini2023}, safeguards and material accountancy sustained across 30 yr of operation~\cite{Ball2025,GlaserGoldston2012,IAEAGlossary2022}, and the economics of production over decades, in which stockpiles build for years and decayed ${}^{238}$Pu realizes value decades after breeding~\cite{parisi2026neutronvalue,parisi2026betaemitters}.

\section{Acknowledgments}

I am grateful for conversations with A. Rutkowski, J. A. Schwartz, and P. F. Peterson; and also to A. Rutkowski for reviewing the manuscript.

\appendix

\section{Production pathways} 

\label{app:pathways}

\Cref{tab:pathways} lists every production pathway proposed in this Article, also shown in \Cref{fig:pathways_chart} on the chart of nuclides. \Cref{fig:decay} draws the decay schemes of the four main chains.

\begin{table*}[!tb]
\centering
\small
\caption{Production pathways proposed in this Article, grouped by product class. Cross sections are at 14 MeV except thermal captures marked (th.); decay-driven generator rows list the parent half-life instead. Photonuclear ($\gamma$,n) reactions can drive the same $A \to A{-}1$ steps as \ntn, shown as the outer arcs of \Cref{fig:pathways_chart}. The prior-work column notes what each cited reference actually demonstrated or proposed; the routes marked ``this work'' have not, to our knowledge, previously been proposed for scalable battery-fuel production.}
\label{tab:pathways}
\footnotesize
\setlength{\tabcolsep}{4pt}
\resizebox{\textwidth}{!}{%
\begin{tabular}{llllp{3.8cm}}
\toprule
Product & Feedstock & Reaction ($\sigma$) & Path to product & Prior work \\
\midrule
${}^{236}$Pu (2.86 yr) & ${}^{237}$Np & \ntn, 0.34 b & ${}^{236\mrm{m}}\mrm{Np} \xrightarrow{\betam,\,22.5\,\mrm{h}} {}^{236}\mrm{Pu}$ (52.5\%) & {\raggedright this work; 1960s surveys, no route~\cite{blanke1960nuclear,Rohrmann1963}; accelerator $\mu$g~\cite{Artun2020,Yamana2001,Aaltonen2003}\par} \\
${}^{232}$U (68.9 yr) & ${}^{231}$Pa & \ngamma, 200 b (th.) & ${}^{232}\mrm{Pa} \xrightarrow{\betam,\,1.31\,\mrm{d}} {}^{232}\mrm{U}$ (42\% maximum) & {\raggedright natural-Pa route, gram scale~\cite{Rohrmann1960,Guillot1965}; fuel tracer~\cite{Rhodes2022}; RTG studies~\cite{Kulikov2020}\par} \\
${}^{232}$U (68.9 yr) & ${}^{233}$U bred in situ & \ntn, ${\sim}0.4$ b & direct; ${}^{233}$U from ${}^{232}$Th\ngamma\ via ${}^{233}$Pa & {\raggedright this work; trace co-production in fission fast blankets noted 1960~\cite{blanke1960nuclear}\par} \\
${}^{232}$U (68.9 yr) & ${}^{236}$Pu & $\alpha$ decay, 2.86 yr & milked from held Pu, 99.9\%+ activity purity (Appendix \ref{app:purity}) & {\raggedright this work\par} \\
${}^{228}$Th (1.9 yr) & ${}^{232}$U & $\alpha$ decay, 68.9 yr & milked from stored ${}^{232}$U at high isotopic purity & {\raggedright milking proposed 1960~\cite{Rohrmann1960}; mCi practice for therapy~\cite{Yong2015,Kokov2022}\par} \\
${}^{238}$Pu (87.7 yr) & ${}^{237}$Np & \ngamma, spectrum & ${}^{238}\mrm{Np} \xrightarrow{\betam,\,2.1\,\mrm{d}} {}^{238}\mrm{Pu}$ & {\raggedright established RTG-fuel route, kg scale~\cite{Roggenkamp1987,Collins2022,urban2021initial}; ICF concept~\cite{Winterberg2017}\par} \\
\midrule
${}^{237}$Np & ${}^{238}$U & \ntn, ${\sim}1$ b & ${}^{237}\mrm{U} \xrightarrow{\betam,\,6.75\,\mrm{d}} {}^{237}\mrm{Np}$ & {\raggedright this work; ${}^{237}$U discovery~\cite{Nishina1940}; weapon signature~\cite{DeGeer1991}\par} \\
${}^{231}$Pa & ${}^{232}$Th & \ntn, 1.5 b & ${}^{231}\mrm{Th} \xrightarrow{\betam,\,25.5\,\mrm{h}} {}^{231}\mrm{Pa}$ & {\raggedright hybrid-blanket breeding for LWR fuel~\cite{Shmelev2015}; burnable absorber~\cite{Kulikov2017}\par} \\
${}^{243}$Am & ${}^{244}$Pu & \ntn, 0.80 b & ${}^{243}\mrm{Pu} \xrightarrow{\betam,\,5.0\,\mrm{h}} {}^{243}\mrm{Am}$ & {\raggedright this work; $\sigma$ evaluation only~\cite{Guo2019}\par} \\
\midrule
${}^{242}$Cm (163 d) & ${}^{241}$Am & \ngamma, 680 b (th.) & ${}^{242\mrm{g}}\mrm{Am} \xrightarrow{\betam,\,16\,\mrm{h}} {}^{242}\mrm{Cm}$ (83\%) & {\raggedright ${}^{242}$Cm RTG design~\cite{Weddell1960}; RPS review~\cite{Lange2008}\par} \\
${}^{242}$Cm (163 d) & ${}^{243}$Am & \ntn, 0.35 b & as above; ${}^{242\mrm{m}}$Am co-produced & {\raggedright this work; isomer-ratio theory~\cite{Maslov2024}\par} \\
${}^{242\mrm{m}}$Am (141 yr) & ${}^{241}$Am, ${}^{243}$Am & isomer branch of the above & IT (99.5\%) $\to {}^{242\mrm{g}}$Am $\to {}^{242}$Cm $\to {}^{238}$Pu & {\raggedright this work; capture route for fission-fuel concepts~\cite{Ronen1988,Ronen2000,Benetti2006}\par} \\
${}^{244}$Cm (18.1 yr) & ${}^{243}$Am & \ngamma, 80 b (th.) & ${}^{244}\mrm{Am} \xrightarrow{\betam,\,10.1\,\mrm{h}} {}^{244}\mrm{Cm}$ & {\raggedright spent-fuel isotopic power~\cite{Posey1973}; HFIR production~\cite{Robinson2020}\par} \\
\midrule
${}^{210}$Pb (22.2 yr) & ${}^{231}$Pa & \ntn, 1.3 b & ${}^{230}\mrm{Pa} \xrightarrow{\betam\,8.4\%,\,17.4\,\mrm{d}} {}^{230}\mrm{U} \xrightarrow{5\alpha,\,\mrm{weeks}} {}^{210}\mrm{Pb}$ & {\raggedright this work; ${}^{210}$Po battery~\cite{blanke1960nuclear}; $\sigma$~\cite{Hashimoto1988}; ${}^{230}$U by protons~\cite{Morgenstern2008}\par} \\
${}^{230}$Th (75.4 kyr) & ${}^{231}$Pa & \ntn, 1.3 b & ${}^{230}\mrm{Pa} \xrightarrow{\mrm{EC}\,92\%,\,17.4\,\mrm{d}} {}^{230}\mrm{Th}$ & {\raggedright this work; natural-source survey~\cite{Figgins1966}\par} \\
${}^{227}$Ac (21.8 yr) & ${}^{231}$Pa stockpile & $\alpha$ decay, 32.8 kyr & direct ingrowth at 21 g yr${}^{-1}$ per tonne of ${}^{231}$Pa & {\raggedright this work; mg from aged ${}^{231}$Pa~\cite{Larsen2003patent}; 125 g stock~\cite{CEN1961Pa}\par} \\
${}^{227}$Ac (21.8 yr) & ${}^{226}$Ra (from ${}^{230}$Th decay) & \ngamma, 13 b (th.) & ${}^{227}\mrm{Ra} \xrightarrow{\betam,\,42\,\mrm{min}} {}^{227}\mrm{Ac}$ & {\raggedright mg isolation~\cite{Hagemann1950}; 70 g yr${}^{-1}$ study~\cite{ANL1950}; heat source built~\cite{Baetsle1971}; ${}^{229}$Th supply~\cite{Hogle2016}\par} \\
\bottomrule
\end{tabular}}
\end{table*}

\section{Blended fuels: shaped power profiles} \label{app:blends}

\begin{figure}[!tb]
\centering
\begin{subfigure}{\columnwidth}
\centering
\includegraphics[width=\columnwidth]{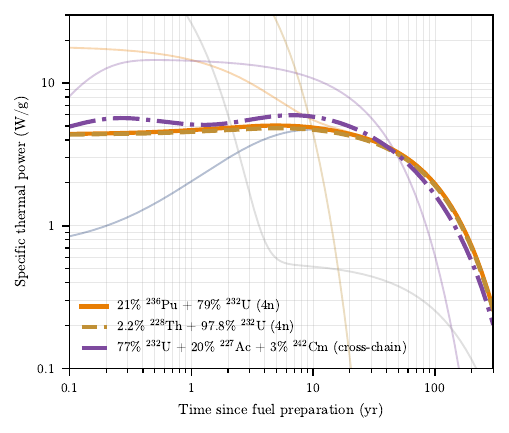}
\caption{}
\label{fig:blends_a}
\end{subfigure}\\[2pt]
\begin{subfigure}{\columnwidth}
\centering
\includegraphics[width=\columnwidth]{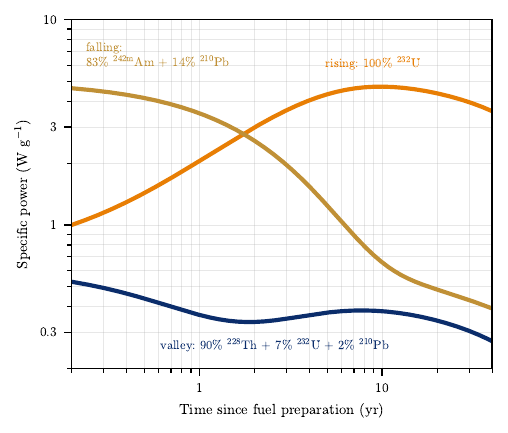}
\caption{}
\label{fig:blends_b}
\end{subfigure}
\caption{Power profiles for blended fuels. (a) Flatness-optimized mixtures (bold; mass fractions in legend) over their faint pure constituents. (b) Shaped profiles from minimax blends of age-staggered fuels, shown as the absolute specific power of each blend (compositions labeled).}
\label{fig:blends}
\end{figure}

Because each chain has a distinct power profile, blending them can shape the power output to a mission's requirement. A fuel blend with individual chain fractions $w_k \geq 0$, where $\sum_k w_k = 1$, gives a power profile
\begin{equation}
P(t) = \sum_k w_k P_k(t),
\end{equation}
for different radionuclides $k$. A mixture can never have a power exceeding its chain with the highest power, and so $P(t) \leq \max_k P_k(t)$. Finding the flattest mix over a target time period is an optimization over the $w_k$. \Cref{fig:blends_a} shows several fuel blends with relatively flat power profiles with different fuel mixes. We find the flattest blends combine different nuclides from the same $^{236}$Pu chain. Every inventory in a chain is a sum of decaying exponentials, so no static mixture can oscillate. However, there can be local minima, as shown in \Cref{fig:blends_b}.

\section{Curium from americium feed} \label{sec:curium}
The curium fuels of \Cref{fig:cascade_search} require a capture, or an \ntn\ followed by americium $\betam$ decay, since \ntn\ alone only steps down in mass. The routes mirror the ${}^{237}$Np pair of \Cref{tab:pathways}. Thermally, ${}^{241}$Am\ngamma\ ($\sigma_\mrm{th}=680$ b) feeds the 16 h ${}^{242\mrm{g}}$Am (isomeric ratio 0.91~\cite{Fioni2001}), which $\betam$-decays to ${}^{242}$Cm at 83\%; the ${}^{241}$Am feedstock is available at scale, already the European RTG fuel~\cite{Ambrosi2019,AMPPEX}. In the fast channel, ${}^{243}$Am\ntn\ (0.35 b at 14 MeV) populates both members of ${}^{242}$Am; their branching at 14 MeV lacks nuclear data measurements, and ${}^{243}$Am, though present in spent fuel, currently lacks a separated stockpile for production.

The key product is ${}^{242}$Cm as a fast-start fuel: 163 d and 120 W g${}^{-1}$ fresh, decaying into ${}^{238}$Pu, so a charge averages $\sim$16 W g${}^{-1}$ over five years, a 100 yr average three times ${}^{238}$Pu alone (\Cref{fig:cascade_search}). The challenges are a months-long shelf life and a strong neutron source ($\sim$$2\times10^{7}$ n s${}^{-1}$ g${}^{-1}$) requiring moderating shields~\cite{Nelson2023}.

${}^{242\mrm{m}}$Am is a 141 yr isomer, mainly decayng through ${}^{242\mrm{g}}$Am to ${}^{242}$Cm (\Cref{fig:decay}). Long proposed as a compact fuel~\cite{Ronen2000,Benetti2006,PereiraMerkelLitz2007}, it is scarce and destroyed by its own $\sim$6400 b thermal fission, favoring a Cd- or Gd-filtered epithermal spectrum~\cite{Benetti2006}.

Simplest of all may be to skip separation entirely: a heat source has no isotopic-purity requirement, so an irradiated ${}^{241}$Am target can be used directly after fission-product cleanup, as a blend of residual ${}^{241}$Am (0.11 W g${}^{-1}$), prompt ${}^{242}$Cm, the ${}^{242\mrm{m}}$Am reservoir that regenerates it, and accumulating ${}^{238}$Pu. Only chemical cleanup is needed, not the isotopic enrichment that stalled ${}^{242\mrm{m}}$Am propulsion concepts~\cite{Benetti2006}.

\begin{figure}[!tb]
\centering
\begin{subfigure}{\columnwidth}
\centering
\includegraphics[width=\columnwidth]{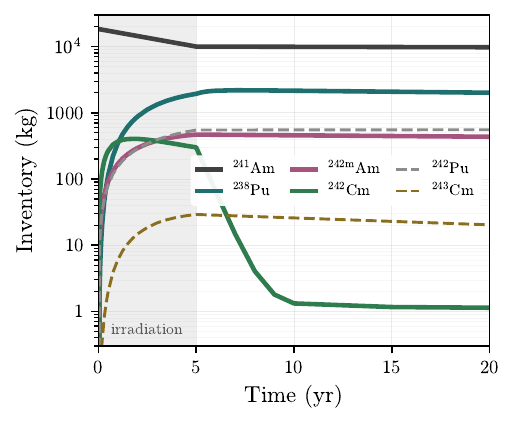}
\caption{}
\label{fig:am241_blend}
\end{subfigure}\\[2pt]
\begin{subfigure}{\columnwidth}
\centering
\includegraphics[width=\columnwidth]{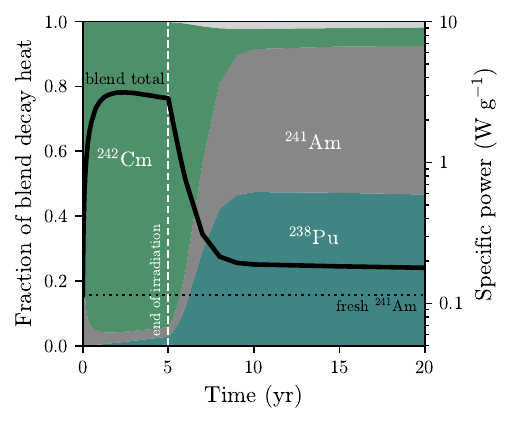}
\caption{}
\label{fig:am241_blend_power}
\end{subfigure}
\caption{Depletion of a 5 mm ${}^{241}$Am metal target (18.4 t) at 1.5 GW$_\mrm{fus}$: 5 yr irradiation (shaded in (a)), then cooling. (a) Blanket inventory. (b) Fraction of decay heat from each nuclide (thin top band is ${}^{243}$Cm and ${}^{244}$Cm), total specific power on the right axis.}
\label{fig:am241_blend_pair}
\end{figure}

We quantify the blend with a depletion simulation (\Cref{fig:am241_blend}): a 5 mm ${}^{241}$Am metal target (18.4 t ) in the geometry of \Cref{fig:geometry}, irradiated for five years at 1.5 GW$_\mrm{fus}$. The target captures 0.17 neutrons per source neutron, and including the ENDF/B-VIII.0 isomeric branching gives $f_\mrm{m} = 0.17$ to the 141 yr isomer, between the thermal (0.10) and fast (0.48) limits. At end of irradiation the target holds 10.0 t of residual ${}^{241}$Am, 1.9 t of ${}^{238}$Pu, 470 kg of ${}^{242\mrm{m}}$Am, and 300 kg of ${}^{242}$Cm: a 2.9 W g${}^{-1}$ blend, 25 times the ${}^{241}$Am feed.

\begin{figure}[!tb]
\centering
\includegraphics[width=\columnwidth]{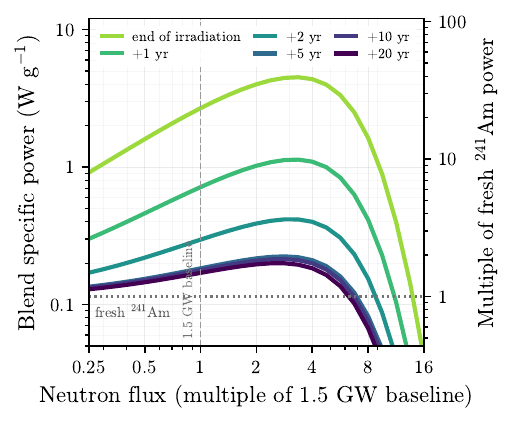}
\caption{Neutron-flux scan of the 5 yr irradiated ${}^{241}$Am blend. Curves show the blend specific power from end of irradiation to 20 yr after shutdown.}
\label{fig:am241_flux_scan}
\end{figure}

\section{Producing \texorpdfstring{${}^{210}$Pb}{Pb-210}} \label{app:pb210}

Reaching ${}^{210}$Pb from stable lead or bismuth is extremely challenging, and the ${}^{230}$Th chain of Appendix \ref{app:pa231} yields only $\sim$8 mg per kg-year. We propose ${}^{231}$Pa\ntn${}^{230}$Pa (1.3 b at 14 MeV), whose 8.4\% $\betam$ branch feeds ${}^{230}$U, whose subsequent five-alpha decay produces ${}^{210}$Pb, depositing $\sim$34 MeV of heat (\Cref{fig:decay}). A similar route, ${}^{231}$Pa from proton irradiation of ${}^{232}$Th feeding ${}^{230}$U/${}^{226}$Th, has been considered for targeted alpha therapy~\cite{Morgenstern2008}.

\begin{figure}[!tb]
\centering
\begin{subfigure}{\columnwidth}
\centering
\includegraphics[width=\columnwidth]{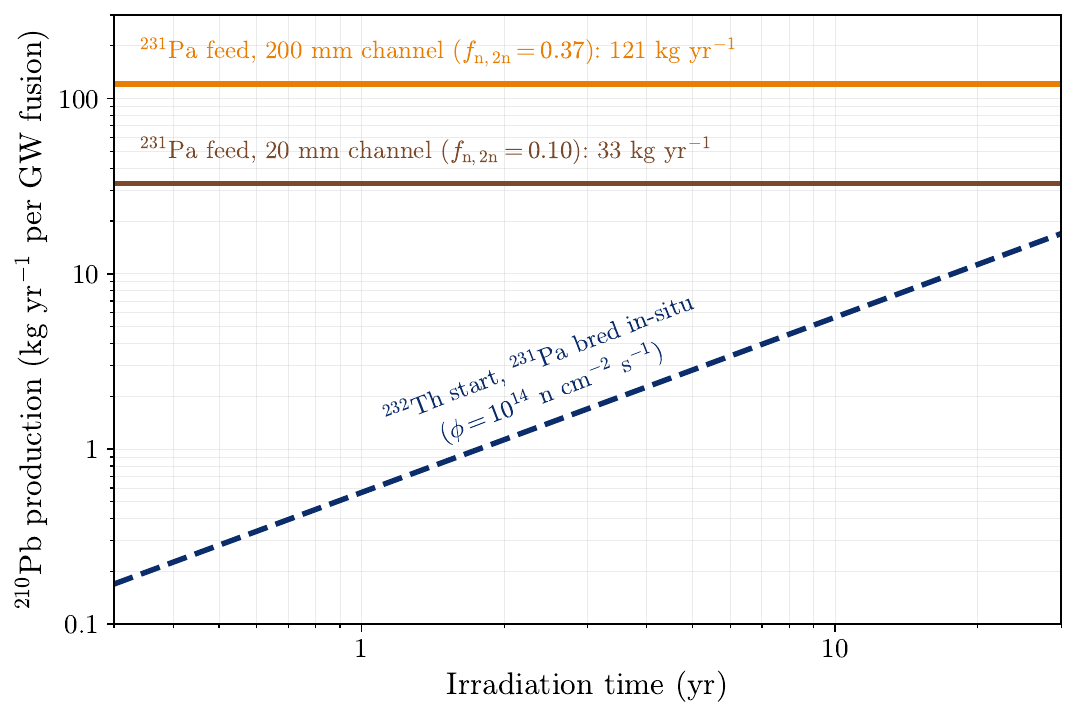}
\caption{}
\label{fig:pb210_yield_a}
\end{subfigure}\\[2pt]
\begin{subfigure}{\columnwidth}
\centering
\includegraphics[width=\columnwidth]{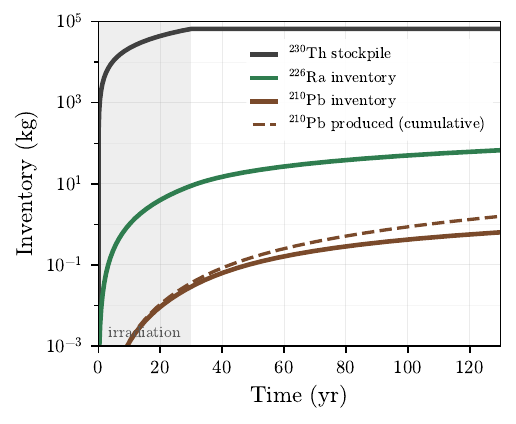}
\caption{}
\label{fig:pb210_yield_b}
\end{subfigure}
\caption{${}^{210}$Pb production from a ${}^{231}$Pa blanket. (a) Rate per GW of fusion power versus irradiation time, with continuous extraction and no losses to ${}^{231}$Pa burnup or ${}^{210}$Pb decay. Solid: direct ${}^{231}$Pa feedstock; dashed: starting from raw ${}^{232}$Th at $\phi=10^{14}$ n cm${}^{-2}$ s${}^{-1}$, with the ${}^{231}$Pa bred in-situ. (b) ${}^{230}$Th accumulated at 2.2 t yr${}^{-1}$ through the 92\% ${}^{230}$Pa branch (200 mm operating point), and its ${}^{226}$Ra and ${}^{210}$Pb decay products after the 30 yr irradiation.}
\label{fig:pb210_yield}
\end{figure}

In a dedicated fast blanket run on a ${}^{231}$Pa inventory, where \ntn\ takes 0.10 to 0.37 of the source neutrons at thicknesses of 20 to 200 mm (\Cref{tab:pa231_blanket}), the 1.5 GW$_\mrm{fus}$ plant produces 50 to 183 kg yr${}^{-1}$ of ${}^{210}$Pb while accumulating 0.6 to 2.2 t yr${}^{-1}$ of ${}^{230}$Th through the 92\% branch. Note that the ${}^{231}$Pa blankets with 20 to 200 mm thickness in \Cref{tab:pa231_blanket} would take decades of a ${}^{232}$Th blanket to breed sufficient ${}^{231}$Pa (in the tens to hundreds of tons) for loading. \Cref{fig:pb210_yield_a} shows both extraction scenarios: from a ${}^{231}$Pa feedstock the production rate is constant, 33 to 122 kg per GW$_\mrm{fus}$-yr; from raw ${}^{232}$Th the production rate increases linearly with time as the ${}^{231}$Pa is first bred, so a practical pathway breeds or buys its protactinium up front (with the caveat that ${}^{231}$Pa is very scarce at present in 2026).

The 92\% branch is not a loss channel as the accumulated ${}^{230}$Th opens a second, slower path to ${}^{210}$Pb through ${}^{226}$Ra (\Cref{fig:pb210_yield_b}). Each tonne of ${}^{230}$Th generates 9.0 g yr${}^{-1}$ of ${}^{226}$Ra, so the 66 t stockpiled over 30 yr of operation holds $\sim$50 kg of ${}^{226}$Ra a century after start-up, yielding ${}^{210}$Pb at tens of grams per year and rising quadratically as the radium builds toward its 1600 yr equilibrium. Because ${}^{230}$Th takes two \ntn\ reactions (${}^{232}$Th to ${}^{231}$Pa, then ${}^{231}$Pa to ${}^{230}$Pa), its production rate scales as the square of the neutron flux, favoring the highest-wall-load machines, so the stockpile is also a feedstock for important medical radioisotopes generated by ${}^{226}$Ra irradiation.

\begin{table}[!tb]
\centering
\caption{Operating points of a dedicated fast ${}^{231}$Pa blanket in the geometry of \Cref{fig:geometry}, in the columns of \Cref{tab:hybrid}; the ${}^{232}$U co-product is in-situ \ngamma\ with a 42\% recoverable maximum. All rates are fresh-blanket tallies at 1.5 GW$_\mrm{fus}$ of D-T fusion.}
\label{tab:pa231_blanket}
\setlength{\tabcolsep}{3pt}
\footnotesize
\begin{tabular}{cccccccc}
\toprule
$t_\mrm{ch}$ & $M_\mrm{ch}$ & $k_\mrm{eff}$ & TBR & $P_\mrm{tot}$ & ${}^{210}$Pb & ${}^{232}$U & ${}^{90}$Sr \\
(mm)         & (t)          &               &     & (GW)          & (kg\,yr${}^{-1}$) & (kg\,yr${}^{-1}$) & (kg\,yr${}^{-1}$) \\
\midrule
20  & 83  & 0.37 & 1.43 & 7.1 & 50  & 1490  & 39 \\
100 & 426 & 0.72 & 1.58 & 39  & 147 & 10700 & 264 \\
200 & 877 & 0.83 & 1.14 & 78  & 183 & 22500 & 535 \\
\bottomrule
\end{tabular}
\end{table}

\section{Millennial heat sources from the thorium blanket} \label{app:pa231}
\begin{figure}[!tb]
\centering
\includegraphics[width=\columnwidth]{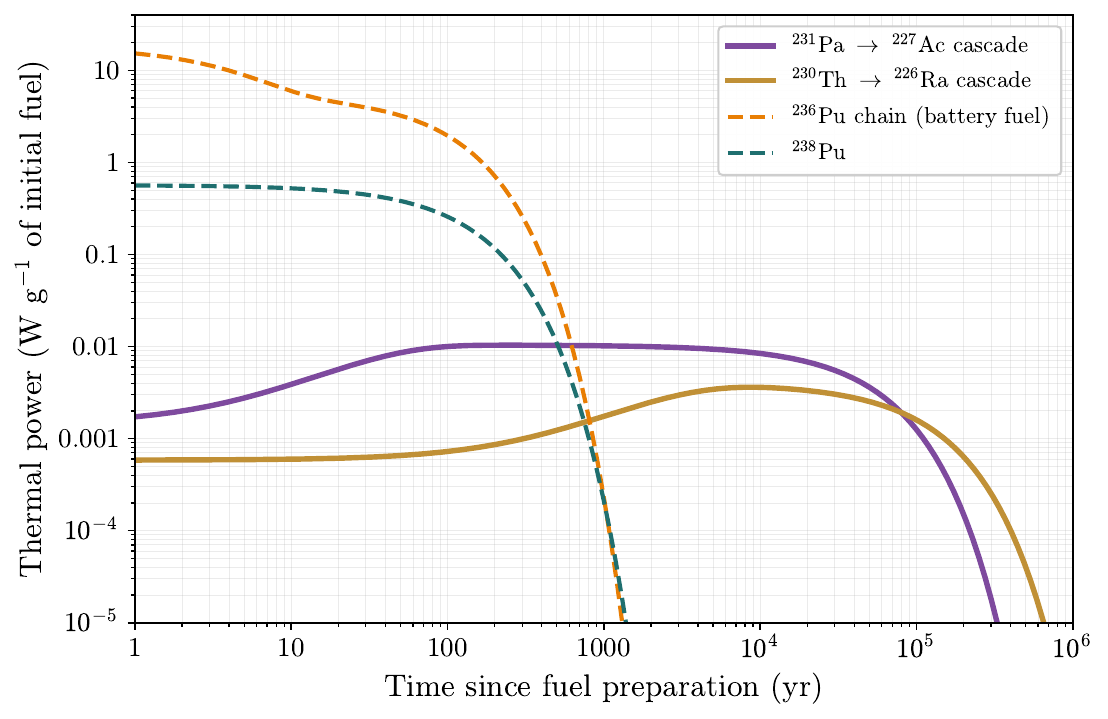}
\caption{Thermal power per gram of initial ${}^{231}$Pa (purple) and ${}^{230}$Th (gold) versus time. Dashed: the ${}^{236}$Pu chain and ${}^{238}$Pu references of the main text.}
\label{fig:pa231_chain}
\end{figure}

The ${}^{232}$Th\ntn\ blanket of \Cref{sec:scan} also opens two heat sources on timescales far beyond the missions of this Article (\Cref{fig:pa231_chain}). Stockpiled ${}^{231}$Pa heads one decay chain: its 32760 yr $\alpha$ decay feeds ${}^{227}$Ac, so each decay ultimately deposits $\sim$40 MeV. A fresh gram of ${}^{231}$Pa produces 1.4 mW, rising to 10 mW g${}^{-1}$ over roughly a century as the 21.8 yr ${}^{227}$Ac equilibrates. The stored energy, 4.6 GWh kg${}^{-1}$, is comparable to the ${}^{236}$Pu chain's 5.1 GWh kg${}^{-1}$, released three orders of magnitude more slowly.

A second \ntn\ reaction on ${}^{231}$Pa reaches still longer timescales: ${}^{231}$Pa\ntn${}^{230}$Pa ($\sigma\approx1.3$ b at 14 MeV~\cite{Brown2018ENDF}) followed by electron capture (92\%, 17.4 d) yields ${}^{230}$Th (75.4 kyr), head of the radium series, whose output rises from 0.6 mW g${}^{-1}$ fresh to $\sim$4 mW g${}^{-1}$ as the 1600 yr ${}^{226}$Ra equilibrates.

The ${}^{231}$Pa and ${}^{230}$Th-headed chains differ at their radon link. The ${}^{230}$Th chain passes through long-lived (for radon) 3.82 d ${}^{222}$Rn: emanation is essentially complete even from coarse fuel forms, and the gas can be piped to a remote hold-up volume where the gamma-heavy ${}^{214}$Pb/${}^{214}$Bi segment decays far from the source. The radon separation of \Cref{sec:radon} therefore transfers directly to the ${}^{230}$Th source, at the cost of exporting roughly two thirds of the chain heat. The ${}^{231}$Pa chain instead passes through 4.0 s ${}^{219}$Rn, too short to escape any practical fuel form, so the ${}^{231}$Pa source keeps its gamma emitters, capped at the soft 832 keV ${}^{211}$Pb line, a factor of three below ${}^{208}$Tl (\Cref{sec:chain}). The choice trades the higher specific power of ${}^{231}$Pa against the separable gamma chain of ${}^{230}$Th; both sit far outside the RTG timescales of \Cref{sec:chain} and are not pursued further, but are included for possible applications that require much longer decay timescales.

\section{\texorpdfstring{Isotopic purity of the ${}^{236}$Pu}{Pu-236 purity}} \label{app:purity}

The two plutonium isotopes ${}^{236}$Pu and ${}^{238}$Pu are produced by neutrons on ${}^{237}$Np with very different neutron energies. ${}^{236}$Pu is via ${}^{237}$Np\ntn, requiring neutrons above 6.7 MeV. ${}^{238}$Pu is via ${}^{237}$Np\ngamma, driven almost entirely by slower neutrons. In the 5 mm channel, 0.070 of the 0.101 \ngamma reactions per source neutron come from neutrons between 1 eV and 0.1 MeV, and none from below 1 eV. The ${}^{236}$Pu/${}^{238}$Pu isotopic ratio is set by the neutron spectrum, which can be modified based on the design.

\begin{figure*}[!tb]
\centering
\begin{subfigure}{0.49\textwidth}
\centering
\includegraphics[width=\textwidth]{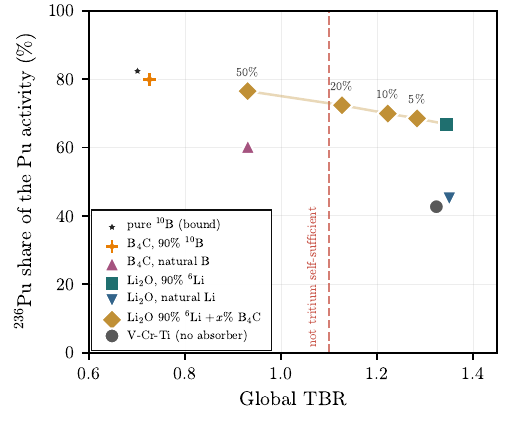}
\caption{}
\label{fig:purity_a}
\end{subfigure}
\hfill
\begin{subfigure}{0.49\textwidth}
\centering
\includegraphics[width=\textwidth]{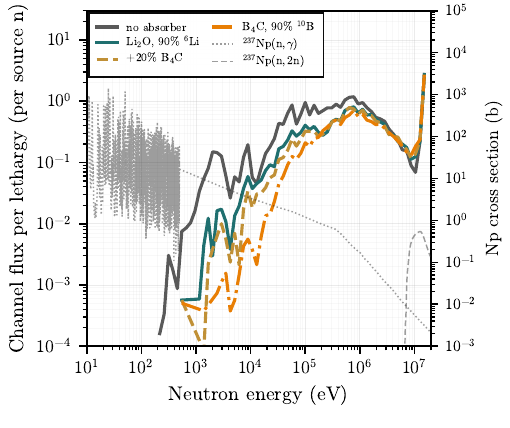}
\caption{}
\label{fig:purity_b}
\end{subfigure}

\vspace{4pt}
\begin{subfigure}{0.49\textwidth}
\centering
\includegraphics[width=\textwidth]{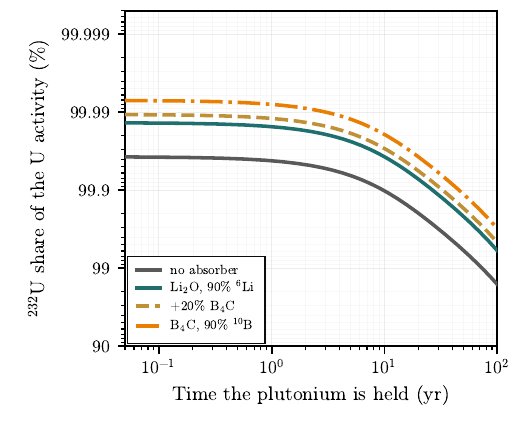}
\caption{}
\label{fig:purity_c}
\end{subfigure}
\hfill
\begin{subfigure}{0.49\textwidth}
\centering
\includegraphics[width=\textwidth]{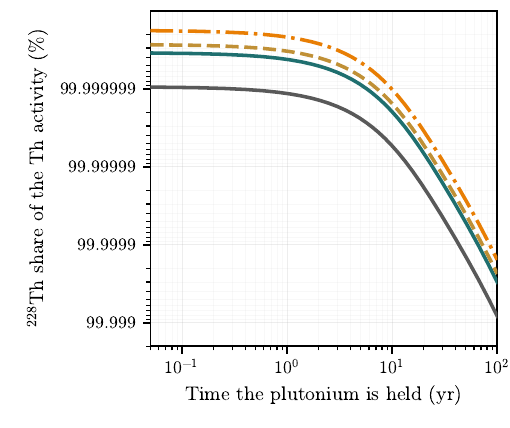}
\caption{}
\label{fig:purity_d}
\end{subfigure}
\caption{Isotopic purity of the ${}^{236}$Pu plutonium from a 5 mm NpO$_2$ blanket with an absorber in the 30 mm layer behind it. (a) ${}^{236}$Pu fraction of the plutonium against TBR. (b) Blanket neutron flux (left y-axis) and cross sections (right y-axis).}
\label{fig:purity}
\end{figure*}

One way to modify the neutron spectrum in the ${}^{237}$Np layer is to change material composition of the 30 mm structural layer between the channel and the FLiBe (\Cref{fig:geometry}). Shown in (\Cref{fig:purity}), Li$_2$O at 90\% ${}^{6}$Li increases the ${}^{236}$Pu fraction of the plutonium from 2.4 to 6.2 at\% while also increasing TBR slightly from 1.32 to 1.34. B$_4$C at 90\% ${}^{10}$B further increases the ${}^{236}$Pu fraction to 11.5 at\% but decreases the TBR to 0.72. 20 vol\% B$_4$C in the Li$_2$O gives 7.9 at\% at a TBR of 1.13.

There is a limit to what an absorber can do. Raising the ${}^{6}$Li enrichment past 90\% achieves almost nothing, and even a layer of pure ${}^{10}$B stops at 13.2 at\%: the surviving captures come mostly from fast neutrons that scatter and capture inside the channel itself, out of reach of any outer layer (\Cref{fig:purity_b}). Purity also favors a thin channel, 6.2 at\% at 5 mm against 2.0 at\% at 30 mm with the same absorber, although thicker blankets have higher ${}^{236}$Pu yield (\Cref{tab:hybrid}).

While the achievable ${}^{236}$Pu purity is relatively low when irradiating ${}^{237}$Np with fusion neutrons, its daughters come out far purer (\Cref{fig:purity_c,fig:purity_d}). Holding the extracted plutonium and stripping its uranium gives ${}^{232}$U from the ${}^{236}$Pu and relatively inert ${}^{234}$U ($t_{1/2}=246$ kyr) from the ${}^{238}$Pu. Measured by activity, the ${}^{232}$U and ${}^{228}$Th fractions are extremely high. The plutonium is already 67\% ${}^{236}$Pu by activity at 6.3 at\%, and 43\% without the absorber, because ${}^{236}$Pu decays 31 times faster than ${}^{238}$Pu. The uranium recovered is 99.98\% ${}^{232}$U by activity if the plutonium is stripped every 1 to 2 yr, and still 99.89\% if it is left for 30 yr, since ${}^{232}$U decays 3600 times faster than the ${}^{234}$U fed by ${}^{238}$Pu. The thorium one step down is 99.999\% ${}^{228}$Th or better over a century, its specific activity being 39000 times that of the ${}^{230}$Th it competes with. In activity terms both daughters are effectively single isotopes.

Given the relatively high value per neutron of $^{232}$U and $^{228}$Th (\Cref{fig:market}), the relatively poor $^{236}$Pu atomic fraction may not be an issue: instead, it could be desirable to allow $^{236}$Pu to decay, extracting its valuable $^{232}$U and $^{228}$Th decay products.

\section{Proliferation and Safeguards} \label{app:u233}

Fusion blankets holding tonnes of actinides pose a proliferation risk. Reference~\cite{Ball2025} recently showed that 5 to 50 t of ${}^{238}$U or ${}^{232}$Th dissolved in an ARC-class~\cite{Sorbom2015} FLiBe blanket breeds one significant quantity (SQ, 8 kg of ${}^{239}$Pu or ${}^{233}$U~\cite{IAEAGlossary2022}) in 14 days to six months at better than 99\% isotopic purity. Reference~\cite{Ball2025} also showed that none of the consequences such as tritium breeding ratio, fission heat at 10 to 15\% of the 500 MW$_\mrm{fus}$, decay heat below 1\%, or a self-protection time of at most one day, is severe enough to make the scenario untenable~\cite{GlaserGoldston2012}. That analysis assumed a plant with no goal of producing isotopes, where fertile $^{232}$Th and/or ${}^{238}$U material had been introduced into FLiBe.

In this work, we instead consider holding fertile material in dedicated layers against the first wall where the flux is highest, possibly with an extraction system that removes actinides on a $\sim$3 h residence time, which~\cite{Ball2025} identifies as the modification that would most shorten a breakout. The systems considered here therefore breed fast: a 5 mm ${}^{238}$U channel produces 3.69 t of essentially isotopically pure ${}^{239}$Pu over 30 yr (\Cref{tab:u238_inventory}), 123 kg yr${}^{-1}$ or one SQ every 24 days, and a 5 mm ${}^{232}$Th channel 2.53 t of ${}^{233}$U. Thicker channels produce more: at 50 mm the ${}^{238}$U channel produces 105 t of ${}^{239}$Pu, one SQ every 20 hours. However, the ${}^{238}$U channel is much more concerning because of the isotopic purity of the ${}^{239}$Pu.

Our purpose is not necessarily to argue against building these blankets, but that a capability of this size requires well-thought safeguards. In this appendix we calculate the co-produced fissile material and methods to reduce proliferation risk. This appendix is not meant to be comprehensive, but rather a first pass on proliferation and safeguards with fusion blankets designed for alpha emitter production. Much more detailed safeguards work is required if this program is developed further check existing work.

Because IAEA full-scope safeguards can be applied to fusion systems that are intentionally breeding special nuclear materials~\cite{Ball2025}, such as those considered in this work, the proliferation-risk profile is different to those considered in~\cite{sievert2010creating,englert2010possible,GlaserGoldston2012,franceschini2013nuclear,Ball2025,diesendorf2023analyzing} where the risks were mainly around clandestine/covert production of weapons-usable material (WUM), whereas here we consider intentional, declared production of large quantities of nuclear battery material. Note that these two profiles are not exclusive: a fusion power plant operator with a blanket for nuclear battery material could still attempt to proliferate by also adding fertile material to FLiBe.

Weapons-grade plutonium is reported to require isotopic purity of $>$93\% ${}^{239}$Pu~\cite{usdoe1997nonproliferation}. While ${}^{233}$U is also a weapons-grade material, very small quantities of $^{232}$U, $\sim$100 ppm, can make handling extremely challenging~\cite{Ball2025} due to $^{208}$Tl decay, and hence $^{233}$U with even $\sim$100 ppm of $^{232}$U could be considered self-protecting. The same reasoning applies to $^{236}$Pu making $^{239}$Pu challenging to handle: given that we are deliberately producing ${}^{236}$Pu in ${}^{238}$U and ${}^{237}$Np blankets, if the $^{236}$Pu concentration is sufficiently high it could render the $^{239}$Pu challenging to handle. However, as we argued earlier in this paper, the $^{208}$Tl radiation takes decades to saturate, and freshly-produced ${}^{236}$Pu being very high up the decay chain would be less self-protecting in Pu material for freshly produced and extracted Pu. This is further exacerbated by ${}^{236}$Pu's relatively short half-life. Additional techniques may also be used to remove $^{236}$Pu daughters from the plutonium product, which would lessen the radiation dose of $^{208}$Tl (and the same for $^{232}$U in the uranium product), so we should also be careful about the extent to which $^{236}$Pu and $^{232}$U can be assumed to be self-protecting once the ${}^{232}$U and $^{236}$Pu concentrations exceed a certain threshold. That being said, as a heuristic for the following analysis, we assume that ${}^{232}$U and $^{236}$Pu concentrations $\gtrsim$100 ppm render the uranium and plutonium products self-protecting.

\begin{figure*}[!tb]
\centering
\includegraphics[width=0.98\textwidth]{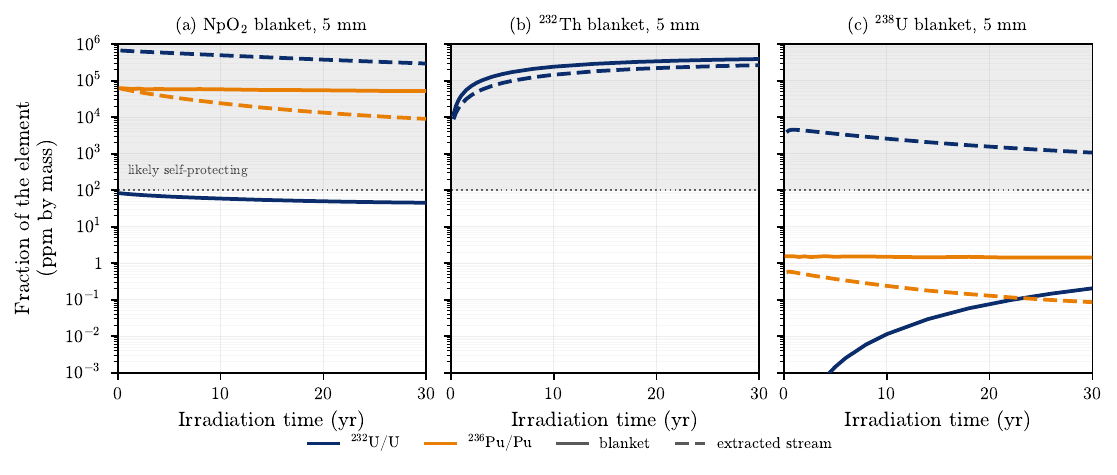}
\caption{${}^{232}$U fraction of the uranium and ${}^{236}$Pu fraction of the plutonium, in the blanket (solid) and in the extracted stream (dashed), over the 30 yr depletion simulations of Appendix \ref{app:depletion}: all three channels at 5 mm: (a) ${}^{237}$Np, (b) ${}^{232}$Th, (c) ${}^{238}$U. Shading shows concentrations above the heuristic $\sim$100 ppm ${}^{232}$U self-protection value.}
\label{fig:selfprotection}
\end{figure*}

\textit{NpO$_2$ blanket:} Our OpenMC depletion simulation of 30 year irradiation of a 5 mm NpO$_2$ blanket shows 8.06 t of $^{238}$Pu, 73 kg of $^{236}$Pu and only 0.45 kg of $^{239}$Pu in the extracted stream at 30 years, a manageable security threat. The plutonium is 99\% $^{238}$Pu and the uranium grown into the stream by decay is 29\% $^{232}$U (\Cref{fig:selfprotection}a). Cumulative $^{236}$Pu over the 30 year irradiation period is much larger, 528 kg, since the 2.86 yr fuel decays fast. The blanket itself holds 305 kg of $^{236}$U and only 0.2 kg of $^{233}$U, and 40 kg of $^{235}$U.

\textit{$^{232}$Th blanket:} Our OpenMC depletion simulation of 30 year irradiation of a 5 mm ${}^{232}$Th metal blanket shows that 2.53 t of $^{233}$U and 0.92 t of $^{232}$U are produced in the extracted stream, posing a manageable security threat because the $^{232}$U concentration is extremely high, 27\% of the uranium (\Cref{fig:selfprotection}b), which makes $^{233}$U handling extremely challenging. The blanket holds almost no uranium, since extraction removes it within hours. It instead accumulates 3.24 t of $^{231}$Pa, retained deliberately as a useful feedstock, with 1.76 t of $^{230}$Th and 249 kg of $^{229}$Th.

\textit{$^{238}$U blanket:} Our OpenMC depletion simulation of 30 year irradiation of a 5 mm 100\%at $^{238}$U metal (not natural U) blanket shows that 3.69 t of $^{239}$Pu is produced in the extracted stream alongside the 5.79 t of $^{237}$Np the channel exists to make. The $^{239}$Pu is the dominant proliferation concern, since the $^{236}$Pu content of that plutonium is negligible and gives no self-protection (\Cref{fig:selfprotection}c). Thinning the channel from 50 mm to 5 mm significantly reduces the $^{239}$Pu stream by a factor of 28 while reducing the $^{237}$Np product only by a factor of 6.6, and it raises the $^{232}$U content of the small uranium stream from 157 ppm to 1070 ppm, an order of magnitude past the heuristic self-protection level. The blanket retains 27.0 t of $^{238}$U and 1.44 t of $^{236}$U with only 1.5 kg of $^{235}$U, bred place by successive \ntn reactions.

\Cref{tab:u233} gives the ${}^{233}$U bred in the ${}^{232}$Th channel after one year at 1.5 GW$_\mrm{fus}$. The buildup is close to linear in time (given that it's a first order neutron reaction on ${}^{232}$Th), and at 200 mm it exceeds the limit of one ${}^{233}$U per source neutron, 6.5 t yr${}^{-1}$: this is because the neutrons from fission reactions themselves neutron capture on ${}^{232}$Th to make more ${}^{233}$U.

\begin{table}[!tb]
\centering
\caption{${}^{233}$U bred in the ${}^{232}$Th channel after one year at 1.5 GW$_\mrm{fus}$, from flux-fixed depletion. ${}^{233}$U includes pending ${}^{233}$Pa; $[{}^{232}\mrm{U}]$ is the ${}^{232}$U fraction of the uranium; ${}^{231}$Pa is the co-bred protactinium of \Cref{sec:scan}.}
\label{tab:u233}
\begin{tabular}{ccccc}
\toprule
$t_\mrm{ch}$ (mm) & TBR & ${}^{233}$U (t\,yr${}^{-1}$) & $[{}^{232}\mrm{U}]$ & ${}^{231}$Pa (t\,yr${}^{-1}$) \\
\midrule
20  & 1.30 & 0.78 & 2.3\% & 0.61 \\
100 & 1.15 & 4.6  & 1.4\% & 2.0  \\
200$^{a}$ & 0.73 & 8.6 & 0.9\% & 2.7 \\
\bottomrule
\multicolumn{5}{l}{\footnotesize $^{a}$TBR $<1$: not tritium self-sufficient.}
\end{tabular}
\end{table}

The ${}^{233}$U can also be suppressed without touching the ${}^{231}$Pa: the ${}^{232}$Th\ngamma\ capture is epithermal, fed by neutrons returning from the FLiBe, whereas \ntn\ is a 14 MeV threshold reaction. A $^{6}$Li liner between channel and breeder absorbs the returning soft neutrons, reducing the ${}^{232}$Th\ngamma\ rate by 2.0, 1.4, and 1.1 times at 20, 100, and 200 mm while leaving the ${}^{231}$Pa yield unchanged and \emph{raising} the tritium breeding ratio in every case (1.30 to 1.34, 1.15 to 1.37, and 0.73 to 0.92). Combined with the ${}^{232}$U and the option of co-loading ${}^{238}$U, the thorium route appears to be compatible with a safeguarded fuel cycle. This is the same approach~\cite{Ball2025} identifies as an effective barrier available to a FLiBe plant: ${}^{6}$Li hardens the neutron spectrum, suppresses capture on the fertile species, and raises the ${}^{232}$U fraction of what is bred. The blanket we study here already uses 90\%at ${}^{6}$Li.

The fission environment is the largest difference from~\cite{Ball2025} that sets fission heat aside as a barrier because 50 to 75 MW (reported in~\cite{Ball2025}) is a $\sim$10-15\% perturbation a proliferator can absorb thermally or possibly conceal by reducing the fusion power. In contrast, the dedicated blanket channels considered here are fusion-fission hybrids, producing 1.5 to 5.7 GW of fission power in the three depletion simulations shown in Appendix \ref{app:depletion}, so their absolute fission product inventory is closer to that of a fission reactor core rather than a perturbation on the fusion power in a FLiBe blanket. Decay heat alone for the blankets in Appendix \ref{app:depletion} after 30 years of irradiation is 224, 576, and 292 MW for the ${}^{237}$Np, ${}^{238}$U, and ${}^{232}$Th channels, compared with the 0.5 to 3.1 MW of~\cite{Ball2025} imply for their 342 m${}^{3}$ blanket. The decay heat does not disappear quickly: the ${}^{238}$U and ${}^{232}$Th channels still dissipate 9.7 and 36 MW a year after shutdown and 3.5 and 27 MW a decade after. Where~\cite{Ball2025} finds a self-protection time of one day, a channel left in place may be more self-protecting for much longer, possibly for its operational life and for decades afterward.

\section{Thirty-year channel depletion} \label{app:depletion}

In this Appendix we discuss the depletion neutronics simulations in more detail. The depletion runs below use the chemical form each channel could plausibly be built in, all at 5 mm. Neptunium metal would have a volumetric power density of 2795 MW m$^{-3}$, five to thirty times the fertile channels and likely incompatible with its low 917 K melting point, so we instead consider NpO$_2$, the form in which ${}^{238}$Pu targets are fabricated. Uranium and thorium metal reach `only' 507 and 89 MW m$^{-3}$ and melt at 1405 and 2023 K, so we keep both as metals, which holds twice the actinide density of the oxides. We perform OpenMC depletion simulations with 30 year irradiations at 1.5 GW$_\mrm{fus}$, then 100 years of cooling.

\begin{figure*}[!tb]
\centering
\begin{subfigure}{0.42\textwidth}
\centering
\includegraphics[width=\linewidth]{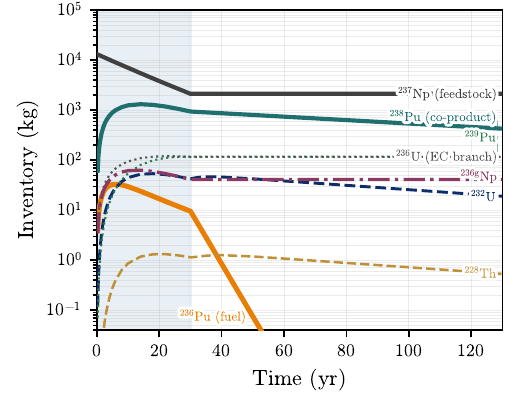}
\caption{NpO$_2$ blanket, no extraction}\label{fig:deptraj_a}
\end{subfigure}\hspace{0.02\textwidth}
\begin{subfigure}{0.42\textwidth}
\centering
\includegraphics[width=\linewidth]{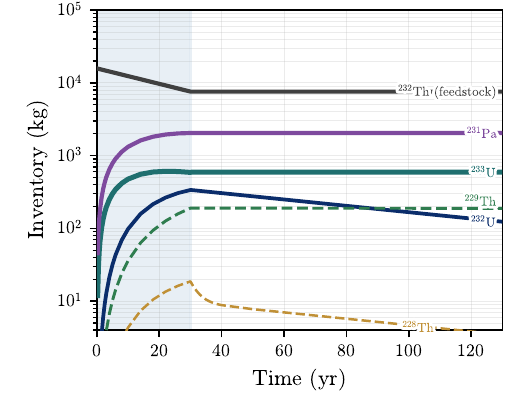}
\caption{${}^{232}$Th blanket, no extraction}\label{fig:deptraj_b}
\end{subfigure}\\[4pt]
\begin{subfigure}{0.42\textwidth}
\centering
\includegraphics[width=\linewidth]{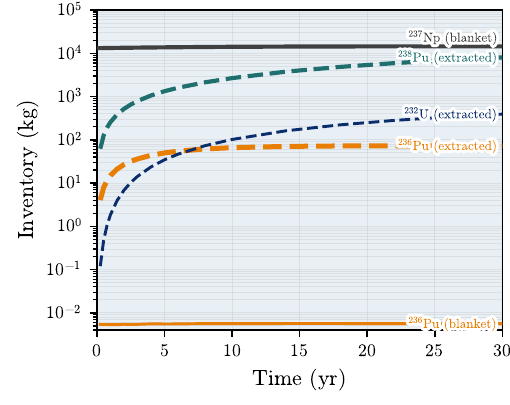}
\caption{NpO$_2$ blanket, with extraction}\label{fig:deptraj_c}
\end{subfigure}\hspace{0.02\textwidth}
\begin{subfigure}{0.42\textwidth}
\centering
\includegraphics[width=\linewidth]{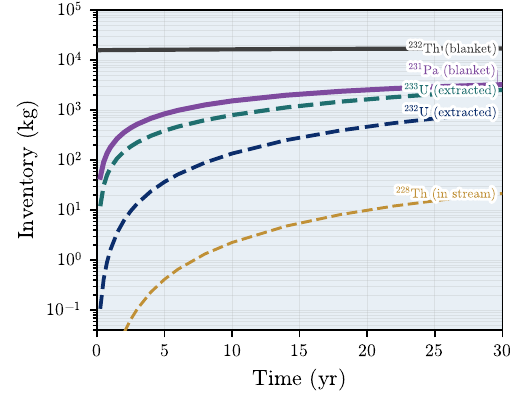}
\caption{${}^{232}$Th blanket, with extraction}\label{fig:deptraj_d}
\end{subfigure}\\[4pt]
\begin{subfigure}{0.42\textwidth}
\centering
\includegraphics[width=\linewidth]{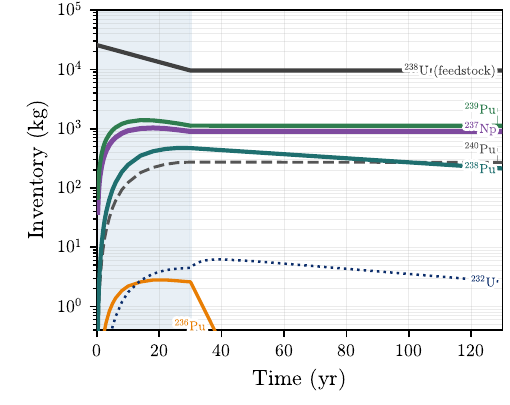}
\caption{${}^{238}$U blanket, no extraction (unlined)}\label{fig:deptraj_e}
\end{subfigure}\hspace{0.02\textwidth}
\begin{subfigure}{0.42\textwidth}
\centering
\includegraphics[width=\linewidth]{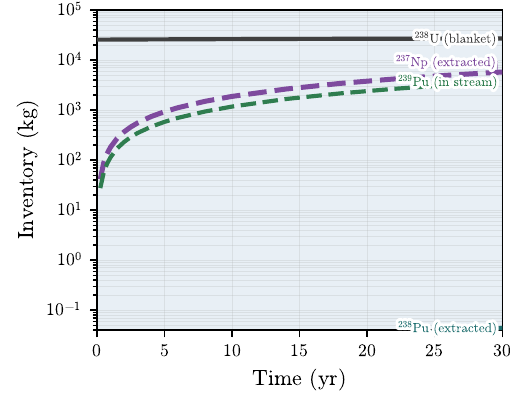}
\caption{${}^{238}$U blanket, with extraction}\label{fig:deptraj_f}
\end{subfigure}
\caption{Inventory of the blankets from OpenMC depletion simulations of 1.5 GW$_\mrm{fus}$. In (a), (b), and (e) the 30 yr irradiation is shaded and 100 yr of decay follow. In (c), (d), and (f) operation is steady for 30 yr with feedstock replenishment and continuous extraction, with solid curves the blanket inventories and dashed curves the extracted stream: plutonium from the ${}^{237}$Np blanket; uranium, actinium, radium, and the lead group from the ${}^{232}$Th blanket, whose protactinium stays in place; neptunium and plutonium from the ${}^{238}$U blanket.}
\label{fig:depletion_trajectories}
\end{figure*}

\Cref{tab:np_inventory,tab:th_inventory,tab:u238_inventory} give the full 30 yr inventories, blanket contents and extracted product streams for the irradiations. \Cref{tab:pa231_inventory} does the same for the ${}^{231}$Pa/D$_2$O layer of \Cref{tab:hybrid}, whose significant products are ${}^{232}$U and ${}^{228}$Th. It breeds very little ${}^{233}$U worth speaking of (47g over 30 years), so it holds none of the fissile concerns of the thorium blanket that feeds it. Sustained feeding keeps the fission rate up, so the ${}^{237}$Np blanket holds a reactor-scale fission isotope product rate led by ${}^{135}$Cs (1.7 t), ${}^{99}$Tc (0.9 t), ${}^{137}$Cs (0.85 t), and ${}^{90}$Sr (0.32 t); the last two are themselves $\beta^-$ battery fuels. 

\begin{table*}[!tb]
\centering
\caption{Nuclide inventory of the 5 mm NpO$_2$ blanket after 30 yr of continuous operation at 1.5 GW$_\mrm{fus}$, with feedstock replenishment and continuous plutonium extraction}
\label{tab:np_inventory}
\scriptsize
\setlength{\tabcolsep}{5pt}
\begin{tabular}{@{}lrrrrrr@{}}
\toprule
Nuclide & \thalf & Atoms & Mass & Produced & Activity (Ci) & Decay heat (W) \\
\midrule
\multicolumn{7}{l}{\emph{Blanket, at 30 yr}} \\
\midrule
\textbf{${}^{237}$Np} & \textbf{2.1\,Myr} & $3.69{\cdot}10^{28}$ & \textbf{14.5\,t} & \textbf{---} & $1.02{\cdot}10^{4}$ & $2.98{\cdot}10^{2}$ \\
${}^{236}$U & 23\,Myr & $7.77{\cdot}10^{26}$ & 305\,kg & --- & $1.97{\cdot}10^{1}$ & $5.24{\cdot}10^{-1}$ \\
${}^{236}$Np & 154\,kyr & $3.34{\cdot}10^{26}$ & 131\,kg & --- & $1.29{\cdot}10^{3}$ & $2.61{\cdot}10^{0}$ \\
${}^{235}$U & 704\,Myr & $1.02{\cdot}10^{26}$ & 39.9\,kg & --- & $8.61{\cdot}10^{-2}$ & $2.39{\cdot}10^{-3}$ \\
${}^{234}$U & 245\,kyr & $1.33{\cdot}10^{25}$ & 5.18\,kg & --- & $3.23{\cdot}10^{1}$ & $9.13{\cdot}10^{-1}$ \\
${}^{235}$Np & 1.1\,yr & $8.53{\cdot}10^{24}$ & 3.33\,kg & --- & $4.67{\cdot}10^{6}$ & $1.55{\cdot}10^{2}$ \\
${}^{238}$Np & 2.1\,d & $6.77{\cdot}10^{24}$ & 2.68\,kg & --- & $6.94{\cdot}10^{8}$ & $3.25{\cdot}10^{6}$ \\
${}^{233}$U & 159\,kyr & $5.50{\cdot}10^{23}$ & 213\,g & --- & $2.05{\cdot}10^{0}$ & $5.85{\cdot}10^{-2}$ \\
${}^{236\mathrm{m}}$Np & 22\,h & $3.19{\cdot}10^{23}$ & 125\,g & --- & $7.38{\cdot}10^{7}$ & $6.46{\cdot}10^{4}$ \\
\textbf{${}^{238}$Pu} & \textbf{88\,yr} & $2.57{\cdot}10^{23}$ & \textbf{101\,g} & \textbf{---} & $1.74{\cdot}10^{3}$ & $5.66{\cdot}10^{1}$ \\
${}^{237}$U & 6.8\,d & $1.51{\cdot}10^{23}$ & 59.3\,g & --- & $4.84{\cdot}10^{6}$ & $9.24{\cdot}10^{3}$ \\
\textbf{${}^{232}$U} & \textbf{69\,yr} & $4.10{\cdot}10^{22}$ & \textbf{15.8\,g} & \textbf{---} & $3.54{\cdot}10^{2}$ & $1.11{\cdot}10^{1}$ \\
\textbf{${}^{236}$Pu} & \textbf{2.9\,yr} & $1.43{\cdot}10^{22}$ & \textbf{5.62\,g} & \textbf{---} & $2.98{\cdot}10^{3}$ & $1.02{\cdot}10^{2}$ \\
${}^{238}$U & 4.5\,Gyr & $1.07{\cdot}10^{22}$ & 4.22\,g & --- & $1.42{\cdot}10^{-6}$ & $3.53{\cdot}10^{-8}$ \\
${}^{231}$Pa & 33\,kyr & $3.44{\cdot}10^{21}$ & 1.32\,g & --- & $6.23{\cdot}10^{-2}$ & $1.85{\cdot}10^{-3}$ \\
\midrule
${}^{16}$O & stable & $6.29{\cdot}10^{28}$ & 1.67\,t & --- & --- & --- \\
${}^{135}$Cs & 2.3\,Myr & $3.36{\cdot}10^{27}$ & 754\,kg & --- & $8.68{\cdot}10^{2}$ & $3.90{\cdot}10^{-1}$ \\
${}^{134}$Xe & stable & $2.95{\cdot}10^{27}$ & 657\,kg & --- & $3.02{\cdot}10^{-14}$ & --- \\
${}^{136}$Xe & stable & $2.89{\cdot}10^{27}$ & 652\,kg & --- & $1.04{\cdot}10^{-14}$ & --- \\
${}^{139}$La & stable & $2.82{\cdot}10^{27}$ & 651\,kg & --- & --- & --- \\
${}^{141}$Pr & stable & $2.77{\cdot}10^{27}$ & 648\,kg & --- & --- & --- \\
${}^{140}$Ce & stable & $2.69{\cdot}10^{27}$ & 625\,kg & --- & --- & --- \\
${}^{133}$Cs & stable & $2.81{\cdot}10^{27}$ & 620\,kg & --- & --- & --- \\
${}^{138}$Ba & stable & $2.44{\cdot}10^{27}$ & 559\,kg & --- & --- & --- \\
${}^{143}$Nd & stable & $2.20{\cdot}10^{27}$ & 521\,kg & --- & --- & --- \\
${}^{144}$Nd & stable & $2.12{\cdot}10^{27}$ & 506\,kg & --- & $5.48{\cdot}10^{-7}$ & $6.01{\cdot}10^{-9}$ \\
${}^{132}$Xe & stable & $2.25{\cdot}10^{27}$ & 492\,kg & --- & --- & --- \\
${}^{99}$Tc & 211\,kyr & $2.61{\cdot}10^{27}$ & 429\,kg & --- & $7.34{\cdot}10^{3}$ & $3.68{\cdot}10^{0}$ \\
${}^{137}$Cs & 30\,yr & $1.71{\cdot}10^{27}$ & 388\,kg & --- & $3.37{\cdot}10^{7}$ & $3.74{\cdot}10^{4}$ \\
${}^{93}$Zr & 1.6\,Myr & $2.26{\cdot}10^{27}$ & 349\,kg & --- & $8.34{\cdot}10^{2}$ & $9.49{\cdot}10^{-2}$ \\
${}^{107}$Pd & 6.5\,Myr & $9.23{\cdot}10^{26}$ & 164\,kg & --- & $8.43{\cdot}10^{1}$ & $4.65{\cdot}10^{-3}$ \\
${}^{90}$Sr & 29\,yr & $9.91{\cdot}10^{26}$ & 148\,kg & --- & $2.04{\cdot}10^{7}$ & $2.37{\cdot}10^{4}$ \\
${}^{129}$I & 16\,Myr & $4.72{\cdot}10^{26}$ & 101\,kg & --- & $1.78{\cdot}10^{1}$ & $9.05{\cdot}10^{-3}$ \\
${}^{151}$Sm & 90\,yr & $2.24{\cdot}10^{26}$ & 56.2\,kg & --- & $1.48{\cdot}10^{6}$ & $1.74{\cdot}10^{2}$ \\
${}^{85}$Kr & 11\,yr & $6.68{\cdot}10^{25}$ & 9.43\,kg & --- & $3.69{\cdot}10^{6}$ & $5.54{\cdot}10^{3}$ \\
\midrule
\multicolumn{7}{l}{\emph{Extracted stream, cumulative at 30 yr}} \\
\midrule
\textbf{${}^{238}$Pu} & \textbf{88\,yr} & $2.04{\cdot}10^{28}$ & \textbf{8.06\,t} & \textbf{9.02\,t} & $1.38{\cdot}10^{8}$ & $4.50{\cdot}10^{6}$ \\
${}^{234}$U & 245\,kyr & $2.43{\cdot}10^{27}$ & 943\,kg & 943\,kg & $5.87{\cdot}10^{3}$ & $1.66{\cdot}10^{2}$ \\
\textbf{${}^{232}$U} & \textbf{69\,yr} & $1.02{\cdot}10^{27}$ & \textbf{394\,kg} & \textbf{448\,kg} & $8.81{\cdot}10^{6}$ & $2.78{\cdot}10^{5}$ \\
\textbf{${}^{236}$Pu} & \textbf{2.9\,yr} & $1.86{\cdot}10^{26}$ & \textbf{73.1\,kg} & \textbf{528\,kg} & $3.87{\cdot}10^{7}$ & $1.32{\cdot}10^{6}$ \\
${}^{228}$Th & 1.9\,yr & $2.56{\cdot}10^{25}$ & 9.7\,kg & 53.6\,kg & $7.95{\cdot}10^{6}$ & $2.56{\cdot}10^{5}$ \\
\textbf{${}^{239}$Pu} & \textbf{24\,kyr} & $1.13{\cdot}10^{24}$ & \textbf{447\,g} & \textbf{447\,g} & $2.77{\cdot}10^{1}$ & $8.47{\cdot}10^{-1}$ \\
${}^{224}$Ra & 3.7\,d & $1.34{\cdot}10^{23}$ & 49.9\,g & 43.1\,kg & $7.95{\cdot}10^{6}$ & $2.68{\cdot}10^{5}$ \\
${}^{230}$Th & 75\,kyr & $6.84{\cdot}10^{22}$ & 26.1\,g & 26.1\,g & $5.38{\cdot}10^{-1}$ & $1.49{\cdot}10^{-2}$ \\
\textbf{${}^{237}$Np} & \textbf{2.1\,Myr} & $8.61{\cdot}10^{21}$ & \textbf{3.39\,g} & \textbf{3.39\,g} & $2.38{\cdot}10^{-3}$ & $6.94{\cdot}10^{-5}$ \\
\bottomrule
\end{tabular}
\end{table*}

\begin{table*}[!tb]
\centering
\caption{Nuclide inventory of the 5 mm ${}^{232}$Th blanket after 30 yr of continuous operation at 1.5 GW$_\mrm{fus}$, with feedstock replenishment and continuous extraction of uranium, actinium, radium, and lead, the protactinium left in blanket.}
\label{tab:th_inventory}
\scriptsize
\setlength{\tabcolsep}{5pt}
\begin{tabular}{@{}lrrrrrr@{}}
\toprule
Nuclide & \thalf & Atoms & Mass & Produced & Activity (Ci) & Decay heat (W) \\
\midrule
\multicolumn{7}{l}{\emph{Blanket, at 30 yr}} \\
\midrule
\textbf{${}^{232}$Th} & \textbf{14\,Gyr} & $4.45{\cdot}10^{28}$ & \textbf{17.1\,t} & \textbf{---} & $1.89{\cdot}10^{0}$ & $4.48{\cdot}10^{-2}$ \\
\textbf{${}^{231}$Pa} & \textbf{33\,kyr} & $8.43{\cdot}10^{27}$ & \textbf{3.24\,t} & \textbf{---} & $1.53{\cdot}10^{5}$ & $4.54{\cdot}10^{3}$ \\
${}^{230}$Th & 75\,kyr & $4.61{\cdot}10^{27}$ & 1.76\,t & --- & $3.63{\cdot}10^{4}$ & $1.01{\cdot}10^{3}$ \\
${}^{229}$Th & 7.3\,kyr & $6.55{\cdot}10^{26}$ & 249\,kg & --- & $5.30{\cdot}10^{4}$ & $1.61{\cdot}10^{3}$ \\
${}^{228}$Th & 1.9\,yr & $3.79{\cdot}10^{25}$ & 14.4\,kg & --- & $1.18{\cdot}10^{7}$ & $3.78{\cdot}10^{5}$ \\
${}^{233}$Pa & 27\,d & $2.47{\cdot}10^{25}$ & 9.56\,kg & --- & $1.99{\cdot}10^{8}$ & $4.94{\cdot}10^{5}$ \\
${}^{230}$Pa & 17\,d & $5.18{\cdot}10^{24}$ & 1.98\,kg & --- & $6.45{\cdot}10^{7}$ & $2.82{\cdot}10^{5}$ \\
${}^{231}$Th & 1.1\,d & $2.31{\cdot}10^{24}$ & 887\,g & --- & $4.72{\cdot}10^{8}$ & $4.66{\cdot}10^{5}$ \\
${}^{232}$Pa & 1.3\,d & $7.86{\cdot}10^{23}$ & 303\,g & --- & $1.29{\cdot}10^{8}$ & $8.37{\cdot}10^{5}$ \\
\textbf{${}^{233}$U} & \textbf{159\,kyr} & $7.35{\cdot}10^{22}$ & \textbf{28.4\,g} & \textbf{---} & $2.74{\cdot}10^{-1}$ & $7.81{\cdot}10^{-3}$ \\
${}^{227}$Th & 19\,d & $7.20{\cdot}10^{22}$ & 27.1\,g & --- & $8.36{\cdot}10^{5}$ & $3.00{\cdot}10^{4}$ \\
\textbf{${}^{232}$U} & \textbf{69\,yr} & $4.76{\cdot}10^{22}$ & \textbf{18.3\,g} & \textbf{---} & $4.10{\cdot}10^{2}$ & $1.29{\cdot}10^{1}$ \\
${}^{233}$Th & 22\,min & $1.42{\cdot}10^{22}$ & 5.51\,g & --- & $2.00{\cdot}10^{8}$ & $5.31{\cdot}10^{5}$ \\
${}^{229}$Pa & 1.5\,d & $9.66{\cdot}10^{21}$ & 3.67\,g & --- & $1.40{\cdot}10^{6}$ & $8.13{\cdot}10^{2}$ \\
${}^{224}$Ra & 3.7\,d & $4.26{\cdot}10^{21}$ & 1.58\,g & --- & $2.52{\cdot}10^{5}$ & $8.51{\cdot}10^{3}$ \\
\midrule
${}^{141}$Pr & stable & $5.91{\cdot}10^{26}$ & 138\,kg & --- & --- & --- \\
${}^{139}$La & stable & $5.97{\cdot}10^{26}$ & 138\,kg & --- & --- & --- \\
${}^{140}$Ce & stable & $5.88{\cdot}10^{26}$ & 137\,kg & --- & --- & --- \\
${}^{144}$Nd & stable & $4.95{\cdot}10^{26}$ & 118\,kg & --- & $1.28{\cdot}10^{-7}$ & $1.41{\cdot}10^{-9}$ \\
${}^{143}$Nd & stable & $4.84{\cdot}10^{26}$ & 115\,kg & --- & --- & --- \\
${}^{138}$Ba & stable & $4.89{\cdot}10^{26}$ & 112\,kg & --- & --- & --- \\
${}^{135}$Cs & 2.3\,Myr & $4.88{\cdot}10^{26}$ & 109\,kg & --- & $1.26{\cdot}10^{2}$ & $5.65{\cdot}10^{-2}$ \\
${}^{136}$Xe & stable & $4.77{\cdot}10^{26}$ & 108\,kg & --- & $1.72{\cdot}10^{-15}$ & --- \\
${}^{142}$Ce & stable & $4.39{\cdot}10^{26}$ & 104\,kg & --- & $5.21{\cdot}10^{-9}$ & --- \\
${}^{134}$Xe & stable & $4.24{\cdot}10^{26}$ & 94.3\,kg & --- & $4.34{\cdot}10^{-15}$ & --- \\
${}^{89}$Y & stable & $5.88{\cdot}10^{26}$ & 86.9\,kg & --- & --- & --- \\
${}^{92}$Zr & stable & $5.63{\cdot}10^{26}$ & 86\,kg & --- & --- & --- \\
${}^{93}$Zr & 1.6\,Myr & $5.39{\cdot}10^{26}$ & 83.3\,kg & --- & $1.99{\cdot}10^{2}$ & $2.26{\cdot}10^{-2}$ \\
${}^{137}$Cs & 30\,yr & $3.34{\cdot}10^{26}$ & 76\,kg & --- & $6.59{\cdot}10^{6}$ & $7.31{\cdot}10^{3}$ \\
${}^{90}$Sr & 29\,yr & $4.01{\cdot}10^{26}$ & 59.9\,kg & --- & $8.27{\cdot}10^{6}$ & $9.60{\cdot}10^{3}$ \\
${}^{99}$Tc & 211\,kyr & $1.92{\cdot}10^{26}$ & 31.6\,kg & --- & $5.40{\cdot}10^{2}$ & $2.71{\cdot}10^{-1}$ \\
${}^{129}$I & 16\,Myr & $5.54{\cdot}10^{25}$ & 11.9\,kg & --- & $2.10{\cdot}10^{0}$ & $1.06{\cdot}10^{-3}$ \\
${}^{85}$Kr & 11\,yr & $5.71{\cdot}10^{25}$ & 8.07\,kg & --- & $3.16{\cdot}10^{6}$ & $4.74{\cdot}10^{3}$ \\
${}^{151}$Sm & 90\,yr & $2.10{\cdot}10^{25}$ & 5.27\,kg & --- & $1.39{\cdot}10^{5}$ & $1.63{\cdot}10^{1}$ \\
${}^{107}$Pd & 6.5\,Myr & $5.12{\cdot}10^{24}$ & 910\,g & --- & $4.68{\cdot}10^{-1}$ & $2.58{\cdot}10^{-5}$ \\
\midrule
\multicolumn{7}{l}{\emph{Extracted stream, cumulative at 30 yr}} \\
\midrule
\textbf{${}^{233}$U} & \textbf{159\,kyr} & $6.55{\cdot}10^{27}$ & \textbf{2.53\,t} & \textbf{2.53\,t} & $2.44{\cdot}10^{4}$ & $6.96{\cdot}10^{2}$ \\
\textbf{${}^{232}$U} & \textbf{69\,yr} & $2.39{\cdot}10^{27}$ & \textbf{920\,kg} & \textbf{1.02\,t} & $2.06{\cdot}10^{7}$ & $6.48{\cdot}10^{5}$ \\
\textbf{${}^{210}$Pb} & \textbf{22\,yr} & $7.36{\cdot}10^{25}$ & \textbf{25.7\,kg} & \textbf{36.6\,kg} & $1.97{\cdot}10^{6}$ & $7.41{\cdot}10^{2}$ \\
${}^{228}$Th & 1.9\,yr & $5.70{\cdot}10^{25}$ & 21.6\,kg & 102\,kg & $1.77{\cdot}10^{7}$ & $5.69{\cdot}10^{5}$ \\
${}^{234}$U & 245\,kyr & $1.42{\cdot}10^{25}$ & 5.52\,kg & 5.52\,kg & $3.44{\cdot}10^{1}$ & $9.72{\cdot}10^{-1}$ \\
\textbf{${}^{227}$Ac} & \textbf{22\,yr} & $2.32{\cdot}10^{24}$ & \textbf{875\,g} & \textbf{1.21\,kg} & $6.33{\cdot}10^{4}$ & $3.15{\cdot}10^{1}$ \\
${}^{226}$Ra & 1.6\,kyr & $6.72{\cdot}10^{23}$ & 252\,g & 253\,g & $2.49{\cdot}10^{2}$ & $7.07{\cdot}10^{0}$ \\
${}^{230}$U & 21\,d & $4.80{\cdot}10^{23}$ & 183\,g & 40.2\,kg & $5.01{\cdot}10^{6}$ & $1.75{\cdot}10^{5}$ \\
${}^{224}$Ra & 3.7\,d & $4.92{\cdot}10^{23}$ & 183\,g & 141\,kg & $2.92{\cdot}10^{7}$ & $9.83{\cdot}10^{5}$ \\
${}^{229}$Th & 7.3\,kyr & $4.16{\cdot}10^{23}$ & 158\,g & 158\,g & $3.36{\cdot}10^{1}$ & $1.02{\cdot}10^{0}$ \\
${}^{223}$Ra & 11\,d & $4.70{\cdot}10^{22}$ & 17.4\,g & 4.64\,kg & $8.91{\cdot}10^{5}$ & $3.10{\cdot}10^{4}$ \\
${}^{227}$Th & 19\,d & $5.36{\cdot}10^{21}$ & 2.02\,g & 328\,g & $6.22{\cdot}10^{4}$ & $2.23{\cdot}10^{3}$ \\
${}^{225}$Ra & 15\,d & $3.61{\cdot}10^{21}$ & 1.35\,g & 264\,g & $5.26{\cdot}10^{4}$ & $4.03{\cdot}10^{1}$ \\
${}^{225}$Ac & 10\,d & $2.74{\cdot}10^{21}$ & 1.02\,g & 316\,g & $5.94{\cdot}10^{4}$ & $2.05{\cdot}10^{3}$ \\
\bottomrule
\end{tabular}

\end{table*}

\begin{table*}[!tb]
\centering
\caption{Nuclide inventory of the 5 mm ${}^{238}$U blanket after 30 yr of continuous operation at 1.5 GW$_\mrm{fus}$, with feedstock replenishment and continuous neptunium and plutonium extraction (${}^{6}$Li$_2$O liner), from the depletion simulation of \Cref{fig:depletion_trajectories}(f).}
\label{tab:u238_inventory}
\scriptsize
\setlength{\tabcolsep}{5pt}
\begin{tabular}{@{}lrrrrrr@{}}
\toprule
Nuclide & \thalf & Atoms & Mass & Produced & Activity (Ci) & Decay heat (W) \\
\midrule
\multicolumn{7}{l}{\emph{Blanket, at 30 yr}} \\
\midrule
\textbf{${}^{238}$U} & \textbf{4.5\,Gyr} & $6.83{\cdot}10^{28}$ & \textbf{27\,t} & \textbf{---} & $9.08{\cdot}10^{0}$ & $2.26{\cdot}10^{-1}$ \\
${}^{236}$U & 23\,Myr & $3.69{\cdot}10^{27}$ & 1.44\,t & --- & $9.34{\cdot}10^{1}$ & $2.48{\cdot}10^{0}$ \\
${}^{235}$U & 704\,Myr & $1.76{\cdot}10^{26}$ & 68.8\,kg & --- & $1.49{\cdot}10^{-1}$ & $4.12{\cdot}10^{-3}$ \\
${}^{234}$U & 245\,kyr & $4.73{\cdot}10^{25}$ & 18.4\,kg & --- & $1.14{\cdot}10^{2}$ & $3.24{\cdot}10^{0}$ \\
${}^{237}$U & 6.8\,d & $1.35{\cdot}10^{25}$ & 5.33\,kg & --- & $4.35{\cdot}10^{8}$ & $8.30{\cdot}10^{5}$ \\
${}^{233}$U & 159\,kyr & $1.14{\cdot}10^{24}$ & 439\,g & --- & $4.24{\cdot}10^{0}$ & $1.21{\cdot}10^{-1}$ \\
\textbf{${}^{237}$Np} & \textbf{2.1\,Myr} & $1.61{\cdot}10^{23}$ & \textbf{63.3\,g} & \textbf{---} & $4.45{\cdot}10^{-2}$ & $1.30{\cdot}10^{-3}$ \\
${}^{239}$Np & 2.4\,d & $9.92{\cdot}10^{22}$ & 39.4\,g & --- & $9.13{\cdot}10^{6}$ & $2.40{\cdot}10^{4}$ \\
${}^{239}$U & 23\,min & $2.08{\cdot}10^{22}$ & 8.26\,g & --- & $2.77{\cdot}10^{8}$ & $7.81{\cdot}10^{5}$ \\
${}^{232}$U & 69\,yr & $1.51{\cdot}10^{22}$ & 5.83\,g & --- & $1.30{\cdot}10^{2}$ & $4.11{\cdot}10^{0}$ \\
${}^{239}$Pu & 24\,kyr & $3.38{\cdot}10^{21}$ & 1.34\,g & --- & $8.32{\cdot}10^{-2}$ & $2.54{\cdot}10^{-3}$ \\
\midrule
${}^{141}$Pr & stable & $1.73{\cdot}10^{27}$ & 406\,kg & --- & --- & --- \\
${}^{139}$La & stable & $1.73{\cdot}10^{27}$ & 399\,kg & --- & --- & --- \\
${}^{143}$Nd & stable & $1.67{\cdot}10^{27}$ & 397\,kg & --- & --- & --- \\
${}^{140}$Ce & stable & $1.66{\cdot}10^{27}$ & 387\,kg & --- & --- & --- \\
${}^{144}$Nd & stable & $1.53{\cdot}10^{27}$ & 366\,kg & --- & $3.97{\cdot}10^{-7}$ & $4.35{\cdot}10^{-9}$ \\
${}^{145}$Nd & stable & $1.37{\cdot}10^{27}$ & 330\,kg & --- & --- & --- \\
${}^{138}$Ba & stable & $1.40{\cdot}10^{27}$ & 320\,kg & --- & --- & --- \\
${}^{142}$Ce & stable & $1.28{\cdot}10^{27}$ & 301\,kg & --- & $1.52{\cdot}10^{-8}$ & --- \\
${}^{135}$Cs & 2.3\,Myr & $1.29{\cdot}10^{27}$ & 288\,kg & --- & $3.32{\cdot}10^{2}$ & $1.49{\cdot}10^{-1}$ \\
${}^{133}$Cs & stable & $1.30{\cdot}10^{27}$ & 287\,kg & --- & --- & --- \\
${}^{146}$Nd & stable & $1.16{\cdot}10^{27}$ & 281\,kg & --- & --- & --- \\
${}^{136}$Xe & stable & $1.24{\cdot}10^{27}$ & 280\,kg & --- & $4.47{\cdot}10^{-15}$ & --- \\
${}^{93}$Zr & 1.6\,Myr & $1.53{\cdot}10^{27}$ & 237\,kg & --- & $5.65{\cdot}10^{2}$ & $6.43{\cdot}10^{-2}$ \\
${}^{99}$Tc & 211\,kyr & $1.35{\cdot}10^{27}$ & 222\,kg & --- & $3.79{\cdot}10^{3}$ & $1.90{\cdot}10^{0}$ \\
${}^{137}$Cs & 30\,yr & $8.95{\cdot}10^{26}$ & 204\,kg & --- & $1.77{\cdot}10^{7}$ & $1.96{\cdot}10^{4}$ \\
${}^{90}$Sr & 29\,yr & $8.76{\cdot}10^{26}$ & 131\,kg & --- & $1.81{\cdot}10^{7}$ & $2.10{\cdot}10^{4}$ \\
${}^{151}$Sm & 90\,yr & $4.17{\cdot}10^{25}$ & 10.4\,kg & --- & $2.75{\cdot}10^{5}$ & $3.23{\cdot}10^{1}$ \\
${}^{107}$Pd & 6.5\,Myr & $3.56{\cdot}10^{25}$ & 6.32\,kg & --- & $3.25{\cdot}10^{0}$ & $1.79{\cdot}10^{-4}$ \\
${}^{129}$I & 16\,Myr & $2.30{\cdot}10^{25}$ & 4.93\,kg & --- & $8.70{\cdot}10^{-1}$ & $4.42{\cdot}10^{-4}$ \\
${}^{85}$Kr & 11\,yr & $1.64{\cdot}10^{25}$ & 2.31\,kg & --- & $9.06{\cdot}10^{5}$ & $1.36{\cdot}10^{3}$ \\
\midrule
\multicolumn{7}{l}{\emph{Extracted stream, cumulative at 30 yr}} \\
\midrule
\textbf{${}^{237}$Np} & \textbf{2.1\,Myr} & $1.47{\cdot}10^{28}$ & \textbf{5.79\,t} & \textbf{5.79\,t} & $4.07{\cdot}10^{3}$ & $1.19{\cdot}10^{2}$ \\
${}^{239}$Pu & 24\,kyr & $9.31{\cdot}10^{27}$ & 3.69\,t & 3.7\,t & $2.29{\cdot}10^{5}$ & $7.00{\cdot}10^{3}$ \\
${}^{235}$U & 704\,Myr & $3.94{\cdot}10^{24}$ & 1.54\,kg & 1.54\,kg & $3.32{\cdot}10^{-3}$ & $9.20{\cdot}10^{-5}$ \\
${}^{239}$Np & 2.4\,d & $2.91{\cdot}10^{24}$ & 1.16\,kg & 3.55\,t & $2.68{\cdot}10^{8}$ & $7.05{\cdot}10^{5}$ \\
\textbf{${}^{238}$Pu} & \textbf{88\,yr} & $1.12{\cdot}10^{23}$ & \textbf{44.1\,g} & \textbf{49.4\,g} & $7.56{\cdot}10^{2}$ & $2.46{\cdot}10^{1}$ \\
${}^{233}$U & 159\,kyr & $6.99{\cdot}10^{22}$ & 27\,g & 27.1\,g & $2.61{\cdot}10^{-1}$ & $7.43{\cdot}10^{-3}$ \\
${}^{240}$Pu & 6.6\,kyr & $4.37{\cdot}10^{22}$ & 17.4\,g & 17.4\,g & $3.95{\cdot}10^{0}$ & $1.21{\cdot}10^{-1}$ \\
${}^{234}$U & 245\,kyr & $1.34{\cdot}10^{22}$ & 5.22\,g & 5.22\,g & $3.25{\cdot}10^{-2}$ & $9.19{\cdot}10^{-4}$ \\
${}^{236}$U & 23\,Myr & $4.87{\cdot}10^{21}$ & 1.91\,g & 1.91\,g & $1.23{\cdot}10^{-4}$ & $3.28{\cdot}10^{-6}$ \\
${}^{236}$Np & 154\,kyr & $4.70{\cdot}10^{21}$ & 1.84\,g & 1.84\,g & $1.81{\cdot}10^{-2}$ & $3.67{\cdot}10^{-5}$ \\
${}^{232}$U & 69\,yr & $4.35{\cdot}10^{21}$ & 1.68\,g & 1.91\,g & $3.75{\cdot}10^{1}$ & $1.18{\cdot}10^{0}$ \\
\bottomrule
\end{tabular}
\end{table*}

\begin{table*}[!tb]
\centering
\caption{Nuclide inventory of the 5 mm ${}^{231}$Pa/D$_2$O converter after 30 yr of continuous operation at 1.5 GW$_\mrm{fus}$, with protactinium replenished and uranium continuously extracted. Masses are full torus. The channel holds 0.33 t of ${}^{231}$Pa, under two months of one ${}^{232}$Th blanket's output.}
\label{tab:pa231_inventory}
\scriptsize
\setlength{\tabcolsep}{5pt}
\begin{tabular}{@{}lrrrrrr@{}}
\toprule
Nuclide & \thalf & Atoms & Mass & Produced & Activity (Ci) & Decay heat (W) \\
\midrule
\multicolumn{7}{l}{\emph{Blanket, at 30 yr}} \\
\midrule
\textbf{${}^{231}$Pa} & \textbf{33\,kyr} & $9.12{\cdot}10^{26}$ & \textbf{350\,kg} & \textbf{---} & $1.65{\cdot}10^{4}$ & $4.91{\cdot}10^{2}$ \\
${}^{230}$Th & 75\,kyr & $1.55{\cdot}10^{26}$ & 59.1\,kg & --- & $1.22{\cdot}10^{3}$ & $3.38{\cdot}10^{1}$ \\
${}^{229}$Th & 7.3\,kyr & $2.55{\cdot}10^{25}$ & 9.7\,kg & --- & $2.06{\cdot}10^{3}$ & $6.25{\cdot}10^{1}$ \\
\textbf{${}^{228}$Th} & \textbf{1.9\,yr} & $1.43{\cdot}10^{24}$ & \textbf{543\,g} & \textbf{---} & $4.45{\cdot}10^{5}$ & $1.43{\cdot}10^{4}$ \\
${}^{232}$Th & 14\,Gyr & $1.12{\cdot}10^{24}$ & 433\,g & --- & $4.76{\cdot}10^{-5}$ & $1.13{\cdot}10^{-6}$ \\
${}^{230}$Pa & 17\,d & $5.82{\cdot}10^{23}$ & 222\,g & --- & $7.25{\cdot}10^{6}$ & $3.17{\cdot}10^{4}$ \\
${}^{232}$Pa & 1.3\,d & $9.62{\cdot}10^{22}$ & 37\,g & --- & $1.58{\cdot}10^{7}$ & $1.02{\cdot}10^{5}$ \\
\textbf{${}^{232}$U} & \textbf{69\,yr} & $5.83{\cdot}10^{21}$ & \textbf{2.24\,g} & \textbf{---} & $5.02{\cdot}10^{1}$ & $1.58{\cdot}10^{0}$ \\
${}^{231}$Th & 1.1\,d & $5.84{\cdot}10^{21}$ & 2.24\,g & --- & $1.19{\cdot}10^{6}$ & $1.18{\cdot}10^{3}$ \\
${}^{227}$Th & 19\,d & $2.83{\cdot}10^{21}$ & 1.07\,g & --- & $3.28{\cdot}10^{4}$ & $1.18{\cdot}10^{3}$ \\
\midrule
${}^{16}$O & stable & $4.01{\cdot}10^{28}$ & 1.07\,t & --- & --- & --- \\
${}^{2}$H & stable & $8.24{\cdot}10^{28}$ & 274\,kg & --- & --- & --- \\
${}^{13}$C & stable & $1.65{\cdot}10^{27}$ & 35.6\,kg & --- & --- & --- \\
${}^{12}$C & stable & $1.05{\cdot}10^{27}$ & 20.9\,kg & --- & --- & --- \\
${}^{139}$La & stable & $3.95{\cdot}10^{25}$ & 9.12\,kg & --- & --- & --- \\
${}^{141}$Pr & stable & $3.74{\cdot}10^{25}$ & 8.75\,kg & --- & --- & --- \\
${}^{140}$Ce & stable & $3.55{\cdot}10^{25}$ & 8.24\,kg & --- & --- & --- \\
${}^{135}$Cs & 2.3\,Myr & $3.51{\cdot}10^{25}$ & 7.86\,kg & --- & $9.05{\cdot}10^{0}$ & $4.06{\cdot}10^{-3}$ \\
${}^{138}$Ba & stable & $3.42{\cdot}10^{25}$ & 7.83\,kg & --- & --- & --- \\
${}^{136}$Xe & stable & $3.44{\cdot}10^{25}$ & 7.77\,kg & --- & $1.24{\cdot}10^{-16}$ & --- \\
${}^{134}$Xe & stable & $2.98{\cdot}10^{25}$ & 6.63\,kg & --- & $3.05{\cdot}10^{-16}$ & --- \\
${}^{142}$Ce & stable & $2.71{\cdot}10^{25}$ & 6.38\,kg & --- & $3.21{\cdot}10^{-10}$ & --- \\
${}^{93}$Zr & 1.6\,Myr & $3.55{\cdot}10^{25}$ & 5.49\,kg & --- & $1.31{\cdot}10^{1}$ & $1.49{\cdot}10^{-3}$ \\
${}^{137}$Cs & 30\,yr & $2.29{\cdot}10^{25}$ & 5.22\,kg & --- & $4.53{\cdot}10^{5}$ & $5.02{\cdot}10^{2}$ \\
${}^{90}$Sr & 29\,yr & $2.32{\cdot}10^{25}$ & 3.47\,kg & --- & $4.79{\cdot}10^{5}$ & $5.56{\cdot}10^{2}$ \\
${}^{99}$Tc & 211\,kyr & $1.17{\cdot}10^{25}$ & 1.92\,kg & --- & $3.29{\cdot}10^{1}$ & $1.65{\cdot}10^{-2}$ \\
${}^{129}$I & 16\,Myr & $5.98{\cdot}10^{24}$ & 1.28\,kg & --- & $2.26{\cdot}10^{-1}$ & $1.15{\cdot}10^{-4}$ \\
${}^{85}$Kr & 11\,yr & $3.71{\cdot}10^{24}$ & 524\,g & --- & $2.05{\cdot}10^{5}$ & $3.08{\cdot}10^{2}$ \\
${}^{151}$Sm & 90\,yr & $1.09{\cdot}10^{24}$ & 272\,g & --- & $7.17{\cdot}10^{3}$ & $8.41{\cdot}10^{-1}$ \\
${}^{107}$Pd & 6.5\,Myr & $4.30{\cdot}10^{23}$ & 76.4\,g & --- & $3.93{\cdot}10^{-2}$ & $2.16{\cdot}10^{-6}$ \\
\midrule
\multicolumn{7}{l}{\emph{Extracted stream, cumulative at 30 yr}} \\
\midrule
\textbf{${}^{232}$U} & \textbf{69\,yr} & $4.61{\cdot}10^{26}$ & \textbf{178\,kg} & \textbf{205\,kg} & $3.97{\cdot}10^{6}$ & $1.25{\cdot}10^{5}$ \\
\textbf{${}^{228}$Th} & \textbf{1.9\,yr} & $1.17{\cdot}10^{25}$ & \textbf{4.43\,kg} & \textbf{27.1\,kg} & $3.64{\cdot}10^{6}$ & $1.17{\cdot}10^{5}$ \\
\textbf{${}^{210}$Pb} & \textbf{22\,yr} & $1.23{\cdot}10^{25}$ & \textbf{4.27\,kg} & \textbf{6.73\,kg} & $3.28{\cdot}10^{5}$ & $1.23{\cdot}10^{2}$ \\
${}^{227}$Ac & 22\,yr & $3.67{\cdot}10^{23}$ & 138\,g & 214\,g & $1.00{\cdot}10^{4}$ & $4.99{\cdot}10^{0}$ \\
\textbf{${}^{233}$U} & \textbf{159\,kyr} & $1.21{\cdot}10^{23}$ & \textbf{46.6\,g} & \textbf{46.6\,g} & $4.49{\cdot}10^{-1}$ & $1.28{\cdot}10^{-2}$ \\
${}^{224}$Ra & 3.7\,d & $6.87{\cdot}10^{22}$ & 25.5\,g & 24.8\,kg & $4.07{\cdot}10^{6}$ & $1.37{\cdot}10^{5}$ \\
${}^{230}$U & 21\,d & $5.41{\cdot}10^{22}$ & 20.6\,g & 7.36\,kg & $5.64{\cdot}10^{5}$ & $1.97{\cdot}10^{4}$ \\
${}^{226}$Ra & 1.6\,kyr & $2.52{\cdot}10^{22}$ & 9.45\,g & 9.49\,g & $9.34{\cdot}10^{0}$ & $2.65{\cdot}10^{-1}$ \\
\bottomrule
\end{tabular}

\end{table*}

\section{Mars logistics with fresh fuel} \label{app:mars}

\begin{figure}[!tb]
\centering
\includegraphics[width=\columnwidth]{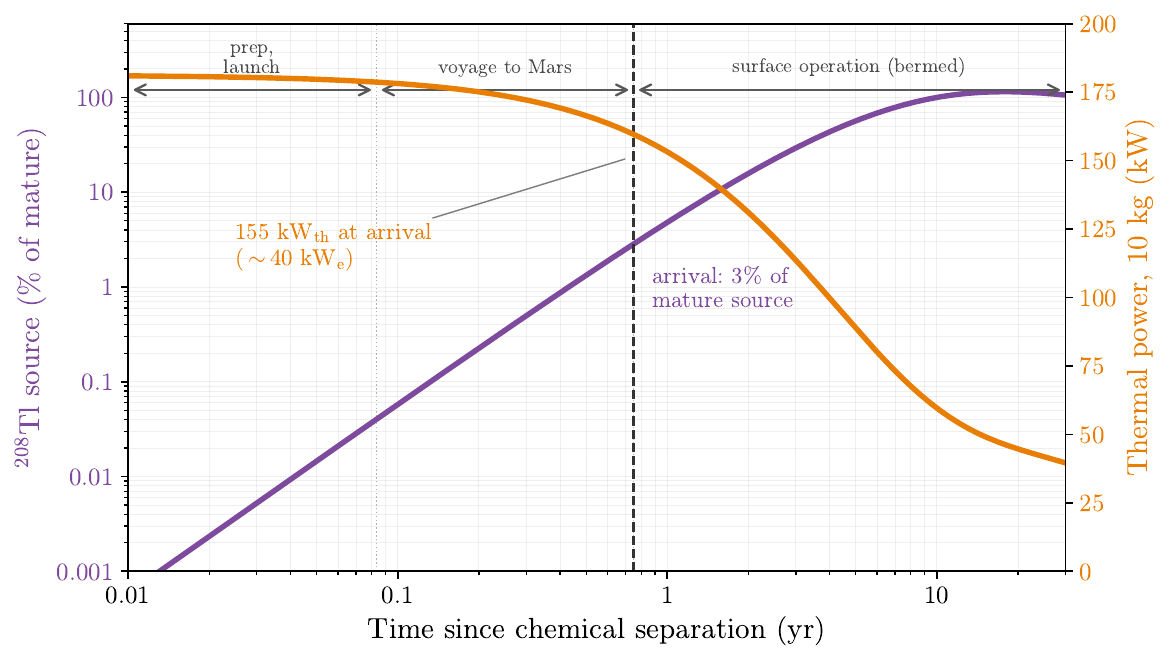}
\caption{Mars base powered by shipped fresh ${}^{236}$Pu. The ${}^{208}$Tl gamma source (purple, \% of mature strength) and the 10 kg shipment's thermal power (orange) versus time since chemical separation. Dashed line is arrival at 9 months.}
\label{fig:mars_base}
\end{figure}

For the ${}^{236}$Pu chain, the ${}^{208}$Tl gamma source's strong time dependence supports a relatively clean transfer of ${}^{236}$Pu on space missions such as to Mars (\Cref{fig:mars_base}). The ${}^{208}$Tl output grows in through the 69 yr ${}^{232}$U and 1.9 yr ${}^{228}$Th bottlenecks, so freshly separated ${}^{236}$Pu emits only 3\% of its mature gamma output after a 9 month voyage. A 10 kg shipment (181 kW$_\mrm{th}$ fresh) is therefore quietest when people are nearest during spaceflight: $\sim$20 mSv/h at 1 m during launch-site handling and 28 Sv/h on arrival, or $\sim$3 Sv/h with the radon vent of \Cref{sec:radon} run through the voyage. Shielding the matured source to the 10 $\mu$Sv/h occupational limit (astronauts already receive $\sim$0.7 mSv/day from cosmic rays~\cite{hassler2014mars}) requires attenuation of $10^{8}$, about 3 m of Martian soil put in place once the fuel arrived on Mars. The buried fuel produces 155 kW$_\mrm{th}$ on arrival and 100 kW$_\mrm{th}$ at year 3, the scale studied for crewed surface power~\cite{Gibson2017}.

\section{Complete decay data for the six chains} \label{app:decaydata}
For completeness, \Cref{tab:chains_full} lists every nuclide of the six chains discussed in this work, with its half-life, decay mode and branching, alpha and mean beta energies, main gamma lines, and spontaneous-fission neutron yield per decay.
\begin{table*}[!tb]
\centering
\footnotesize
\caption{Complete decay data for the six chains discussed in this work. $E_\alpha$ is the main alpha energy and $\bar{E}_\beta$ the mean beta kinetic energy (the antineutrino carries the balance of the beta $Q$); $E_\gamma$ lists the up to five strongest gamma lines of each decay branch as energy in keV with absolute intensity in percent per decay of the parent, however weak (ENSDF evaluations via the IAEA Live Chart; atomic X-rays excluded); $n_\mrm{SF}$ is the spontaneous-fission neutron yield per decay (SF branching $\times$ $\bar\nu$). Branching $b$ is shown in parentheses where a nuclide has more than one mode. Half-lives, modes, and branchings are from NUBASE2020~\cite{Kondev2021NUBASE}.}
\label{tab:chains_full}
\begin{tabular}{llcccp{8.9cm}c}
\toprule
Nuclide & Mode ($b$) & $\thalf$ & $E_\alpha$ & $\bar{E}_\beta$ & $E_\gamma$: keV (\%) & $n_\mrm{SF}$ \\
 & & & (MeV) & (MeV) & & (/dec.) \\
\midrule
\multicolumn{7}{l}{\textit{${}^{236}$Pu chain (4n $\rightarrow$ ${}^{208}$Pb)}} \\[1pt]
${}^{236}$Pu & $\alpha$ & 2.86 yr & 5.77 &  & 47.6 (0.065), 109 (0.022), 166 ($7{\times}10^{-4}$), 644 ($2{\times}10^{-4}$), 516 ($2{\times}10^{-4}$) & $2.9{\times}10^{-9}$ \\
${}^{232}$U & $\alpha$ & 68.9 yr & 5.32 &  & 57.8 (0.2), 129 (0.068), 270 ($3{\times}10^{-3}$), 328 ($3{\times}10^{-3}$), 332 ($5{\times}10^{-5}$) &  \\
${}^{228}$Th & $\alpha$ & 1.91 yr & 5.42 &  & 84.4 (1.2), 216 (0.25), 132 (0.13), 166 (0.1), 206 (0.019) &  \\
${}^{224}$Ra & $\alpha$ & 3.66 d & 5.69 &  & 241 (4.1), 293 ($6{\times}10^{-3}$), 646 ($5{\times}10^{-3}$), 422 ($3{\times}10^{-3}$), 404 ($2{\times}10^{-3}$) &  \\
${}^{220}$Rn & $\alpha$ & 55.6 s & 6.29 &  & 550 (0.11) &  \\
${}^{216}$Po & $\alpha$ & 0.145 s & 6.78 &  & 805 ($2{\times}10^{-3}$) &  \\
${}^{212}$Pb & $\betam$ & 10.6 h &  & 0.10 & 239 (43.6), 300 (3.3), 115 (0.6), 177 (0.052), 415 (0.013) &  \\
${}^{212}$Bi & $\alpha$ (0.36) & 60.6 min & 6.05 &  & 39.9 (1.1), 453 (0.36), 288 (0.34), 328 (0.12), 473 (0.05) &  \\
${}^{212}$Bi & $\betam$ (0.64) & 60.6 min &  & 0.49 & 727 (6.7), 1620 (1.5), 785 (1.1), 1079 (0.56), 893 (0.38) &  \\
${}^{208}$Tl & $\betam$ (0.36) & 3.05 min &  & 0.56 & 2615 (99.8), 583 (85), 511 (22.6), 861 (12.5), 277 (6.6) &  \\
${}^{212}$Po & $\alpha$ (0.64) & 0.3 $\mu$s & 8.78 &  & none &  \\
\midrule
\multicolumn{7}{l}{\textit{${}^{227}$Ac chain (4n+3 $\rightarrow$ ${}^{207}$Pb)}} \\[1pt]
${}^{227}$Ac & $\betam$ (0.986) & 21.8 yr &  & 0.010 & 37.9 (0.049), 28.6 (0.042), 24.5 (0.028), 15.2 ($6{\times}10^{-4}$), 9.3 ($1{\times}10^{-4}$) &  \\
${}^{227}$Ac & $\alpha$ (0.014) & 21.8 yr & 4.95 &  & 99.6 ($6{\times}10^{-3}$), 160 ($5{\times}10^{-3}$), 69.3 ($4{\times}10^{-3}$), 86.7 ($3{\times}10^{-3}$), 148 ($3{\times}10^{-3}$) &  \\
${}^{227}$Th & $\alpha$ & 18.7 d & 6.04 &  & 236 (12.9), 50.1 (8.4), 256 (7), 330 (2.9), 300 (2.2) &  \\
${}^{223}$Fr & $\betam$ & 22 min &  & 0.37 & 50.1 (33.9), 79.7 (8.7), 235 (3), 49.8 (2.8), 20.3 (1.6) &  \\
${}^{223}$Ra & $\alpha$ & 11.4 d & 5.72 &  & 269 (13.3), 154 (6), 324 (3.6), 144 (3.5), 338 (2.6) &  \\
${}^{219}$Rn & $\alpha$ & 3.96 s & 6.82 &  & 271 (10.8), 402 (6.6), 131 (0.13), 294 (0.073), 518 (0.044) &  \\
${}^{215}$Po & $\alpha$ & 1.78 ms & 7.39 &  & none &  \\
${}^{211}$Pb & $\betam$ & 36.1 min &  & 0.45 & 405 (3.8), 832 (3.5), 427 (1.8), 767 (0.62), 705 (0.46) &  \\
${}^{211}$Bi & $\alpha$ (0.997) & 2.14 min & 6.62 &  & 351 (13) &  \\
${}^{207}$Tl & $\betam$ & 4.77 min &  & 0.49 & 898 (0.26), 570 ($2{\times}10^{-3}$), 328 ($1{\times}10^{-3}$) &  \\
${}^{211}$Po & $\alpha$ (0.003) & 0.52 s & 7.45 &  & 898 (0.55), 570 (0.54), 328 ($3{\times}10^{-3}$), 1064 ($7{\times}10^{-4}$) &  \\
\midrule
\multicolumn{7}{l}{\textit{${}^{210}$Pb chain (4n+2 tail $\rightarrow$ ${}^{206}$Pb)}} \\[1pt]
${}^{210}$Pb & $\betam$ & 22.2 yr &  & 0.004 & 46.5 (4.2) &  \\
${}^{210}$Bi & $\betam$ & 5.01 d &  & 0.32 & none &  \\
${}^{210}$Po & $\alpha$ & 138 d & 5.30 &  & 803 ($1{\times}10^{-3}$) &  \\
\midrule
\multicolumn{7}{l}{\textit{${}^{242\mrm{m}}$Am chain (isomer)}} \\[1pt]
${}^{242\mrm{m}}$Am & IT (0.995) & 141 yr &  &  & 48.6 ($1{\times}10^{-4}$) &  \\
${}^{242\mrm{m}}$Am & $\alpha$ (0.005) & 141 yr & 5.20 &  & 49.4 (0.13), 86.7 (0.023), 110 (0.02), 163 (0.015) &  \\
${}^{242}$Am & $\betam$ (0.827) & 16.0 h &  & 0.19 & 42.1 (0.036) &  \\
${}^{242}$Am & EC (0.173) & 16.0 h &  &  & 44.5 (0.015) &  \\
${}^{242}$Cm & $\alpha$ & 162.8 d & 6.11 &  & 44.1 (0.033), 102 ($3{\times}10^{-3}$), 157 ($1{\times}10^{-3}$), 561 ($1{\times}10^{-4}$), 605 ($1{\times}10^{-4}$) & $1.6{\times}10^{-7}$ \\
${}^{238}$Pu & $\alpha$ & 87.7 yr & 5.50 &  & 43.5 (0.039), 99.9 ($7{\times}10^{-3}$), 153 ($9{\times}10^{-4}$), 766 ($2{\times}10^{-5}$), 743 ($5{\times}10^{-6}$) & $4.2{\times}10^{-9}$ \\
${}^{234}$U$^{a}$ & $\alpha$ & 245 kyr & 4.77 &  & 53.2 (0.12), 121 (0.034), 455 ($2{\times}10^{-5}$), 508 ($2{\times}10^{-5}$), 582 ($1{\times}10^{-5}$) &  \\
${}^{242}$Pu$^{a}$ & $\alpha$ & 375 kyr & 4.90 &  & 44.9 (0.038), 104 (0.025), 159 ($2{\times}10^{-4}$) & $1.2{\times}10^{-5}$ \\
\midrule
\multicolumn{7}{l}{\textit{${}^{238}$Pu (reference)}} \\[1pt]
${}^{238}$Pu & $\alpha$ & 87.7 yr & 5.50 &  & 43.5 (0.039), 99.9 ($7{\times}10^{-3}$), 153 ($9{\times}10^{-4}$), 766 ($2{\times}10^{-5}$), 743 ($5{\times}10^{-6}$) & $4.2{\times}10^{-9}$ \\
${}^{234}$U$^{a}$ & $\alpha$ & 245 kyr & 4.77 &  & 53.2 (0.12), 121 (0.034), 455 ($2{\times}10^{-5}$), 508 ($2{\times}10^{-5}$), 582 ($1{\times}10^{-5}$) &  \\
\midrule
\multicolumn{7}{l}{\textit{${}^{241}$Am (reference)}} \\[1pt]
${}^{241}$Am & $\alpha$ & 432 yr & 5.49 &  & 59.5 (35.9), 26.3 (2.3), 33.2 (0.13), 43.4 (0.073), 99.0 (0.02) & $1.2{\times}10^{-11}$ \\
${}^{237}$Np$^{a}$ & $\alpha$ & 2.14 Myr & 4.79 &  & 29.4 (14.5), 86.5 (12.2), 94.7 (0.59), 143 (0.42), 57.1 (0.36) &  \\
\bottomrule
\multicolumn{7}{l}{\footnotesize $^{a}$Long-lived terminal daughter; effectively stable on battery timescales.}
\end{tabular}
\end{table*}

\bibliography{references}

\end{document}